\documentclass[preprint]{aastex}

\newcommand{\tcc}	{t_{\rm cc}}
\newcommand{\calr}	{{\cal R}}
\newcommand{\mct}	{\multicolumn{2}{c}}
\newcommand{\beq}	{\begin{equation}}
\newcommand{\eeq}	{\end{equation}}

\newcommand{\miso}	{M_{\rm iso}}

\shorttitle{Interstellar Shock-Cloud Interaction}

\begin{document}

\title{On the Hydrodynamic Interaction of Shock Waves 
with Interstellar Clouds.  II. The Effect of Smooth Cloud 
Boundaries on Cloud Destruction and Cloud Turbulence}

\author{Fumitaka Nakamura\altaffilmark{1,2},
Christopher F. McKee\altaffilmark{3,4}, Richard I. Klein\altaffilmark{3,5}, 
Robert T. Fisher\altaffilmark{6}}

\altaffiltext{1}{Faculty of Education and Human Sciences, Niigata University,
8050 Ikarashi-2, Niigata 950-2181, Japan}
\altaffiltext{2}{Division of Theoretical
Astrophysics,
National Astronomical Observatory, Mitaka, Tokyo 181-8588, Japan}
\altaffiltext{3}{Department of Astronomy, 
University of California, Berkeley, CA 94720}
\altaffiltext{4}{Department of Physics, 
University of California, Berkeley, CA 94720}
\altaffiltext{5}{Lawrence Livermore National Laboratory, P.O. Box 808,
Livermore, CA 94550}
\altaffiltext{6}{The ASC FLASH Center, The University of Chicago, 5640
S. Ellis, Chicago, IL 60637}

\begin{abstract}
The effect of smooth cloud boundaries on
the interaction of steady planar shock waves with interstellar clouds 
is studied using a high-resolution local adaptive mesh refinement 
technique with a second-order accurate axisymmetric 
Godunov hydrodynamic scheme.
A three-dimensional calculation is also done to confirm
the results of the two-dimensional ones.
We consider an initially spherical cloud whose 
density distribution is flat near the cloud center 
and has a power-law profile in the cloud envelope.
Our model can be specified by three model parameters: the power law
index of the density profile ($n$), the density contrast ($\chi$), and the
Mach number of an incident shock relative to the sound speed of the
preshock ambient medium ($M$).
The evolution of the shocked cloud is divided into four stages:
(1) the initial transient stage, in which the initial shock is 
transmitted to the cloud, and at the same time is reflected 
at the cloud surface to form a bow shock or bow wave;
(2) the shock compression stage, in which 
the shocked cloud is compressed in the direction of shock propagation; 
(3) the postshock expansion stage, in which the cloud expands
 in the radial direction; and 
(4) the cloud destruction stage, in which 
the shocked cloud is shredded into small fragments and is 
eventually destroyed. 
This evolution is qualitatively consistent with previous simulations 
of shocks interacting with 
uniform clouds with sharp cloud boundaries.
However, smooth cloud boundaries significantly affect the morphology of
 the shocked clouds and cloud destruction timescales.
When an incident shock is transmitted into a smooth cloud, 
velocity gradients in the cloud envelope 
steepen the smooth density profile
at the upstream side, resulting in a sharp density jump 
having an arc-like shape.
We refer to this density jump as a ``slip surface'' because the velocity is 
sheared parallel to its surface.
Such a slip surface forms immediately when a shock strikes
a cloud with a sharp boundary.
For smoother boundaries,
the formation of slip surface and therefore the onset of 
hydrodynamic instabilities are delayed.
Since the slip surface is subject to the Kelvin-Helmholtz and 
Rayleigh-Taylor instabilities, the shocked cloud is eventually destroyed 
in $\sim 3-10$ cloud crushing times,  
where the cloud crushing time is the characteristic time
for the shock to pass through the cloud.
The ratio of the
cloud destruction time to the cloud crushing time depends
mainly on the smoothness of the density profile and only weakly on the
 density contrast for strong shocks.
We construct analytic models of cloud drag and vorticity generation 
that compare well with the numerical results.
After complete cloud destruction (in several cloud crushing times), 
small blobs formed by fragmentation due to hydrodynamic
instabilities have significant velocity dispersions of the order of 
0.1 $v_b$, where $v_b$ is the shock velocity in the ambient medium.
Our simulations show that, irrespective of the smoothness of the 
initial density profile, it is difficult to sustain the vortical 
motions inside the cloud because 
the cloud is being destroyed by hydrodynamic instabilities.
This suggests that turbulent motions generated by shock-cloud interaction
are directly associated with cloud destruction.
The interaction of a shock with a cold H~I cloud should
lead to the production of a spray of small H~I shreds, which
could be related to the small cold clouds recently observed
by Stanimirovic \& Heiles (2005).
We also calculate the dependence of the velocity dispersion 
($\Delta v$) on region size ($R$), 
the so-called linewidth-size relation, 
based on a three-dimensional simulation. 
The linewidth-size relation obtained from our 
simulation is found to be time-dependent. In the early stages of 
cloud destruction, the small-scale fluctuations dominate 
because of the nonlinear growth of the Kelvin-Helmholtz instability, 
and thus the linewidth-size relation is more or less flat. 
In the later stages, the small-scale fluctuations tend to damp, 
leading to a linewidth that increases with size.
A possibility for gravitational instability triggered 
by shock compression is also discussed.
Only radiative shocks with $\gamma < 4/3$ can
lead to induced gravitational collapse.
We show that in the absence of significant 
nonthermal motions behind the shock, star formation can be induced in an
initially uniform cloud by a radiative shock 
only if it is not too strong.
Our results indicate that the postshock gas
has significant nonthermal motions, and this 
can further limit the onset of gravitational collapse.
\end{abstract}

\section{Introduction}
\label{sec:introduction}

The interaction of shock waves with interstellar clouds is a very common
event in interstellar medium (ISM). 
For example, supernova explosions produce large blast waves 
that propagate into the ISM.   
The blast waves strike interstellar clouds,
compressing them and destroying them.
This process plays an important role in the formation of multi-phase 
structures in the ISM \citep{DCox74, CMcKee77}. 
Recently, several authors have proposed it as a mechanism for
generating the supersonic turbulence observed in the ISM
\citep[e.g.,][]{PKornreich00, MMacLow03, EVazquez00}.
Compression by shocks may be able to trigger star formation in
interstellar clouds \citep{BElmegreen77}.
In the early universe, supernova explosions of the first stars 
may be responsible 
for the first significant generation of turbulence and the
first metal enrichment 
in the intergalactic medium after the Big Bang.  
Other energetic events, such as stellar winds of massive
stars, bipolar outflows of young stellar objects, cloud-cloud collisions, 
and spiral density waves can also produce interstellar shocks that 
interact with the surrounding clumpy ISM.
Hence, understanding 
how shock waves interact with interstellar clouds is an 
important step in understanding the structure and evolution of the ISM.

	  Many authors have studied the interaction of shock waves with
interstellar clouds, either analytically 
\citep[e.g.,][]{CMcKee75, LSpitzer82, SHeathcote83, CMcKee87,
PNulsen82} or numerically
(Sgro 1975; Woodward 1976; Nittman et al. 1982; 
Tenorio-Tagle \& Rozyczka 1986; Bedogni \& Woodward 1990; 
Stone \& Norman 1992; Klein, McKee \& Colella 1990, 1994 (hereafter KMC), 
Mac Low et al. 1994; Dai \& Woodward 1995; Xu \& Stone 1995).
Such calculations are closely related to the interaction of
a cloud with a wind (KMC; Schiano, Christiansen, \& Knerr 1995;
Gregori et al. 2000).
Analytic studies have focused mainly on the
structure and evolution of a shock transmitted into a cloud 
before cloud destruction.
After the shock passes through the cloud, hydrodynamic 
instabilities such as Rayleigh-Taylor, Richtmyer-Meshkov, and 
Kelvin-Helmholtz begin to develop and may eventually destroy the
entire cloud.  It is difficult to follow such highly nonlinear stages 
of evolution analytically, and thus numerical calculations are
needed.  KMC performed 
comprehensive two-dimensional numerical studies of 
the interaction of shocks with uniform clouds based on 
the adaptive mesh refinement (AMR) technique. 
They showed that small, nonradiative clouds are destroyed 
in several cloud crushing times, where the cloud crushing time 
is the timescale for a shock to propagate through the cloud
(see \S \ref{sec:formulation} for the definition).
They also demonstrated that good spatial resolution 
(at least $10^2$ cells per cloud radius) is needed
in numerical
calculations in order to follow the cloud destruction accurately.
\citet{JStone92}, \citet{RKlein94b}, and \citet{JXu95} performed 
three-dimensional numerical simulations
on this problem, showing that overall evolution of the cloud destruction 
is in good agreement with the two-dimensional results of KMC, 
although vortex rings observed in the two-dimensional calculations 
are unstable in three dimensions and the cloud destruction proceeds 
more violently in all directions.

Recently, Klein et al. (2000, 2003) performed a series of
Nova laser experiments investigating the evolution of a high density
sphere embedded in a low density medium after the passage of a strong
shock (with a Mach number of $\sim $10).
The results of these experiments can be compared directly with 
interstellar shock-cloud interaction 
under certain scale transformations \citep{DRyutov99}.
The comparison of their experimental results with two-dimensional
simulations can reproduce the morphological 
evolution of the experimental shocked spheres reasonably well. 
In later stages of cloud destruction, the three-dimensional instability 
of vortex rings plays a significant role, and fully three-dimensional 
calculations are required to follow the evolution. 
Follow-up experiments on the Omega laser confirmed the presence 
of vortex ring instabilities, which were shown to be in good agreement 
with the predictions made by the Widnall instability \citep{HRobey02}.

Previous numerical studies have assumed that the initial clouds 
have a uniform density and sharp boundaries.  
However, real interstellar clouds have internal density gradients.  
For example, in molecular cloud cores with no embedded young stars, 
the density distributions are flat near the center and 
decrease with the distance from the cloud center \citep{ABacmann00}.
Observations of NGC 7293 (the Helix nebula) imply 
the presence of large radial density gradients in the cometary cloud cores 
there \citep{ABurkert98,CODell00}.
The actual structure of the clouds is likely to be complex,
but observations suggest that it would be worthwhile to
study the effect of smooth density gradients on the cloud-shock
interaction.
\citet{PKornreich00} have carried out the first
such study.
They derived an approximation for the distribution of vorticity
generated in a shocked cloud with a smooth density profile.
They argued that the vortex motions generated by the shock 
could be converted  into turbulent motions observed in interstellar
clouds provided  that the motions are confined in the parent cloud.
However, from their work, it was unclear whether such a cloud can survive 
after the shock passage.  
In fact, many numerical calculations have shown that uniform clouds, 
which have sharp density jumps at the cloud-intercloud boundaries, are 
easily destroyed primarily under the influence of 
Kelvin-Helmholtz and Rayleigh-Taylor instabilities (e.g., KMC).
Recently, \citet{APoludnenko02} observed complete destruction for a
cloud with a smooth but steep density gradient.
It therefore remains an open question whether
shocks can maintain turbulent motions in 
interstellar clouds.

In this paper, we study the effects of density 
gradients on the cloud destruction process, using 
our higher-order Godunov AMR hydrodynamic code
\citep{KTruelove98,RKlein99}.  
We first present two-dimensional axisymmetric calculations.
A three-dimensional simulation is also done to
clarify the property of turbulent motions generated by the shock-cloud
interaction. 
In \S \ref{sec:formulation} we summarize the formulation of our problem. 
We show that at least $10^2$ grids per cloud 
radius are needed to follow the evolution of smooth clouds accurately, 
just as in the uniform case.
An overall description of shocked clouds is presented in 
\S \ref{sec:result}, in which we demonstrate that 
the initial smooth density profile steepens after shock passage, and 
a density discontinuity is formed at the upstream side of the shocked cloud. 
A significant velocity shear develops across the discontinuity.
The velocity shear makes this ``slip surface'' unstable to the 
Kelvin-Helmholtz instability, leading to complete cloud destruction
even for smooth clouds.
We also investigate the dependence of cloud evolution on the initial model 
parameters such as the density profile, shock strength, 
equation of state, and cloud shape.
In \S \ref{sec:cloud drag}, we estimate how the cloud is accelerated 
by the shock, which affects the growth rate of the Kelvin-Helmholtz
instability, and therefore the timescale of cloud destruction.
In \S \ref{sec:slip surface} and \S \ref{sec:cloud destruction}, 
the formation of the slip surface 
and the subsequent cloud destruction are examined in more detail.  
Since the entire cloud is eventually destroyed by the shock, 
the vorticity generated by the shock is not confined to the original
cloud.  Instead, the shear motions are converted into random motions
of small fragments after cloud fragmentation.
In \S \ref{sec:vorticity}, the vorticity generated by shock passage 
is compared with a simple analytic model.
In \S \ref{sec:discussion}, we summarize our results, and discuss 
how density gradients affect the observed morphologies 
of shocked clouds.  
We discuss several properties of turbulent motions generated 
by shock-cloud interaction, computing synthesized spectra 
 and the velocity-size relation.
Finally, we  derive the condition under
which shock waves are capable of triggering gravitational
contraction of shocked clouds resulting in star formation.

\section{Formulation of the Problem}
\label{sec:formulation}

\subsection{A Cloud with a Smooth Boundary}

In this paper, we investigate the hydrodynamic interaction of a planar
shock with an interstellar cloud, focusing on the effect of a
smooth cloud boundary.
We consider a steady planar shock moving through the
intercloud medium at a velocity $v_b$ at infinity,
corresponding to a Mach number $M\equiv v_b/C_{i0}$, where
$C_{i0}=(\gamma_i P_{i0}/\rho_{i0})^{1/2}$ is the adiabatic sound speed
in the intercloud medium.
The subscript $i$ denotes the values in the intercloud medium.
The subscript $0$ denotes the values in the preshock state. 
(In the following, the subscript $1$ is 
used for the quantities in the post-shock state.)
This shock hits a spherical cloud whose density 
distribution is given in a cylindrical coordinate system $(\varpi, z)$ 
centered on the cloud by 
\begin{equation}
\rho_0(\varpi,z)=\rho_{i0} + \frac{\rho_{c0} - \rho_{i0}}{1+(r/r_{\rm co})^n}  \ , 
\end{equation}
where $r=(\varpi^2+z^2)^{1/2}$, $r_{\rm co}$ is a characteristic radius, 
$\rho_{c0}$ is the density at the cloud center, and $\rho_{i0}$ is the 
density at infinity.  
The initial density contrast between the
cloud center and the intercloud medium is 
\begin{equation}
\chi \equiv {\rho_{c0}} / {\rho_{i0}}  \  .
\end{equation}
This density distribution is essentially the same as \citet{PKornreich00},
although they applied this profile to a cylindrical cloud with a 
two-dimensional Cartesian geometry.
For comparison, we shall follow a couple of runs with Cartesian geometry.
For $n\rightarrow \infty$, this model cloud reduces 
to a uniform cloud with a sharp boundary.
Because we do not include the effect of gravity in our simulations,
the cloud is assumed to be initially in pressure equilibrium with
the ambient medium with a constant pressure $P_0 $ 
everywhere in front of the shock.
We also focus on nonradiative shocks, and therefore
the ratio of specific heats is set to $\gamma= 5/3$ for most runs,
although we calculate a couple of models with $\gamma = 1.1$ to examine
the effects of radiative cooling.

	  In order to minimize the effects of the density variation 
along the initial shock front, we must begin our calculation
with the shock in a region where the density variation 
is less than a few percent of the density at infinity ($\rho_{i0}$).
Thus, the distance from the cloud center to the initial shock
front is larger for clouds with smoother density gradients and/or larger
density contrasts.

  To measure the degree of mass concentration
in the model cloud,
we define two characteristic radii: a core radius, $r_{\rm co}$, 
and a cloud radius, $r_{\rm cl}$.
The core radius denotes the size of the central relatively flat 
region and is equal to the characteristic radius, $r_{\rm co}$.
The density at $r_{\rm co}$ is $(\chi^{-1}+1)\rho_{c0}/2$, which
approaches $\rho_{c0}/2$ for large $\chi$.
The cloud radius, $r_{\rm cl}$, denotes 
the overall size of the cloud, and is defined to be the radius 
at which the density is twice that of the ambient medium ($2 \, \rho_{i0}$),  
\begin{equation}
r_{\rm cl} = (\chi -2)^{1/n} r_{\rm co}.
\end{equation}
We assume that $r_{\rm cl}> r_{\rm co}$, which is true for $\chi>3$.
These two characteristic radii divide the model cloud into two
parts: the core ($r\le r_{\rm co}$) and the cloud envelope 
($r_{\rm co} < r\le r_{\rm cl}$).
The region $r > r_{\rm cl}$ is
the intercloud medium.
A third characteristic radius, the rms radius, is introduced
in \S \ref{sec:global} below.

     We define the core mass and the cloud mass in terms of density:
The core mass, $m_{\rm co}$, is the mass of gas with 
$\rho\geq 0.5(1+\chi^{-1}) \rho_{c0}$, so that initially it is
the mass inside $r_{\rm co}$.
The cloud mass 
is the mass with $\rho\geq 2\rho_{i1}$, so that initially
it is the mass inside $r_{\rm cl}$.
These masses begin with values $m_{\rm co,0}$ and $m_{\rm cl, 0}$,
respectively, and then
decrease due to shock heating and to
mixing with the shocked intercloud medium.
The degree of mass concentration in the initial cloud is measured by 
the ratio
\begin{equation}
\calr\equiv \frac{m_{\rm cl,0}}{\frac 43 \pi\rho_{c0}r_{\rm co}^3},
\label{eq:calr}
\end{equation}
which is unity for a uniform cloud, where
$m_{\rm cl,0}$ is the initial cloud mass. 
In the following, the initial core mass, $m_{\rm co,0}$, 
is defined as the mass contained within $r\le r_{\rm co}$. 
For example, for $n=2$ and large
$\chi$, we have $\calr\rightarrow 4\chi^{1/2}-3\pi/4$.

	The thickness of the cloud boundary 
($h_{\rm bd}$) can be approximated
as the density scale height at $r=r_{\rm co}$,
\begin{equation}
h_{\rm bd} = \left(\frac{\chi +1}{\chi -1}\right)\frac{2}{n} r_{\rm co} \ .
\label{eq:hbd}
\end{equation}

	In real interstellar clouds, the value of $n$ is likely 
to vary widely.
For example, in dense molecular cloud cores with 
no embedded young stars, the density profiles are flat near the center, 
follow a power-law profile of $\rho \propto r^{-2}$
in the intermediate radii, 
and fall steeply as $\rho \propto r^{-4}$ or even $r^{-5}$ 
near the edges of the cores at $r \sim 10^4$ AU \citep{ABacmann00}.
Observations of the Helix nebula imply 
that the density distributions in the cloud cores can be fit with
power-law profiles 
with $n\simeq 3-9$
\citep{ABurkert98,CODell00}.
In the following, we vary $n$ in the range of $2 \le n \le \infty$.

We do not take into account the effect of thermal conduction, which 
can play a role in evolution of non-magnetized clouds.  
According to KMC, non-magnetized clouds with $\chi \sim 10^2$ are 
ablated by evaporation in a time comparable to the cloud crushing time
if the shocked intercloud medium is hot enough that the 
mean-free path is comparable to the cloud radius, so that the
conduction is saturated \citep{Cowie77}.
To neglect the effect of thermal conduction, we assume 
either that the intercloud medium is not that hot, or that
there is a magnetic field that is too weak to be dynamically significant, 
but is strong enough to inhibit thermal conduction.

\subsection{Time Scales}
\label{subsec:time scales}

Following KMC, we define two characteristic time scales governing the
evolution of the shocked cloud: 
(1) the shock-crossing time,
\begin{equation} 
t_{\rm sc} \equiv \frac{2r_{\rm co}}{v_b} , 
\label{eq:t_sc}
\end{equation}
which is the time scale for the shock in the intercloud medium 
to sweep across the cloud, and (2)
the cloud-crushing time,
\begin{eqnarray}
t_{\rm cc} &\equiv& \frac{\chi^{1/2} r_{\rm co}}{v_b} = 
\frac{\chi^{1/2}r_{\rm co}}{M C_{i0}} = \frac{r_{\rm co}}{M C_{c0}} \\
&=& 0.98\times 10^5 r_1 M_{10}^{-1} C_{c0,\, 5}^{-1} ~~~~ {\rm yr}
\label{eq:t_cc}
\end{eqnarray}
where 
$r_1\equiv r_{\rm co}/1\, {\rm pc}$, 
$M_{10}\equiv M/10$, $C_{c0,\, 5}\equiv C_{c0}/10^5 \, {\rm cm \, s}^{-1}$,
$C_{c0}$ is the sound speed in the preshock cloud, 
$C_{i0}$ is the sound speed in the preshock ISM, and $t_{cc}$
is the time scale for the cloud to be crushed by the shock 
transmitted into the cloud.
Here, the cloud-shock velocity is approximated as 
$v_s \simeq v_b / \chi^{1/2}$, since the pressure behind 
the cloud shock is comparable to that behind the 
intercloud shock for a strong shock 
($\rho_{c0}v_s^2 \simeq \rho_{i0}v_b^2$, where
$v_s$ is the velocity of the cloud shock).
We note that for smooth clouds, the actual mean velocity of the cloud shock 
is estimated as $v_s \simeq v_b / \left<\chi\right>^{1/2}$, where
the mean density contrast, $\left<\chi\right>$, is generally smaller than $\chi$. Therefore,
for a smooth cloud, the cloud shock can reach the initial cloud
center by the time of $1 \, \tcc \left<\chi\right> / \chi$ ($< 1 \, \tcc$).

      According to \citet{CMcKee87}, the typical pressure variation
timescale for a dense cloud in a Sedov-Taylor blast wave is given by 
$t_P\simeq 0.1 \, {R_c}/{v_b}$, where $R_c$ is the distance from the origin 
of a blast wave.
Therefore, our assumption of a steady shock requires that $\tcc\ll t_P$, or
equivalently, that the cloud radius be smaller than 
$0.1 \, R_c/\chi^{1/2}$.
This is known as the small cloud approximation
 (see \S 2 of KMC for more detail).

To investigate the overall evolution of the shocked clouds, we also define 
the following four characteristic timescales:
(1) {\it the drag time}, $t_{\rm drag}$, which is the time when
 the cloud velocity relative to the velocity of the post shock ambient
 flow has decreased by a factor of $1/e$ (see \S \ref{sec:cloud drag}),
(2) {\it the maximum expansion time}, $t_m$, which is the time when 
the effective cloud radius in the $\varpi$-direction ($a$) 
has increased to 90 \% of its maximum value (see \S \ref{sec:cloud drag}),
(3) {\it the cloud destruction time}, $t_{\rm dest}$, which is 
 the time when the mass of the main axial fragment has become
smaller than the initial cloud mass by a factor of $1/e$ (see \S
\ref{sec:cloud destruction}),  and
(4) {\it the mixing time}, $t_{\rm mix}$, which is the time when
the mass of the region with $\rho \ge 2 \, \rho_{i1}$ 
has reached half the initial cloud mass 
(see \S \ref{sec:cloud destruction}).

\subsection{Global Quantities}
\label{sec:global}

Following KMC, we define several integrated physical quantities
to study the cloud evolution.  
In KMC, a two-fluid treatment that followed both the cloud and
intercloud matter allowed the global quantities to be
evaluated by integrating the physical quantities 
only in the cloud material.
In this paper, we use a density threshold, $\rho _{\rm th}$, to determine
which material can be considered to be ``cloud material.''
The corresponding threshold mass is 
\begin{equation}
m_{\rm th} = \int _{\rho \ge \rho_{\rm th}} \rho dV \ .
\label{eq:threshold mass}
\end{equation}
The threshold density $\rho_{\rm th}$ 
is chosen such that $m_{\rm th}$ corresponds to
either the characteristic initial core mass 
($m_{\rm co,0}\equiv m[r\leq r_{\rm co}, \; t=0]$) or 
the initial cloud mass ($m_{\rm cl,0}$).
The regions with $m_{\rm co,0}$ and $m_{\rm cl,0}$ are then regarded 
as the core and cloud, respectively.
The threshold densities ($\rho_{\rm th, core}$
and $\rho_{\rm th, cl}$) vary with time, and 
are obtained from eq. (\ref{eq:threshold mass}) 
by means of a bisection method.
Unless noted otherwise, we will choose $m_{\rm cl}$ as 
the threshold mass to obtain the global quantities.

   To measure the evolution of the cloud shape, we define the effective
radii normal to and along the cylinder axis, $a$ and $c$, respectively, as 
\begin{equation}
a = \left(\frac 52 \left<\varpi^2\right>\right)^{1/2},  ~~~~~
c = \left[5 (\left<z^2\right>-\left<z\right>^2)\right]^{1/2} \ ,
\label{eq:ac}
\end{equation}
where
\begin{eqnarray}
\left<z\right> &=& m_{\rm th}^{-1}\int _{\rho \ge \rho_{\rm th}}
\rho z dV , \ ~~~~~
\left<z^2\right> = m_{\rm th}^{-1}\int _{\rho \ge \rho_{\rm th}}
\rho z^2 dV, \\ 
\left<\varpi^2\right> &=& m_{\rm th}^{-1}\int _{\rho \ge \rho_{\rm th}}
\rho \varpi^2 dV  \ .
\end{eqnarray}

To measure the cloud destruction and turbulence, we define 
the velocity dispersions in the axial and radial directions,
respectively, as 
\begin{equation}
\delta v_z = (\left<v_z^2\right>-\left<v_z\right>^2)^{1/2} 
\ \ \ {\rm and} \ \ \ 
\delta v_\varpi = \left<v_\varpi^2\right>^{1/2} \ , 
\label{eq:dv}
\end{equation}
where
\begin{equation}
\left<v_z^2\right> = m_{\rm th}^{-1} \int  _{\rho \ge \rho_{\rm th}}
\rho v_z^2 dV  \ , ~~~~~
\left<v_\varpi^2\right> = m_{\rm th}^{-1} \int _{\rho \ge \rho_{\rm th}}
\rho v_\varpi^2 dV  \ \  .
\end{equation}
The mean velocity, $\left<v_z\right> $, is defined as
\begin{equation}
\left<v_z\right> = m_{\rm th}^{-1} \int _{\rho \ge \rho_{\rm th}}
\rho v_z dV  \  
\label{eq:cloud velocity}
\end{equation}
where $v_z$ is measured in the frame of the unshocked cloud.
Finally, the mean density, $\left<\rho\right> $, is defined as
\begin{equation}
\left<\rho\right> = V_{\rm th} ^{-1} \int _{\rho \ge \rho_{\rm th}}
\rho dV  \  , 
\label{eq:mean density}
\end{equation}
where $V_{\rm th}$ is the volume of the region with $\rho \ge \rho_{\rm th}$.
In the following, the initial mean density is refered to as
$\left<\rho_0\right> $ and the mean initial density contrast 
is defined as $\left<\chi\right> = \left<\rho_0\right> / \rho_{i0}$.

\section{Numerical Simulations}

We solve the two-dimensional Euler equations 
of hydrodynamics using an Eulerian adaptive mesh refinement (AMR) 
hydrodynamic code with a second-order Godunov, 
dimensionally-split approach
(Truelove et al. 1998; Klein 1999).
In this code, as a computation proceeds and one wishes to resolve certain 
features within the flow, fine grids are dynamically created 
at locations meeting a pre-determined criterion, 
initialized from interpolated existing coarse grid data, 
and evolved. This entire process is continually repeated, 
allowing us to create and destroy fine grids in a general 
and controllable fashion, wherever and whenever the need arises
\citep{KTruelove98}.

In this paper, the entire computational domain is covered with
a base grid (level 1) and we use 2 additional levels of
refinement (levels 2 and 3).
The entire computation domain is covered with a base
grid (level 1).
Finer grids (levels 2 and 3) are created or destroyed by using 
two refinement criteria, a density criterion and 
density gradient criterion.
The density criterion leads to refinement
of cells in the base grid to level 2 when
the density in these cells exceeds
a threshold density, $\rho_{\rm ref}$.  
In all the calculations presented in this paper, 
we set $\rho_{\rm ref} = 1.2 \, \rho_{i1}$, where $\rho_{i1}$ is the
density of the post-shock ambient gas.
This guarantees that the initial cloud 
($\rho_0 \ge 2 \, \rho_{\rm i0}$) is refined to 
at least level 2 after passage of the shock.
We followed several models with different $\rho_{\rm ref}$ and found
that the numerical results are essentially identical for 
$\rho_{\rm ref} \lesssim 1.5 \, \rho_{i1}$
(i.e., the global quantities defined in the previous subsection are 
converged to within a few percent).
The density gradient criterion leads
to refinement of levels 1 or 2 
when the local density scale height normalized by the cell size 
becomes smaller than a threshold value, $\tilde{l}_{\rm th}$,
\begin{equation}
\tilde{l_j} \equiv \frac{l_j}{\Delta x_j} = 
	    \frac{\rho_j}{(d\rho/dx|_j)\Delta x_j}<\tilde{l}_{\rm
	    th}\ ,
\end{equation}
where $\Delta x_j$ denotes the grid spacing of the $i$-th grid
and the density gradient $d\rho/dx|_j$ is computed at the cell center.
We typically adopt $\tilde{l}_{\rm th} = 20$ in this paper because 
the global quantities are converged to within a few percent for 
$\tilde{l}_{\rm th} \ge 10 $.
This criterion guarantees that shock fronts and contact discontinuities 
are refined to level 2 in the region of $\rho < \rho_{\rm ref}$
and to level 3 in the region of $\rho \ge \rho_{\rm ref}$.

Our grids are Cartesian and properly nested; i.e., 
a fine grid must be surrounded by at least one layer of 
coarser cells. The only exception occurs at domain boundaries, 
where fine and coarse interfaces may coincide. 
The increase in resolution between levels, 
referred to as the refinement ratio, is taken to be 4 
for every level.

The spatial resolution of the calculation is specified by
the number of finest grid points that covers an initial core radius
$r_{\rm co}$.  We use 120 finest grid points per core radius $r_{\rm co}$ 
as our standard resolution
because we have demonstrated that with this resolution 
the evolution of the global quantities is well-converged. 
We refer to this resolution as $R_{120}$.
More generally, we define the resolution of a calculation by
$R_N$, where the subscript $N$ is the number of cells per $r_{\rm co}$ at the finest level 
of refinement.
The spatial resolutions we used in this
paper and several characteristics of the initial clouds 
are summarized in Table \ref{tab:model}.

Since our model cloud does not have a sharp boundary between the cloud and 
intercloud medium, we do not explicitly separate the cloud 
and intercloud medium as different fluids; i.e., 
we use a single-fluid version of the AMR code in which
the ratio of the specific heats, $\gamma$, is always 
identical for both the cloud and intercloud medium.
By contrast, KMC used 
a two-fluid code that treated the cloud and the intercloud medium as 
different fluids.  
When we approximate the effect of radiative losses by
setting $\gamma\sim 1$, we must assume that 
both the cloud and the intercloud medium are radiative, whereas
KMC were able to consider a radiative cloud in an
adiabatic intercloud medium. As a result, our
results for the radiative case differ from theirs
(see \S \ref{subsec:gamma}).

\subsection{Convergence Tests}
\label{sec:convergence}

In hydrodynamic simulations, it is very important to demonstrate
that the calculations are performed at spatial resolutions that 
resolve important features of the hydrodynamic evolution
such as the growth of hydrodynamic instabilities
(see also KMC and Klein \& Woods 1998).
Comparing the time evolution of the global quantities 
at different spatial resolutions, KMC showed that 
at least $10^2$ cells per cloud radius are needed to 
follow the interaction of shock waves with uniform clouds accurately.
Here, we carry out a similar study for smooth clouds.

\begin{figure}
\plotone{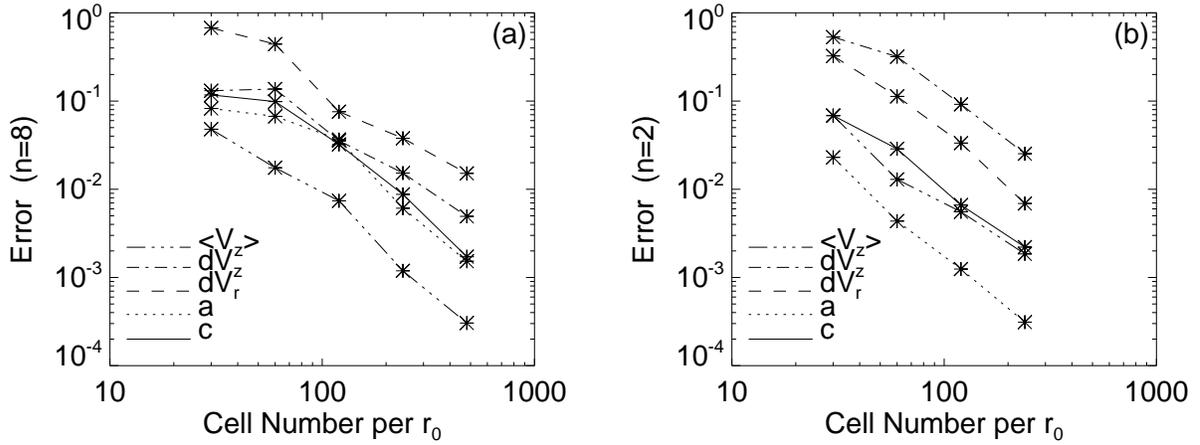}
\caption{
Results of convergence studies for (a) $(n,\chi,M)=(8,10,10)$ and
(b) (2, 10, 10). The ordinate denotes the relative error, while the
 abscissa denotes the number of grid cells per radius,
 which is a measure of the
 spatial resolution. See the text for the definition of the relative
 errors.  The global quantities are compared at $t=3 \, t_{\rm cc}$ and 
$6 \, t_{\rm cc}$ for $n=8$ and 2, respectively.
\label{fig:convergence}}
\end{figure}

    To perform the convergence study, we calculate two models 
for the shock-cloud interaction
with $(n, \chi, M) = (2, 10, 10)$ and $(8, 10, 10)$ 
with different spatial resolutions 
($R_{30}$, $R_{60}$, $R_{120}$, $R_{240}$, $R_{480}$, 
and $R_{960}$ for $n=8$ and  
$R_{30}$, $R_{60}$, $R_{120}$, $R_{240}$, and $R_{480}$ 
for $n=2$) and compare the global quantities.
Figure \ref{fig:convergence} shows how the errors at
the end of the calculation 
($t=3 \, t_{\rm cc}$ for $n=8$ and $t=6 \, t_{\rm cc}$ for $n=2$) depend on
the spatial resolution.  
The errors at a resolution $R_N$ 
are defined as the differences between the global quantities
at $R_N$ and those at the finest resolution, $R_f$. In other
words, for a global quantity $Q_N$ measured at $R_N$, the error is
\begin{equation}
\Delta Q_N = \frac{|Q_N-Q_f|}{|Q_f|} \ .
\end{equation}
The convergence test indicates that the errors of the 
velocity dispersions generally decrease in proportion to 
the inverse of the cell number $N$, which implies
first-order accuracy.
On the other hand, the convergence of the other quantities 
is better than that of the velocity dispersions;
e.g., the errors in the radii $a$ and $c$ scale
as $N^{-2}$.
The slow convergence of the velocity dispersions 
may be related to the fact that the code accuracy is reduced to 
first order at discontinuities such as 
shocks and contact discontinuities, 
where substantial velocity differences are generated.
All the quantities converge to within about 10 \% 
for the spatial resolutions
of $R_{120}$ or finer.
Hence, we conclude that at least $\sim 10^2$ cells per $r_{\rm co}$ 
are necessary for a converged calculation.
In the following, we adopt $R_{120}$ as
the standard spatial resulution.

\section{Numerical Results}
\label{sec:result}

In this section, we describe the overall evolution of 
clouds with smooth density profiles.
In \S \ref{subsec:n=8} and \S \ref{subsec:n=2}, we consider both clouds with 
a relatively steep density gradient ($n=8$) and with  
a shallow density gradient ($n=2$) 
as representative cases of nonradiative ($\gamma = 5/3$) clouds
with smooth density profiles.
In these models, the other parameters are fixed 
at a density contrast of $\chi = 10$ and a Mach number $M=10$.
The dependence of the results on the parameters $n$ and $\chi$ 
is summarized in \S \ref{subsec:dependence}.
In \S \ref{subsec:mach scaling} we discuss the applicability
of Mach scaling 
(see \S 4 of KMC) to smooth clouds.
The effect of $\gamma$ is studied in \S \ref{subsec:gamma}.
Finally, the evolution of cylindrical clouds is compared with that of 
spherical clouds in \S \ref{subsec:cylinder}.

In describing our results, we shall use $r_{\rm co}$ as a unit 
of length, $\rho_{c0}$ as a unit
of density, and $t_{\rm cc}$ as a unit of time.
Thus, our model is specified by four dimensionless parameters: 
$n$, $\chi$, $M$, and $\gamma$.
The evolution time is measured from the epoch at which the
initial shock has reached to the point $z=-r_{\rm co}$ on the $z$-axis.
In Table \ref{tab:model}, we summarize the parameters of the models 
presented in this paper.

	  According to previous studies, the overall evolution of 
shocked uniform clouds can be divided into four different 
stages: (1) the {\it initial transient stage}, 
in which an incident shock is transmitted into 
the cloud, and at the same time a bow shock or bow wave 
propagates into an upstream region;
(2) the {\it shock compression stage}, 
in which the cloud is compressed mainly in the $z$-direction;
(3) the {\it postshock expansion stage}, in which the shock-compressed cloud
expands laterally, and (4) the {\it cloud destruction stage},
in which the shocked cloud is destroyed primarily 
by Kelvin-Helmholtz and Rayleigh-Taylor instabilities.
As shown below, while the overall evolution of smooth clouds 
can also be divided into these four stages, there
are significant differences in
the morphology and evolution.

\subsection{A Cloud With A Steep Density Gradient: 
AS8 ($n=8$, $\chi=10$, and $M=10$)}
\label{subsec:n=8}

To begin with, we consider a cloud with $n=8$,
$\chi=10$, and $\gamma=5/3$ interacting with a Mach 10 shock.
In this model, about 70 \% of the cloud mass is concentrated in the 
central core ($m_{\rm co,0} \sim 0.7 \, m_{\rm cl,0}$). 
There are several significant 
differences between the evolution of a uniform shocked cloud and a 
smooth shocked cloud, as we now describe.

\begin{figure}
\plotone{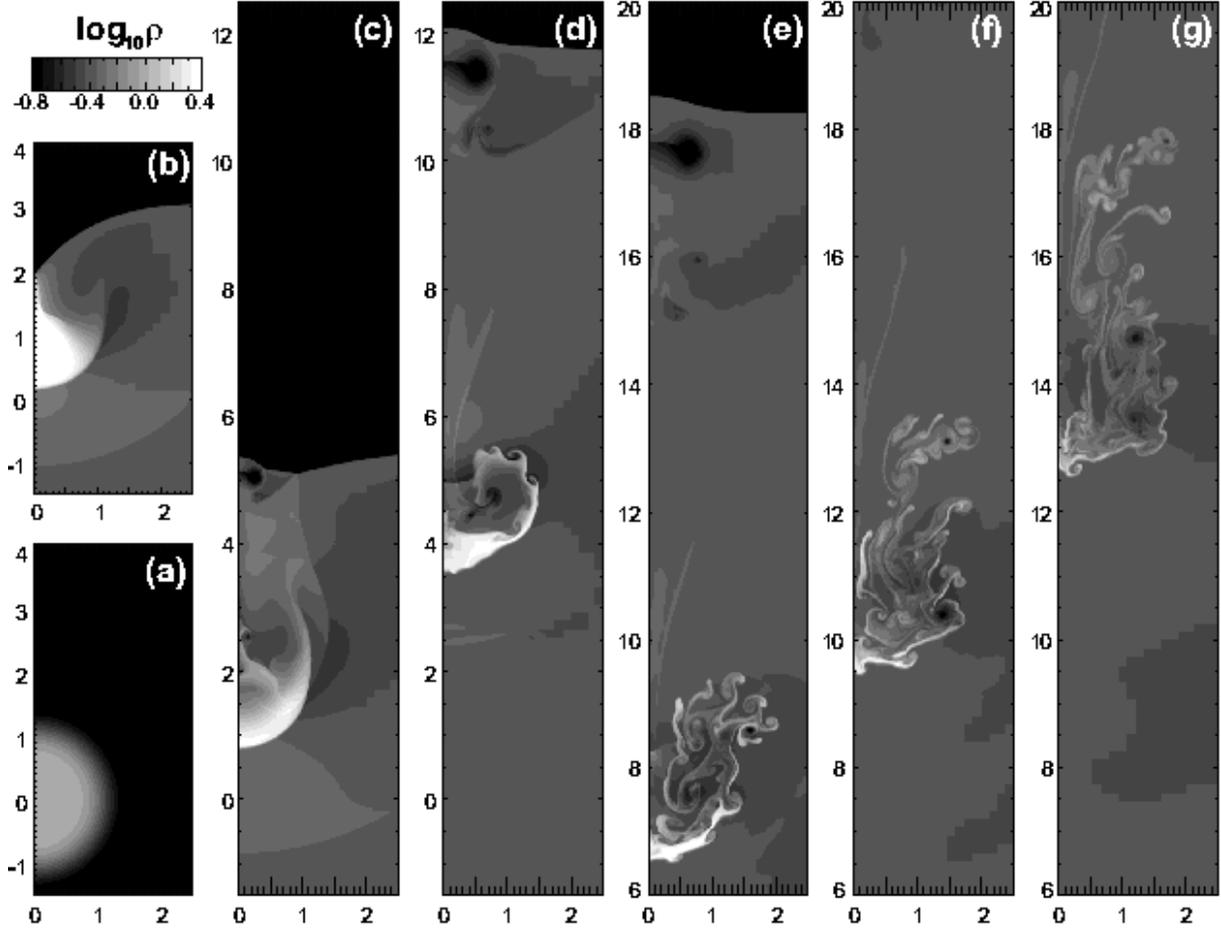}
\caption{
Snapshots of the density distribution for  model AS8 with 
$n=8$, $\chi=10$, and $\gamma=5/3$ interacting with a strong shock of $M=10$ 
at seven different times: (a) initial state ($t=- 0.39 \, \tcc$), 
(b) 1.24 $\tcc$, (c) 2.00 $\tcc$, (d) 4.00 $\tcc$, (e) 6.00 $\tcc$, 
(f) 8.00 $\tcc$, and (g) 10.00 $\tcc$, where the evolution
 time is measured from the stage at which the shock has reached at
 $z/r_{\rm co}=-1$ on the $z$-axis.  Tha abscissa and ordinate denote 
the $\varpi$- and $z$-coordinates, respectively,
 normalized by the core radius, $r_{\rm co}$.
The initial planar shock was initiated at $z/r_{\rm co}=-2.1$.
The arch-like density jump appearing at $z/r_{\rm co}\sim -1$ and $-0.85$ 
in panels (b) and (c), respectively, is a bow shock, which is 
formed by the shock reflected at the cloud surface.
The black spot appearing at $(\varpi/r_{\rm co}, z/r_{\rm co})\sim (0.2, 5)$ in panel (c)
 is a supersonic vortex ring, which propagates, with almost the same
 speed as the intercloud shock, up to $(\varpi/r_{\rm co}, z/r_{\rm co})\simeq (0.5, 11.4)$ 
and $(0.6, 17.6)$ in panels (d) and (e), respectively.
\label{fig:snapn8}}
\end{figure}

In Figure \ref{fig:snapn8}, we present snapshots of the density
distributions at seven different times.
At the first encounter with the incident shock, the cloud is compressed
mainly in the $z$-direction (the {\it initial transient stage}).
In this stage, a shock reflects from the front of the cloud, 
propagating away from the cloud into the upstream postshock flow. 
The reflected shock develops into a bow shock, which can be recognized 
as a density jump in the upstream flow 
in Figures \ref{fig:snapn8}b and \ref{fig:snapn8}c.
The bow shock eventually becomes a weak acoustic wave 
at later stages of evolution \citep{LSpitzer82}.
In fact, the bow shock is not so prominent in Fig. \ref{fig:snapn8}d
as in Figures \ref{fig:snapn8}b and \ref{fig:snapn8}c.
In Figure \ref{fig:snapn8}b, the shock passing over the cloud envelope 
 converges on the $z$-axis behind the cloud 
(at $z/r_{\rm co}\simeq 1.5 - 2$ in Fig. \ref{fig:snapn8}b), 
reflects off the $z$-axis, and by a double Mach reflection
drives a shock back into the shocked cloud.
Thereafter, the shock transmitted into the 
front of the cloud (the cloud shock)
interacts with the reflected shock after 
the cloud shock
has passed through the high density
region. (For the uniform case, this process occurs 
inside the cloud. Compare Fig. \ref{fig:snapn8}b and Fig. 2 of 
\citet{RKlein90}.)
This reflected shock compresses the shocked cloud from the rear
in the $z$-direction (the {\it shock compression stage}).  
At this stage, the shocked cloud is deformed into an arc-like
structure without any sign of the growth of 
instabilities (Fig. \ref{fig:snapn8}c).  
The cloud then expands laterally 
because the pressure is lower at the sides of the cloud compared to 
that on the axis
(the {\it postshock expansion stage}, see also \citet{JNittman82}).
At the time shown in Figure 
\ref{fig:snapn8}d, the Kelvin-Helmholtz instability 
begins to develop along the slip surface, and eventually destroys 
the shocked cloud (the {\it cloud destruction stage}, 
see Figs. \ref{fig:snapn8}e, \ref{fig:snapn8}f, and \ref{fig:snapn8}g).
The cloud has broken up into small fragments by the time
shown in Figure \ref{fig:snapn8}f.

Since the velocity of the cloud shock is smaller in the higher density 
region (since $v \propto \rho ^{-1/2}$),
a significant velocity gradient develops in the cloud envelope.
In the shock compression and postshock expansion stages, 
this velocity gradient in the cloud envelope 
steepens the initial smooth density profile at the front of the cloud, 
resulting in a sharp density jump (Fig. \ref{fig:snapn8}b).
We refer to this density jump as a ``{\it slip surface}'' because 
the velocity is sheared parallel to its surface.
The evolution of the slip surface is crucial 
in understanding the evolution of smooth clouds,
because the growth rate of the Kelvin-Helmholtz instability
across the surface depends on the thickness of the surface itself.

The convergence of the intercloud shock on the $z$-axis creates 
another strong reflected shock, which propagates downstream. 
It interacts with the intercloud shock to produce 
a Mach-reflected shock, which develops into a double Mach-reflected 
shock with two triple points.
A powerful supersonic vortex ring forms just 
behind the Mach-reflected shock, and is carried away 
from the cloud with almost the same speed as the incident shock.
The vortex ring propagates up to $(\varpi/r_{\rm co}, z/r_{\rm co})\simeq (0.2, 5)$ in 
Figure \ref{fig:snapn8}c, 
and moves up to $(\varpi/r_{\rm co}, z/r_{\rm co})\simeq (0.5, 11.4)$ and $(0.6, 17.6)$ 
in Figures. \ref{fig:snapn8}d and \ref{fig:snapn8}e, respectively.

Immediately after the shock passage, the morphology of the shocked 
cloud is significantly different from the uniform case, 
for which the lateral expansion occurs more rapidly and 
the Kelvin-Helmholtz instability 
starts to grow immediately after the shock passage. 
For example, for the uniform cloud, the ripple generated by the 
instability already apprears along the cloud surface at $t\sim 1 \, \tcc$
[see Figs. 1- 4 of \citet{RKlein90}], and the arm-like structure 
starts to break up away from the cloud by $t\sim 3 \, \tcc$.  
For the smooth cloud, the cloud is deformed into an arch-like structure
 due to the shock compression, without any signs of the Kelvin-Helmholtz 
instability
even at $2 \, \tcc$.
The retarded growth of the instability is 
due to the fact that the instability grows faster
for shorter wavelengths (the growth rate $\propto k^{-1}$,
where $k$ is the wavenumber), but
is strongly suppressed for wavelengths 
shorter than the thickness of the shear layer
[See Chapter 11 of \citet{SChandrasekhar61}]. 
{\it Therefore, cloud destruction is retarded until a 
density discontinuity develops; in this
case, the density discontinuity is associated with the slip
surface.}
Also, the Richtmyer-Meshkov instability, which
grows linearly in time for small amplitudes, as opposed to the exponential 
growth  of the Kelvin-Helmholtz and Rayleigh-Taylor
instabilities in the linear regime, 
does not develop appreciably on the $z$-axis 
(Compare Fig. \ref{fig:snapn8}b and 
Fig. 5 of \citet{RKlein90}).
In \S \ref{sec:slip surface} and \ref{sec:cloud destruction}
we shall discuss in more detail 
the formation of the slip surface and the subsequent 
destruction of the cloud by the Kelvin-Helmholtz instability.

\subsection{A Cloud With A Shallow Density Gradient: AS2
($n=2$,  $\chi=10$, and $M=10$)}
\label{subsec:n=2}

Next, we consider a very smooth cloud with $n=2$.  The
other parameters are
the same as those of the previous model: $\chi=10$, $\gamma=5/3$, and
$M=10$.  The density profile of this cloud resembles that of a 
Bonner-Ebert sphere, or isothermal equilibrium sphere, which
has a density
profile $\rho \propto r^{-2}$ in its envelope.
The cloud mass is less centrally concentrated than in the previous case 
($m_{\rm co,0} = 0.10m_{\rm cl,0}$); only about 10 \% of the 
total cloud mass is in the central core. 
Therefore, the smooth density profile has a greater effect for this
model than for the $n=8$ model presented in \S \ref{subsec:n=8}.

\begin{figure}
\plotone{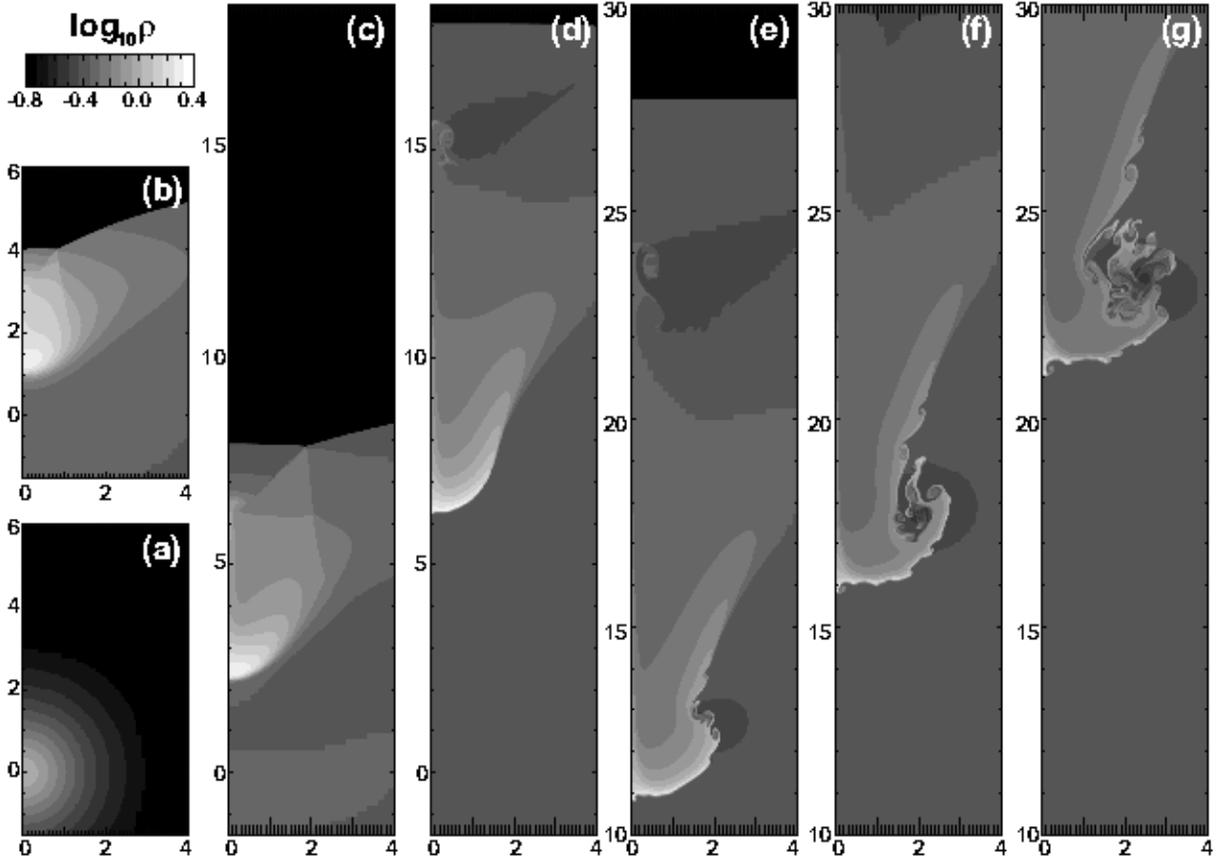}
\caption{
Snapshots of the density distribution for model AS2 with 
$n=2$, $\chi=10$, and $\gamma=5/3$ interacting with a shock of $M=10$ 
at seven different times: (a) initial state ($t=-5.76 $ $\tcc$), 
(b) 1.93 $\tcc$, (c) 3 $\tcc$, (d) 6 $\tcc$, (e) 9 $\tcc$, 
(f) 12 $\tcc$, and (g) 15 $\tcc$, where the evolution
 time is measured from the stage at which the shock has reached at
 $z/r_{\rm co}=-1$ on the $z$-axis. 
The initial planar shock was initiated at $z/r_{\rm co}=-18$.
}
\label{fig:snapn2}
\end{figure}

      The shallow density gradient makes the interaction of the shock with
the cloud milder than in the previous case.  
In Figure \ref{fig:snapn2} we show 
snapshots of the density distributions of the shocked cloud.
At the initial transient stage, a bow wave forms in the upstream
flow, as can be seen at $z/r_{\rm co} \simeq -2 \sim -1$ in
Figs. \ref{fig:snapn2}b and \ref{fig:snapn2}c.
However, it does not evolve into a shock, in contrast to the previous 
case and to the uniform cloud case.  
The reflected shock in the downstream flow is also weak. 
Because the effective density contrast between the cloud and 
intercloud medium is small  
($\left<\rho_{\rm cl}\right>/\rho_{i0} \sim 2.9 $)
due to the small mass concentration, the mean relative velocity
between the cloud and intercloud medium is also small. 
As a result, the intercloud shock does not converge on the $z$-axis
at the rear of the cloud but instead interacts directly with the 
cloud shock 
(compare Figs. \ref{fig:snapn8}b and \ref{fig:snapn2}b).
Thus, shock compression from the downstream side is weak, resulting in
a slow lateral expansion.
The interaction also produces a single Mach-reflected shock 
with a single triple
point, which propagates away into the downstream flow 
from the cloud with almost the same speed
as the intercloud shock 
(Figs. \ref{fig:snapn2}c, \ref{fig:snapn2}d, and \ref{fig:snapn2}e).
Although the intercloud shock does not reflect from the $z$-axis, the 
postshock flow tends to converge on the $z$-axis, producing 
a vortex ring near the $z$-axis. 
This vortex ring propagates with a speed slower than the Mach 
reflected shock by a factor of $\sim 0.75$.

Another effect of the shallow density gradient is a significant 
retardation in the formation of the slip surface. 
Because of the smaller mean relative velocity between the cloud and intercloud
medium, it takes more time to form the slip surface. 
The Kelvin-Helmholtz instability grows very slowly 
due to both the 
small velocity shear and the retardation of the slip surface formation.
An undulation along the slip surface does not appear 
until $t \sim 9 \, t_{\rm cc}$ (Fig. \ref{fig:snapn2}e),
whereas in the $n=8$ case, it alreadly appears at
$t \sim 4 \, t_{\rm cc}$ (Fig. \ref{fig:snapn8}d).
Thus, it takes much more time for the cloud to be 
completely destroyed for $n=2$ than for $n=8$.

\subsection{Dependence on $n$ and $\chi$}
\label{subsec:dependence}

As mentioned in \S \ref{sec:formulation}, our model is specified by
four parameters: the power index of the density profile ($n$), 
the density contrast $(\chi)$, the strength of the incident shock ($M$), and 
the ratio of specific heats ($\gamma$).
Here, we consider the effects of varying $n$ and $\chi$,
which determine the mass distribution of the initial cloud.
The dependence on $M$ and $\gamma $ will be discussed in 
subsequent subsections.

First, we consider the effect of the density gradient at the cloud
boundary
(see Figs. \ref{fig:snapn8} and \ref{fig:snapn2}
and Tables \ref{tab:dependence chi = 10} and 
\ref{tab:dependence chi = 100}).
For smaller $n$, the interaction of the shock with the cloud is milder.  
For example, during the initial transient stage, the bow shock 
is weaker for smaller $n$, and the reflected shock and supersonic vortex
ring formed at the downstream are less powerful.
This is in part due to the fact that the mean relative velocity 
between the cloud and the intercloud medium decreases with $n$.
For smaller $n$, the initial mean density contrast 
between the cloud and the intercloud medium 
($\left<\chi\right> = \left<\rho_{\rm cl}\right>/\rho_{i0}$)
is smaller 
(see the seventh column of Table \ref{tab:model}), 
and thus the relative velocity just after the initial shock passage 
is smaller, since
\begin{equation}
v_{\rm rel}\equiv v_b - v_s \sim \left(1- \left<\chi\right>^{-1/2}
       \right) v_b
\label{eq:vrel}
\end{equation} 
for strong shocks.
This in turn delays the formation of the slip surface and therefore
the onset of the Kelvin-Helmholtz and Rayleigh-Taylor instabilities.
As a result, it takes longer for clouds 
with smaller $n$ to be destroyed:  
The cloud destruction times
$t_{\rm dest}$ for the $n=2$ models are longer than 
for the $n=8$ and $n=\infty$ models
by factors of $\sim 2$ and $\sim 3$, respectively
(see \S \ref{sec:cloud destruction} for the definition of $t_{\rm
dest}$).

The radial expansion is milder for smoother clouds;
the radial expansion speed decreases with $n$ 
because the relative velocity between the cloud and intercloud medium
decreases with $n$ (see \S \ref{subsec:drag}).
Also, the maximum effective radius 
normalized by the initial radius, $a/a_0$, 
decreases monotonically with $n$,
reflecting the fact that the shock-cloud interaction is
milder for smaller $n$.

\begin{figure}
\plotone{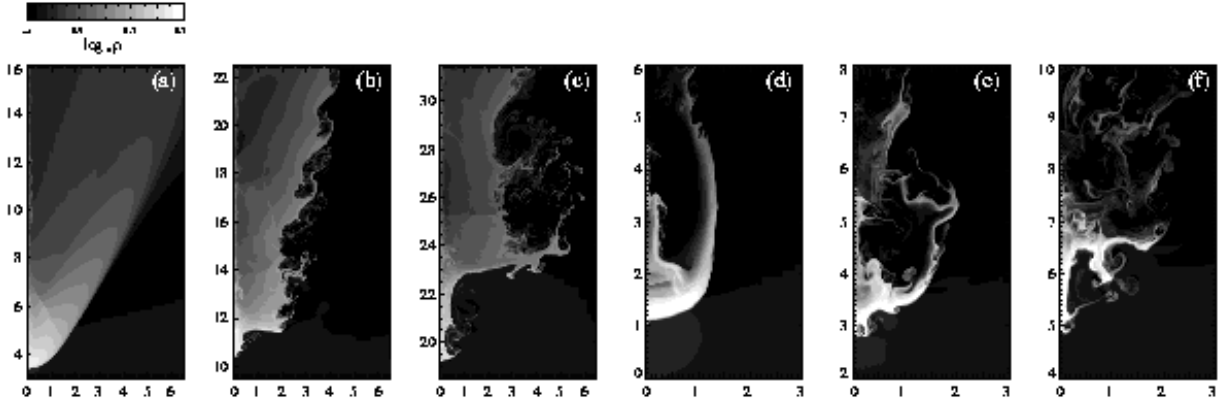}
\caption{
(a), (b), and (c): Snapshots of the density distribution
for model AS4 with $(n, \chi, M) = (2, 100, 10)$ 
at (a) $t=3 \, \tcc$, (b) $t= \, 6 \tcc$, and (c) $t= \, 9\tcc$.
(d), (e), and (f): Snapshots of the density distribution
for model AS11 with $(n, \chi, M) = (8, 100, 10)$ 
at (d) $t=2 \, \tcc$, (e) $t=4 \, \tcc$, and (f) $t=6 \, \tcc$.
\label{fig:dependence} } 
\end{figure}

Next, we consider the effect of varying the density contrast $\chi$.
Figure \ref{fig:dependence} compares the snapshots of 
the $\chi=100$ clouds with $n=8$ and $n=2$, and
Table \ref{tab:dependence chi = 100} summarizes the global
quantities for $\chi=100$ over a range of $n$.  
The most important effect of increasing the density contrast $\chi$ is 
that it leads to an increase in
the relative velocity between the cloud and the intercloud medium
($v_{\rm rel}$ increases with $\left<\chi\right>$
according to eq. \ref{eq:vrel} and $\left<\chi\right>$
increases with $\chi$ according to Table 1).
This leads to a longer drag time (see eq. [2.6] of KMC).
More significantly,
because the Kelvin-Helmholtz and Rayleigh-Taylor instabilities
both depend on the relative velocity between the cloud and intercloud
medium, 
(for the Rayleigh-Taylor instability, this is because
the higher relative velocity leads to a greater acceleration)
the higher relative velocities at higher $\chi$ lead to faster
growth of the instabilities there.  
For example, for $n=2$, a 
small-scale undulation of the slip surface due to the 
Kelvin-Helmholtz instability already appears at $t\sim 4 \, \tcc$ 
for $\chi =100$, 
while it does not appear until $t\sim 8 \, \tcc$ for $\chi =10$
(compare Figs. \ref{fig:dependence}b and \ref{fig:snapn2}d).
At $\chi=100$, the Rayleigh-Taylor instability 
leads to substantial destruction of the cloud near the $z$-axis, 
whereas the Rayleigh-Taylor instability does have
significant effects at $\chi=10$ (see Figs. \ref{fig:snapn8} and
\ref{fig:snapn2}).
Our simulations also show that the cloud destruction timescale 
measured in units of cloud-crushing times depends only weakly on $\chi$.
For the $\chi=10$ and 100 clouds with the same values of $n$, 
the results for $t_{\rm dest}/t_{cc}$ coincide within errors 
of $\sim$ 10 percent 
(see Tables \ref{tab:dependence chi = 10}
and \ref{tab:dependence chi = 100}).
This is consistent with the analytic estimate of KMC, 
which suggests that the growth times of the Kelvin-Helmholtz 
and Rayleigh-Taylor instabilities are comparable to 
the cloud-crushing time 
(see \S 2 of KMC).

	Another important effect of increasing the density
contrast is that the intercloud shock is able to converge
onto the $z$-axis for very smooth clouds ($n=2$) at $\chi=100$, whereas
it cannot do so at $\chi=10$.
As a result, at $\chi=100$ a
strong reverse shock is driven back into the cloud from the rear,
compressing the cloud in the axial direction.
After the reflected shock collides with the transmitted
shock in the cloud,
the cloud undergoes a steady expansion in the axial direction
that increases with $\chi$. 
At later times, the Rayleigh-Taylor instability also contributes to the
larger effective radius in the axial direction.
On the other hand, the radial expansion seems to reach an asymptotic
value (see \S \ref{sec:cloud drag}).

\subsection{Dependence on $M$}
\label{subsec:mach scaling}

\subsubsection{Strong Shocks: Mach Scaling}
\label{subsubsec:strong shock}

For strong shocks ($M\rightarrow \infty$), 
the hydrodynamic equations for an inviscid gas are 
invariant under the transformation 
\begin{equation}
t\rightarrow tM,~~~ v\rightarrow v/M,~~~ P\rightarrow P/M^2,
\label{eq:mach}
\end{equation}
with the position and density unchanged.
This is referred to as ``Mach scaling''.
For uniform clouds, KMC demonstrated that Mach scaling is 
reasonably valid.  Here, we investigate whether Mach scaling 
is applicable to smooth clouds.

\begin{figure}
\plotone{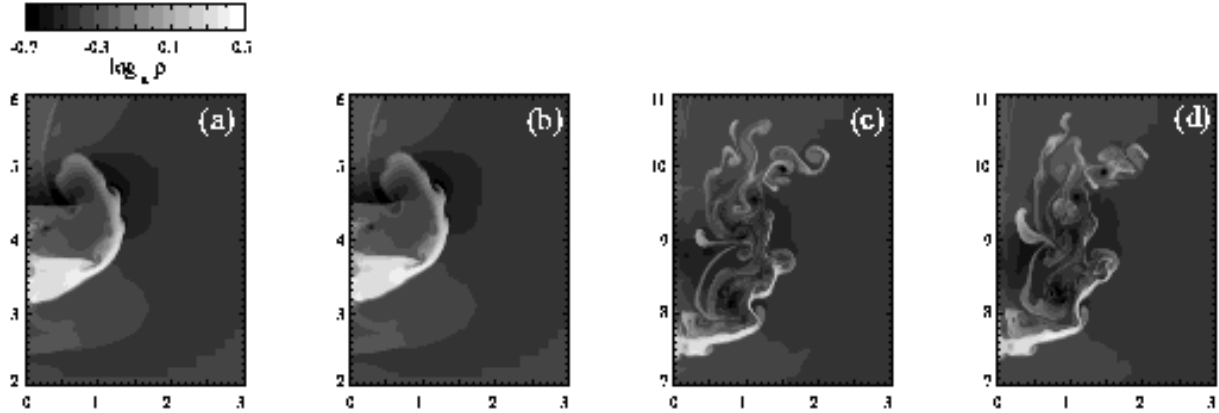}
\caption{Mach scaling. 
Comparison of the density distributions between models AS8 and AS9 with 
$n=8$ and $\chi =10$:
(a) model AS9 ($M=100$) at $t = 3.5 \, \tcc$, 
(b) model AS10 ($M=1000$) at $t = 3.5 \, \tcc$ , 
(c) model AS9 ($M=100$) at $t = 6.4 \, \tcc$,  
(d) model AS10 ($M=1000$) at $t = 6.4 \, \tcc$. 
The morphology of the cloud is remarkably similar at the first time,
 whereas small differences in the small-scale structures begin to 
appear at the second time.  
Note that the global quantities agree very well even at the later times 
(see the text for more detail).
}
\label{fig:mach}
\end{figure}

In Figure \ref{fig:mach}, we compare the density distributions
of a cloud with $n=8$, $\chi=10$ and $\gamma = 5/3$ for 
two different Mach numbers, $M=100$ and $M=1000$.
The clouds are depicted at two different times: $t=3.5 \, t_{\rm cc}$ 
(Figs. \ref{fig:mach}a and \ref{fig:mach}b) 
and $t=6.4 \, t_{\rm cc}$ (Figs. \ref{fig:mach}c and \ref{fig:mach}d).
At the first epoch (Figs. \ref{fig:mach}a and \ref{fig:mach}b), 
the cloud morphologies are essentially the same.
At the second epoch (Figs. \ref{fig:mach}c and \ref{fig:mach}d), 
morphological differences begin to appear at small scales,
even though the overall morphologies remain similar.
Those morphological differences become more significant at later stages. 
These deviations most likely originate from 
truncation errors and the artificial viscosity, 
which do not scale as equation (\ref{eq:mach}).

In spite of the deviations in the cloud morphology, 
the global quantities agree very well 
even at late times if all the variables are 
scaled as in equation (\ref{eq:mach}).
For example, for the $M=100$ and $M=1000$ cases, 
the effective cloud radii and velocity dispersions
coincide within about 5 percent and 10 percent,
respectively, until the final stage of the computation ($t\sim 10 \, 
\tcc$).  
(Values of these quantities
at $t=t_m \sim 5 \, \tcc$ are listed in Table \ref{tab:mach}.)
The characteristic timescales such as $t_{\rm drag}$, $t_{\rm dest}$,
$t_{\rm mix}$, and $t_m$ are in excellent agreement within errors
of a few \% for strong shocks.

\subsubsection{Weak Shocks}

Here we consider the weak shock case, in which Mach scaling 
is not applicable.  We follow two cases
with initial parameters that are the same as those of \S 
\ref{subsec:n=8} and \S \ref{subsec:n=2}, except that
the Mach number is $M=1.5$.

In Table \ref{tab:mach}, we summarize the global quantities of the
models with $(n, \chi, M) = (8, 10, 1.5)$ and (2, 10, 1.5).
Our numerical simulations show that the overall evolution of the 
weak shock case is qualitatively similar to that of the strong shock case.
An important difference between the weak and strong 
shock cases is that cloud drag is less efficient in the weak-shock case,
in part due to slower lateral expansion of the shocked cloud.
For example, for $n=8$ and 2, the drag times are 17.2 $\tcc$ 
and 21.4  $\tcc$, respectively, for the weak-shock case,
whereas they are 3.3  $\tcc$ and 7.5  $\tcc$, respectively,
for the strong-shock case.

Furthermore, the relative velocity between the cloud and 
intercloud medium is smaller in the weak-shock case,
even after allowing for Mach scaling.  
This smaller relative velocity leads to slower growth 
of the Kelvin-Helmholtz instability (and thus slower cloud destruction). 
For example, for $n=8$, the ripple along the slip surface due to 
the Kelvin-Helmholtz instability begins to appear at $t\sim 12 \, \tcc$ in
the $M=1.5$ case, while it already appears at $t \sim 4 \, \tcc$ for $M=10$.
Thus, the cloud destruction time is about 2 $-$ 3 times longer 
in the weak-shock case.

The supersonic vortex ring is weak or absent for weak shocks.
For $M=1.5$ we did not observe the formation of a supersonic 
vortex ring for either $n=8$ or $n=2$, 
although the interaction between the reflected shock and the 
incident shock did generate a single triple point.
Also, for the weak shock, a bow wave forms instead of a bow shock. 
Despite these qualitative differences between the strong and weak
shock cases, the overall evolution of the weak shock case 
is quite similar to that of the strong shock case.
We thus conclude that varying the strength of the incident shock 
does not alter our main conclusions on the shock-cloud interaction.

\subsection{Dependence on $\gamma$: Effects of Radiative Losses}
\label{subsec:gamma}

In the above subsections, we considered the evolution 
of nonradiative clouds, in which $\gamma$ was set to $5/3$.
Here, we consider the effects of radiative losses,
which we model by setting $\gamma=1.1$.
Such a model is applicable for relatively large HI clouds 
and molecular clouds (e.g., KMC, Larson 2005)
We recalculate the evolution of two models with ($n, \chi, M$) = 
(8, 10, 10) and (2, 10, 10). 
The numerical results are summarized in Table \ref{tab:radiative}.
Note that in our calculations, $\gamma$ is  
1.1 for both the cloud and the intercloud 
gas.  In contrast, KMC set $\gamma=5/3$ for the 
intercloud gas in their radiative calculations.

The softer equation of state leads to substantially greater compression;
e.g., for the radiative model with $n=8$, the effective radii, $a$ and $c$, 
are reduced by as much as factors of 2.3 and 2.6, respectively, whereas 
for $\gamma = 5/3$, $a$ and $c$ are reduced by factors of 1.3 and 1.8, 
respectively.
Thus, the cross section of the radiative cloud is smaller than that of the
nonradiative cloud, which
lessens the efficiency of the cloud drag 
by the ambient flow.
On the other hand, the post-shock intercloud density is higher than  
in the nonradiative case by a factor of about 4.7 (for $M=10$). 
This higher intercloud density gives a greater drag force 
on the shocked cloud than in the nonradiative case (see
eq. \ref{eq:drag} below).
In our calculations, the latter effect is more important, and 
as a result, the radiative drag times are much shorter than 
the nonradiative ones
for both the $n=8$ and $n=2$ models.
Because of this greater acceleration by the ambient gas, 
the relative velocity at the cloud boundary is rapidly reduced,
leading to milder cloud destruction by the Kelvin-Helmholtz and 
Rayleigh-Taylor instabilities 
(see Table \ref{tab:radiative}).
The higher compression in the radiative case
may facilitate gravitational collapse 
in the shocked cloud (\S \ref{sec:discussion}).


\subsection{Effect of Cloud Geometry}
\label{subsec:cylinder}

To investigate the effect of the cloud geometry, we consider 
cylindrical clouds.
We adopt a Cartesian coordinate system ($x, y, z$)
and assume that the physical quantities are uniform 
along the $z$-axis, which is identical to the cylinder axis.
Mirror symmetry is assumed with respect to the $x$-$z$ plane.
The shock propagates in the $y$-direction.
Hence, the $x$- and $z$-axes in this case correspond to the $\varpi$ axis
in the spherical case, and the $y$-axis in this case
corresponds to the $z$-axis in the spherical case.
In Table \ref{tab:cylinder}, we summarize the global quantities 
of the cylindrical clouds with $(n,\chi,M)=$ (2, 10, 10) and
(8, 10, 10).
Comparison between the cylindrical and spherical clouds 
indicates that the cloud shape becomes more flattened 
and longer in the transverse direction for the cylindrical cloud.  
For example, for the $n=8$ case, the effective radius in the 
transverse direction reaches $a/r_{\rm co} \simeq 2.5$ in the cylindrical 
cloud, which is about 20 \% larger than that of the spherical cloud
with the same $n$.
The somewhat greater expansion of the cylindrical cloud 
causes more efficient drag. Therefore, the cloud is 
accelerated more rapidly for the cylindrical cloud, leading
to a slightly slower cloud destruction due to smaller
relative velocities between the cloud and intercloud medium.
This slower cloud destruction also causes a slower increase
in the axial cloud size for the cylindrical cloud;
at $t = 12 \, \tcc$, it reaches about $c/r_{\rm co}\sim 4.0$ in the 
cylindrical cloud, while it reaches about $c/r_{\rm co}\sim 4.7$ in the 
spherical cloud. 
In spite of these quantitative differences, the overall evolution 
of the cylindrical clouds is quite similar to that of the spherical clouds.
Thus, we conclude that varying the cloud geometry does not alter our 
main conclusions obtained from the spherical calculations.

\section{Cloud Drag}
\label{sec:cloud drag}

As shown above, the slip surface is subject to the Kelvin-Helmholtz
and Rayleigh-Taylor instabilities, 
and as a result, the entire cloud is eventually destroyed.
The acceleration of the cloud is responsible for the growth of 
the Rayleigh-Taylor instability, and the time history of
the velocity, which is determined by the acceleration, governs
the evolution of the Kelvin-Helmholz instability.
Therefore, it is very important to quantitatively understand 
the cloud acceleration.
For uniform clouds, KMC constructed a simple analytic formula for 
the cloud velocity that reproduced the numerical results quite well.
Here, we extend their analytic model to smooth clouds.

\subsection{Analytic Formula for the Cloud Velocity}
\label{sec:analytic}

According to \S 2 of KMC, the equation of motion for a shocked cloud 
can be expressed as
\begin{equation}
m_{\rm cl,0} \frac{dv_c}{dt} = - \frac{1}{2} C_D \rho_{i1} v_c^2 A (t)  \ ,
\label{eq:drag}
\end{equation}
where $m_{\rm cl,0}$ is the cloud mass, $v_c$ is the mean velocity of the cloud 
relative to the velocity of the shocked intercloud medium, 
$\rho_{i1}$ is the density of the post-shock ambient medium, 
$C_D$ is the cloud drag coefficient, and $A(t)$ is the 
time-dependent cross section of the
cloud (see Figs. \ref{fig:snapn8}
and \ref{fig:snapn2}).
As shown below in Figure \ref{fig:cloud shape1}, 
the cloud radius ($a$) 
is reduced slightly in the shock compression stage, and 
then increases significantly with time during the post-shock expansion stage. 
Thereafter, it does not change much after it reaches its maximum.
Here, we approximate the evolution of the cross section of the cloud as 
\begin{eqnarray}
A(t) = \pi a^2 &=& \pi \left[a_0^2 + (C_c t)^2\right]  
\hspace{0.6cm}  t \le t_m  \\
&=& \pi \left[a_0^2 + (C_c t_m)^2\right] \hspace{0.3cm} t > t_m  \ ,
\label{eq:cloud drag}
\end{eqnarray}
where $a_0$ is the effective value of the initial cloud radius and 
$C_c$ is the expansion velocity in the $\varpi$-direction.
In this equation, we assume that the cloud expands 
at a constant speed, $C_c$, until it reaches its maximum transverse
extent.
We define the maximum expansion time, $t_m$, to be 
the time at which the effective radius in the $\varpi$-direction
($a$) has increased to 90 \% of its maximum value.
The maximum expansion times obtained from the numerical calculations 
are listed in Tables \ref{tab:dependence chi = 10} $-$ \ref{tab:cylinder}.
Equation (\ref{eq:cloud drag}) 
is identical in form to the expression given by KMC, but
there is an important difference:
$C_c$ is {\it not} necessarily equal
to the sound speed in the cloud, but instead 
depends on model parameters.

    In order to obtain a result that is directly comparable
to that of KMC, we work in the frame of the shocked intercloud
medium, denoted by a prime.
Integrating equation (\ref{eq:drag}) with respect to time, 
we obtain the following form of the cloud velocity,
\begin{eqnarray}
\frac{v_c'}{v_{c0}'} &=& \left\{1 + \left(\frac{9C_D'}{8\chi^{1/2}}\right)
\frac{t}{\tcc}\left[\frac{a_0^2}{r_{\rm co}^2} + 
\frac{1}{3}\left(\frac{C_c}{v_s}\frac{t}{\tcc}\right)^2\right]\right\}^{-1}
 ~~~~\hspace{2 cm} (t \le t_m) \label{eq:velocity0} \\
&=& \left\{1 + \left(\frac{9C_D'}{8\chi^{1/2}}\right)
\left[\frac{a_0^2}{r_{\rm co}^2}\frac{t}{\tcc} + \left(\frac{C_c}{v_s}
\frac{t_m}{\tcc}\right)^2 \left(\frac{t}{\tcc} - \frac{2}{3}
\frac{t_m}{\tcc} \right)\right]\right\}^{-1}~~~~  (t > t_m)
\label{eq:velocity}
\end{eqnarray}
where 
\begin{equation}
C_D'\equiv \frac{2}{3(\gamma_i-1)\calr}\; C_D.
\end{equation}
We have assumed that the shock is
strong ($M\gg 1$) and made the approximation $r_{\rm co}\simeq v_s \, \tcc$, where
$v_s\simeq v_b/\chi^{1/2}$ is the velocity of the transmitted
cloud shock. 
This result is identical to that of KMC except that the
drag coefficient $C_D$ is replaced by an effective drag
coefficient $C_D'$ that can be much less than
$C_D$ for very smooth clouds (see below eq. \ref{eq:calr}).
In the case considered
by KMC (a uniform cloud [$\calr=1$] in an intercloud medium with 
$\gamma_i=\frac 53$), $C_D'=C_D$.
The initial cloud velocity relative to the shocked ambient gas
is  $v_{c0}'=-2v_b/(\gamma+1)$.
In the following, we set the drag coefficient $C_D = 1$ (see KMC).

\subsection{Comparison with Numerical Simulations}
\label{subsec:drag}

\subsubsection{Cloud Shape}

In the left panels of Fig. \ref{fig:cloud shape1}, 
we show the time evolution of the rms cloud radii, $a$ and $c$.
The cloud radius in the transverse direction, $a$, decreases with time
in the shock compression stage, and then increases in the 
post-shock expansion stage, finally reaching a maximum.
Thereafter, it shrinks by about 5\%-30\%, depending on 
$n$ and $\chi$, with the greater shrinkage occurring for the
larger values of $\chi$.
On the other hand, the axial radius, $c$, keeps increasing with time.
Thus, the shocked cloud becomes elongated in the direction of shock
propagation.  
The initial reduction in the transverse radius 
due to shock compression happens at
a later time for smaller $n$; 
for $(\chi, M)=(10, 10)$, the radius 
$a$ takes its minimum value at $t\sim 1.7$ $\tcc$ 
for $n=8$ and 3.2 $\tcc$ for $n=2$.
As $n$ and $\chi$ increase,
the cloud expands more rapidly,
the aspect ratio of the shocked cloud, $c/a$, 
increases more rapidly, and the
maximum value of $a/a_0$ increases.

For a uniform cloud, the expansion speed of the cloud can be
approximated as the sound speed of the shocked cloud material (KMC), 
\begin{equation}
\frac{C_c}{v_b} \simeq \frac{C_c}{\chi ^{1/2}v_s}
= \left[\frac{2\gamma(\gamma-1)}{\chi(\gamma+1)^2}\right]^{1/2}\ .
\end{equation}
For $\chi=10$, this gives $C_c=0.56v_s$ for $\gamma=5/3$.
For smooth clouds, the expansion speeds are smaller:
$C_c/v_s \sim $ 0.07 , 0.23, and 0.25
for $n=2$, 4, and 8 with $\chi=10$, respectively.
The increase of expansion speed with $n$ is 
due to the increase in relative velocity between the
cloud and the intercloud medium as the cloud boundary
becomes sharper and the mean density contrast larger; this in turn leads to
a larger pressure difference between the shocked cloud and
the intercloud medium.

	The expansion speed also tends to increase
with $\chi$. In addition to the effect just
described, the shock driven into the rear of the cloud augments
the pressure in the cloud interior due to the 
transmitted shock.
As a result, the
radial expansion due to the pressure difference between the
cloud and the side of the cloud is somewhat faster.
Indeed, for $n\geq 8$ and $\chi \gg 10$, the expansion
speed $C_c$ is approximately equal to the adiabatic
sound speed in the cloud, just as in the case of uniform
clouds.

\subsubsection{Cloud Velocity}

In the right panels of Fig. \ref{fig:cloud shape1}, 
we compare the analytic result for the cloud
velocity given in equation (\ref{eq:velocity}) with 
the value obtained from the numerical calculations.
As shown in the figures, the expansion velocity $C_c$ depends 
on $n$ and $\chi$.
Since we determine $C_c$
by fitting the evolution of the cloud radius 
$a$ in the time interval $ t \le t_m$, our result for the cloud
velocity is really only semi-analytic.

For clouds with a large density gradient, i.e., $n=8$, 
our analytic model of the cloud velocity agrees 
well with the numerical results for both $\chi =10$
and 100.  
For $\chi=10$, the analytic cloud velocities 
are within about 25 \% and 10 \% of the numerical values
for $t \le t_{\rm drag}$ and $t>t_{\rm drag}$, respectively.
The cloud velocity of the $n=8$ cloud drops more slowly 
than the uniform case because its radial expansion is
slower, and thus its cross section is smaller.
For $\chi=100$, the numerical values and analytic model agree to within
10 \% until the radius $a$ reaches its maximum. Thereafter,
the deviation 
increases as the cloud undergoes a brief bout of acceleration,
presumably due to the effects of the reverse shock.
A similar effect 
is observed in some other models [see e.g., panel (d)],
where it leads to a deviation from our model that
is only temporary.

	For smooth clouds with $n = 2$, the initial reduction in 
the effective cloud radius associated with the cloud compression
significantly affects the cloud acceleration.
Therefore, for such clouds we replace $a_0$ in eq. (\ref{eq:cloud drag}) 
with the time-averaged radius over the time interval 
$t_0 \le t \le t_1$, where
$t_0$ is the time at which the shock first reaches the cloud surface 
on the $z$-axis, and $t_1$ is the time at which the cloud radius 
expands back to its initial value.
This procedure reduces $a_0$ in eq. (\ref{eq:cloud drag}) by about 10\%
and 17\% for $\chi=10$ and 100, respectively.
Even with this adjustment, the agreement between the 
analytic model and the numerical simulations at early times 
is not as good as 
for the $n=8$ cases. This is 
because a significant fraction of the mass of 
the cloud is not much denser than the intercloud medium
(the mean density contrast between the 
cloud and the intercloud medium  is only
$\left<\rho\right>/\rho_{i0} \sim 3$ 
for the $n=2$ case),
and as a result the transmitted shock gives this gas
a significant velocity.
However, in deriving equation (\ref{eq:velocity}),
we assumed that the cloud velocity is essentially zero 
immediately after the initial shock passage.
At later times, the agreement between the 
analytic model and the numerical 
solution becomes much better in the $n=2$ case: 
for $5 \, \tcc \lesssim t \lesssim 20 \, \tcc$,
the agreement is within
about 20 \% for both $\chi=10$ and 100.
Note that the cloud was still expanding at the
end of the calculation for $\chi=100$ ($t<t_m$), so
we were unable to fully test the model in this case.

In summary, our analytic model of the cloud drag can reproduce
the numerical results reasonably well over a  wide range of initial 
conditions.

\section{Formation of the Slip Surface}
\label{sec:slip surface}

As shown in \S \ref{sec:result}, smooth clouds are completely destroyed
by Kelvin-Helmholtz and Rayleigh-Taylor instabilities, whose 
growth rates depend on the wavelength of the perturbation.
According to linear theory, the growth timescales of 
these instabilities 
are estimated as $t_{\rm KH}/\tcc \sim v_b/(v_{\rm rel}kr_{\rm co})$ 
and $t_{\rm RT}/\tcc=1/(kr_{\rm co})^{-1/2}$, respectively
(see \S 2 of KMC), where $k$ is the wavenumber of the perturbation. 
In both cases, the instability grows faster on smaller scales.
Furthermore, linear theory shows that 
perturbations with wavelengths smaller 
than the thickness of the shear layer are stabilized significantly. 
Thus, the growth of the instabilities is closely related to 
the development of the slip surface.
In this section, we investigate how a slip surface develops 
in a smooth cloud.

In Figures \ref{fig:slip}a $-$ \ref{fig:slip}c, 
we represent the distributions of $\rho$, $v_r$, and $v_z$
along $r=0.5r_{\rm co}$ at five different times 
for the cloud with small density gradient, i.e., $n=8$ and $\chi=10$.
When the transmitted shock passes through the cloud (dash-dot-dot-dot
curves), the maximum density increases by a factor of $\sim 4$,
as inferred from the jump condition, and then decreases as the cloud 
expands laterally.
At the first encounter of the shock, the velocity shear layer is formed 
at the cloud envelope with a thickness $h_{\rm bd}$.
As can be seen in panel (c), the cloud envelope at the upstream side 
moves faster than the high density core, catching up with the 
high density core.  As a result, the radial velocity profile
bcomes steep there [panel (b)], and thus the slip surface forms.
At the same time, the density profile becomes steep and therefore
the slip surface is also identical to a contact discontinuity.
Here, we define the formation time of the slip surface, $t_{\rm slip}$, 
as the time at which the radial velocity gradient 
reaches a first local maximum at the upstream side of the shocked cloud.
In this model, $t_{\rm slip}$ is estimated 
as $t_{\rm slip}\simeq 1.71 \, \tcc$.
At $r= 0.5r_{\rm co}$, where we measured the physical quantities, 
the slip surface is almost perpendicular to the $z$-axis, 
and thus the radial velocity difference across the slip
surface can be regarded as the total velocity difference there.
We note that the definition of $t_{\rm slip}$ depends on the position 
where the radial velocity is measured. However, as long as we measure
the radial velocity gradient in the range of 
$0.3 \lesssim r/r_{\rm co} \lesssim 0.7 $, 
we get almost the same value. Thus, we measure $t_{\rm slip}$ 
at $r=0.5r_{\rm co}$ for all models.
At $t\simeq 3 \, \tcc$, the reverse shock compresses the cloud from the rear,
and the cloud is squeezed from both sides by the reverse shock 
and the post-shock intercloud flow.
Thus, the density gradient of the slip surface 
becomes very steep as can be seen in 
the solid curve in panel (a).
The slip surface forms close to the core, and it can form
only after the transmitted shock has reached the core.

For the cloud with small density gradient, i.e., $(n, \chi)$ = (2, 10),
the shear layer has a wider thickness than for $n=8$,
at the first encounter of the shock.
The lateral expansion of the core is a little faster for the smoother
cloud, which leads to a smaller velocity difference between the core 
and intercloud medium.
Because of this smaller velocity difference, 
the formation of the slip surface is retarded for the smooth cloud.
In this model, $t_{\rm slip}$ is evaluated as $t_{\rm slip} = 4.92 \, \tcc$.
Since the reverse shock is weak in this model, the additional
compression due to the reverse shock observed in the $n=8$ case 
does not happen.

When the slip surface forms, the Kelvin-Helmholtz instability
begins to develop and produce ripples along it. 
For all the models calculated in this paper, 
the velocity dispersions in the axial and radial directions, 
$\delta v_\varpi$ and $\delta v_z$, begin to rise
at the epoch of ripple formation due to the Kelvin-Helmholtz instability.
The time evolution of the velocity dispersions
is shown in Figure \ref{fig:dispersion} for two models with
$(n, \chi, M)=(2, 10, 10)$ and $(8, 10, 10)$.
For the $n=8$ case, the velocity dispersions in the $\varpi$- and 
$z$-directions begin to increase at $t\simeq 2.3 \, \tcc$ 
and $3.2 \, \tcc$, respectively.
For the $n=2$ case, they 
begin to increase at  $t\simeq 7 \, \tcc$ and $8 \, \tcc$, respectively.
These times are in good agreement with the times of ripple formation
(3.5 $\tcc$ and 8.4 $\tcc$ for $n=8$ and 2, respectively).
This indicates that the small-scale perturbations due to the 
Kelvin-Helmholtz instability indeed generate random motions 
of the small fragments.
We note that the velocity dispersions in the axial and radial directions
reach about 10 \% of the incident shock velocity over a wide range of
initial model parameters.
They tend to be slightly larger for clouds with 
steeper density profiles (larger $n$---
see Tables \ref{tab:dependence chi = 10} through \ref{tab:cylinder}).

\section{Cloud Destruction}
\label{sec:cloud destruction}

The transmitted shock deforms a smooth cloud into an
arc-like structure bounded by
a slip surface.
The slip surface is subject to the Kelvin-Helmholtz 
and Rayleigh-Taylor instabilites, which produce
fragments that are swept away by the flow. 
Although their growth rates are smaller, fluctuations with larger 
wavelengths break the cloud into relatively large
fragments, destroying the cloud more efficiently.
On the other hand, fluctuations with smaller wavelengths contribute 
more efficiently to mixing the cloud material with the ambient gas.
To evaluate these two processes of cloud destruction, 
we consider two different timescales: the cloud destruction time 
($t_{\rm dest}$) and the mixing time ($t_{\rm mix}$),
which are defined in \S \ref{subsec:cloud destruction}
and \S \ref{subsec:cloud mixing}, respectively.

\subsection{Cloud Destruction Time: $t_{\rm dest}$}
\label{subsec:cloud destruction}

When fluctuations with wavelengths comparable to the cloud size
grow in time, they break the cloud into relatively large
fragments.  This large-scale fragmentation decreases the mass of 
the main body of the shocked cloud efficiently.
To make a quantitative estimate of large-scale fragmentation, 
we follow KMC and define the cloud destruction time, $t_{\rm dest}$,
as the time at which the mass of the main axial fragment has become
smaller than the initial cloud mass by a factor of $1/e$.
The mass of the main axial fragment is evaluated as the mass of the gas
that is physically connected to the main axial fragment by gas
with $\rho> 2\, \rho_{i1}$. (Recall that the initial cloud is 
defined as the gas whose density is greater than $2 \, \rho_{i0}$.)
In practice, we identified the gas in the main axial fragment 
by drawing a density contour at a level of $2 \, \rho_{i1}$ and then
integrating the density in the region surrounded by the contour line.

	    In Tables \ref{tab:dependence chi = 10} through 
\ref{tab:cylinder}, 
we summarize the cloud destruction times obtained from our numerical
simulations.
The cloud destruction time is longer for smoother clouds
because both the Kelvin-Helmholtz and Rayleigh-Taylor instabilites 
grow more slowly in such clouds. 
For example, for $n =2$,  
$t_{\rm dest}/\tcc \sim 10 -12$, which is about three times 
larger than for the uniform case.
On the other hand, 
$t_{\rm dest}/t_{\rm cc}$ depends only weakly on the initial density 
contrast, $\chi$,   
because the growth rates of the Kelvin-Helmholtz and 
Rayleigh-Taylor instabilities scale with $\tcc$ (see the previous section).

We note that our cloud destruction times of the uniform clouds tend to 
be smaller than those of KMC. 
For $\chi=10$, we find $t_{\rm dest}=3.45 \, \tcc$ whereas they obtained
$t_{\rm dest}=3.79 \, t_{\rm cc}$; for $\chi=100$, we find
$t_{\rm dest}=3.02 \, \tcc$ vs. their value of $3.90 \, t_{\rm cc}$.
This is due to the fact that
our cloud destruction time depends weakly on the threshold density 
of the main axial fragment, especially for large $\chi$.
In this paper, the threshold density of the main axial fragment 
is set to twice the shocked ambient density, $2\, \rho_{i1}$, for all the models.
For smaller values of the threshold density, the cloud destruction
time increases: For example, at $\chi=100$, 
$t_{\rm dest} \simeq 3.6 \, t_{\rm cc}$ for 
$\rho_{\rm th}=1.5 \, \rho_{i1}$, and $t_{\rm dest} \simeq 3.86\, \tcc$
for $\rho_{\rm th}=1.2 \, \rho_{i1}$. 
Thus, our cloud destruction times seem to approach those of KMC
as the threshold density decreases, and we
expect them to coincide with those of KMC 
if we were able to count in all the cloud material.

\subsection{Mixing Time: $t_{\rm mix}$}
\label{subsec:cloud mixing}

	Fluctuations with relatively short wavelengths,
which have relatively large growth rates, are
more effective at mixing cloud material with the 
the ambient gas than the long wavelength fluctuations
that break up the cloud.
To evaluate the efficiency of mixing, we define the mixing time, 
$t_{\rm mix}$, as the time at which the mass of the region whose density 
is greater than $2\rho_{i1}$ ($m_{\rm cl}$) reaches half the initial 
cloud mass ($m_{\rm cl,0}$).

In Figure \ref{fig:mixing}, we depict the time evolution of the 
cloud mass normalized to its initial value, $m_{\rm cl}/m_{\rm cl,0}$.
The cloud mass decreases more rapidly for smooth clouds than
for uniform clouds 
at early times, whereas at late times, it decreases more slowly.
This indicates that the lower density gas in the cloud envelope 
tends to be mixed more rapidly, but the high density gas in 
the central core tends to be mixed only after the formation of
the slip surface, which is a prerequisite for the onset
of the Kelvin-Helmholtz instability and  
which is retarded for a smooth clouds.
For the $n=2$ models, the mixing time is 
determined by mixing of the envelope because of the large fractional mass
of low density gas, whereas for models with 
larger $n$, the mixing time is determined by the KH instability.

     Because the mixing time depends on relatively small-scale
fluctuations, one might expect it to depend sensitively on the
spatial resolution of the computation.
However, comparison with different spatial resolutions 
[$R_{30}$, $R_{60}$, $R_{120}$, and $R_{240}$ for
$(n,\chi,M)=(8,10,10)$]
shows that the mixing time at $R_{120}$ is within 8 \% of 
that at $R_{240}$, indicating that the mixing is 
reasonably accurate at $R_{120}$.
Also, the errors of the mixing time decreases 
nearly in proportion to the inverse of the cell number,
which implies first order accuracy.

Figure \ref{fig:mass fraction} shows 
how the cloud material is mixed with the ambient gas
for two models: $(n,\chi,M)=(8,10,10)$ and (2, 10, 10).
The histograms indicate the fraction of mass $\Delta m/m_{\rm cl,0}$ 
within a corresponding density bin with a width of $0.2 \, \rho_{\rm c0}$.
For $n=8$, about 50 \% of the total cloud mass has density 
of $\simeq \rho_{c0}$ at the initial state. 
When the shock is transmitted into the cloud, the cloud density 
near the center increases by a factor of 4. 
At the same time, however, 
the shocked cloud starts lateral expansion, which reduces the density of 
the shocked cloud material.  At $t\simeq t_{\rm cc}$, 
the distribution of the mass fraction is roughly flat 
over the range $0.4 \, \rho_{c0} \le \rho \le 4.4 \, \rho_{c0}$.
At this time, the cloud mass (the mass of gas with
$\rho>2 \, \rho{i1}$) has dropped to about 85\% of the initial
cloud mass. Gas from the initial cloud that is at lower densities
corresponds to envelope gas that has been shocked by both 
an incident shock and a reverse shock.
Two shocks increase the entropy of the shocked gas above
that that due to one shock. As a result the final
density is only about twice the initial value (KMC),
and some of the gas in the shocked envelope has a density
below $2 \, \rho_{i1}$.
Mixing due to hydrodynamic instabilities starts 
at $t\sim 4 \, \tcc$.
As the cloud material mixes with the intercloud medium, 
the density of the mixed gas decreases with time, 
approaching the intercloud density with $\rho \simeq 0.4 \, \rho_{c0}$.

     At $t=6.04 \, \tcc$ (Fig. \ref{fig:mass fraction} c), 
the mass distribution has two peaks, at 
$\rho/\rho_{\rm c0} \sim 0.7$ and 2.5.
The higher-density peak at $\rho/\rho_{\rm c0} \sim 2.5$ corresponds to
the mass in the main axial fragment, which decreases with time.
The lower-density peak at $\rho/\rho_{\rm c0} \sim 0.7$ corresponds to
the mass that is mixed by the shocked ambient medium.
The low density peak shifts towards lower density with time, gradually 
approaching the density of the shocked ambient medium,
$\rho_{i1}\sim 0.4 \, \rho_{c0}$.
By $t\simeq 10 \, t_{\rm cc}$, most of the cloud material has densities 
lower than 0.8 $\rho_{c0}$.
On the other hand, for $n=2$, most of the mass 
(about 80 \% of the total mass) is already contained  
in the lowest density bin,
$0.4 \, \rho_{c0}\le \Delta \rho < 0.6 \, \rho_{c0}$.  
After passage of the shock, the density range of the cloud material 
widens immediately, although 
more than half the mass has a density less than $\rho_{c0}$.
The Kelvin-Helmholtz instability mixes 
cloud material with the ambient medium, thereby reducing the mass fraction 
of the high density material.  By the time shown in 
panel (h) of Fig. \ref{fig:mass fraction}, most of the cloud material
has a density  $\sim 0.5 \, \rho_{c0}$ owing to efficient mixing.

Our simulations demonstrate that the destruction and mixing 
of smooth clouds depend sensitively on the initial density distribution.  
In clouds with relatively steep density
gradients, the core is almost completely destroyed 
by $t_{\rm mix}$;
for example, for $n=8$ and $\chi=100$, the core mass 
is 
0.10 $m_0$ at $t=t_{\rm mix}$ (recall that in this
case, the core mass is the mass with $\rho> 0.51 \, \rho_{c0}$). 
Thus, at this point
the high density core is almost destroyed, 
and cloud mixing proceeds very efficiently throughout the entire cloud.
By contrast, significantly more of the core survives
for clouds with shallow density 
gradients: $m_{\rm co}(t_{\rm mix})= 0.29 m_{\rm co,0}$ for
the model with $n=4$ and $\chi=100$, and 
$m_{\rm co}(t_{\rm mix})= 0.25 m_{\rm co,0}$ 
for the model with $n=2$ and $\chi=100$.
For the case of a shallow density gradient, the cloud envelope is 
stripped from the main body of the cloud by $t=t_{\rm mix}$, but
much of the core remains intact.
In such a core, the velocity dispersion is also relatively small,
which indicates that random motions produced by the shock-cloud interaction 
are directly associated with cloud destruction, and do not significantly 
affect the cloud core before the core is destroyed.

\section{Vorticity}
\label{sec:vorticity}

The interaction of a shock wave with a cloud generates powerful 
vortex rings, which play an important role in the destruction of the cloud.
For uniform clouds, KMC constructed a simple analytic model for vorticity
production by the shock-cloud interaction that
was in good agreement with their numerical results.
In this section, we extend their model to 
the smooth clouds, and compare it with the numerical results 
presented in the previous sections.

\subsection{Vorticity Production: Analytic Model}

In the course of the shock-cloud interaction, 
vorticity is produced by the baroclinic term 
(the right-hand side of eq. [\ref{eq:vorticity}]) in 
the vorticity equation for an inviscid fluid, 
\begin{equation}
\frac{\partial \mbox{\boldmath$\omega$}}{\partial t} + 
\nabla \mbox{\boldmath$\times$}
\left(\mbox{\boldmath$\omega \times v$}\right) = \nabla P \times 
\nabla \left(\frac{1}{\rho}\right)    \, .
\label{eq:vorticity}
\end{equation}
Integrating the above equation over a cross section $\mbox{\boldmath$A$}$,  
we obtain the increase rate of circulation 
($\Gamma \equiv \int \mbox{\boldmath$\omega \cdot dA$} $), 
\begin{equation}
\frac{d\Gamma}{dt} = \int \left(\nabla P\times \nabla \frac{1}{\rho}\right) 
\mbox{\boldmath$\cdot dA$}   \, .
\end{equation}
According to KMC, the vorticity production can be classified into 
four components.  Two are associated with the cloud-intercloud boundary
and are produced by the initial passage of the shock 
($\Gamma_{\rm shock}$) and by the subsequent post-shock flow 
($\Gamma _{\rm post}$).   The third is associated with the triple points 
associated with the Mach reflected shocks behind the cloud 
($\Gamma _{\rm ring}$).  The fourth is the vorticity produced 
in the cloud ($\Gamma _{\rm cloud}$), which is smaller than 
those in the other components by a factor of order $\chi^{-1/2}$.
As shown below, the vorticity production in smooth clouds 
is qualitatively similar to that in uniform clouds.

First, we consider 
the circulation produced by the initial passage of the shock,
$\Gamma_{\rm shock}$, defined
as the circulation generated 
by the time at which the shock sweeps over the cloud and 
converges on the $z$-axis.
For the uniform case, this component is generated only in a thin layer 
between the cloud and the intercloud medium.
In contrast, for smooth clouds, the layer where this component is produced 
is extended in the radial direction because of the smooth cloud boundary.
Indeed, as we shall 
see below, for smooth clouds such as those with $n=2$ and $\chi=10$, 
this circulation is produced 
throughout most of the cloud envelope, 
not just at the cloud surface.
The time for the shock to sweep over the cloud is 
$t_{\rm sweep}\sim 3 a_0/v_b$.
For uniform clouds, this estimate of the time scale
coincides with that of KMC. 
KMC estimated the increase rate of $\Gamma_{\rm shock}$ as 
\begin{equation}
\frac{d\Gamma}{dt} \simeq -\frac{3}{4}
\left(\frac{v_b \cos \theta}{h}\right)
\left(\frac{dz/\cos \theta}{dt}\right) h \;(1-\chi^{-1/2})
= -\frac{3}{4}v_b^2\;(1-\chi^{-1/2}) \ ,
\label{eq:gammashk}
\end{equation}
where $\theta$ is the angle at which the shock impacts the surface and 
$h$ is the thickness of the shear layer, which is approximately
the grid-cell size for uniform clouds.
A strong shock is also assumed (see eq. 7.8 of KMC for more details).
KMC allowed for a reduction in the shear
due to the motion of the shocked cloud
by inserting a factor
$(1-\chi^{-1/2})$
in their final result for $\Gamma_{\rm shock}$; here we have included
this factor in the original equation as well.

   Equation (\ref{eq:gammashk})
indicates that the rate of increase of $\Gamma_{\rm shock}$
is  independent of the thickness of the shear layer.
Thus, the fact that
smooth clouds have a shear layer with a thickness
$h\sim h_{\rm bd}$, the scale height in the boundary
(eq. \ref{eq:hbd}), should not affect the growth of the circulation.
Thus, we find 
\begin{equation}
\Gamma_{\rm shock} \simeq \frac{d\Gamma}{dt} \times t_{\rm sweep} 
\simeq -\frac{9}{4} v_b a_{0} (1 - \chi^{-1/2})  \   .
\label{eq:gamma_shock}
\end{equation}
In the next subsection, we show that this simple model
reproducesour numerical results  reasonably well.

Next, we consider the vorticity production due to the interaction
of the shocked cloud with the post-shock flow, $\Gamma_{\rm post}$.
When the initial shock passes through the cloud and 
the slip surface is formed at the upstream side of the cloud, 
the vorticity continues to be generated at the slip surface 
by the baroclinic term, which is non-zero when the density
and pressure gradients are not parallel, 
and the equation of state is not barotropic.
The pressure is approximately constant {\it across} the slip surface, 
but it varies {\it along} the slip surface, being a maximum at the 
stagnation point at the front of the cloud and smaller along
the sides of the cloud.  On the other hand, 
density varies primarily across the slip surface.
Since the slip surface resembles the surface of the uniform cloud,
we adopt KMC's equation for the growth of the post-shock circulation:
\begin{equation}
\frac{d\Gamma}{dt} \simeq \Delta P \Delta \left(\frac{1}{\rho}\right)
\simeq -\frac{\Delta P}{\rho_{i1}} \simeq -\frac{1}{2}v_c'^2  \  .
\end{equation}
Then, the circulation produced by the post-shock flow can be estimated as 
\begin{equation}
\Gamma_{\rm post} \simeq -\frac{1}{4}v_{c0}'^2 t_{\rm drag} 
\simeq -\frac{9}{64}\left(\frac{\chi^{1/2}t_{\rm drag}}{\tcc}\right)
(1-\chi^{-1/2})^2 v_b r_{\rm co}  \  ,
\label{eq:gamma_post}
\end{equation}
where we have approximated the cloud velocity as
$v_c'=v_{c0}' \exp (-t/t_{\rm drag})$ with
$v_{c0}'=[2v_b/(\gamma+1)] (1-\chi^{-1/2}) $.
Here we have followed KMC in
adopting a simpler approximation
for the cloud velocity than that given in
\S \ref{sec:global} (eqs. [\ref{eq:velocity0}]
and [\ref{eq:velocity}]).

Third, we consider the vorticity production 
associated with the triple points that 
are formed by the interaction of the initial shock 
and the reflected shock behind the cloud ($\Gamma _{\rm ring}$).
Following KMC, we estimate the circulation of the supersonic vortex ring as
\begin{equation}
\Gamma_{\rm ring} = \frac{3}{4} v_b a_{0}  \, .
\label{eq:gamma_ring}
\end{equation}
Note that this component has a positive sign, which is the opposite of 
$\Gamma_{\rm shock}$ and $\Gamma_{\rm post}$
(see \S 7.1 of KMC for more detail).

Finally, the vorticity produced in the core $\Gamma_{\rm core}$,
is expected to be smaller than $\Gamma_{\rm shock}$ and $\Gamma_{\rm post}$ 
 by a factor of order $\chi^{-1/2}$.
Since this is small, we follow KMC in not attempting 
to model this component.

Recently, \citet{PKornreich00} constructed 
an analytic formula for the vorticity 
generated by the shock-cloud interaction for smooth clouds,
which corresponds to our $\Gamma_{\rm shock}$.
They assumed that (1) the post-shock flow is steady and parallel
to the $z$-axis, and 
(2) 
the velocity induced by the shock is inversely proportional 
to the square root of the pre-shock density.
In order to make a comparison with our analytic model, we 
derive the circulation estimated from their model.
From their eq. (40), 
\begin{eqnarray}
\Gamma_{\rm KS} &=& \int \mbox{\boldmath$\omega \cdot dA$}= 
-\frac{2n r_{\rm co} v_b}{\gamma+1} (1-M^{-2})(\chi-1) 
\nonumber \\
&&  \times\int^{\infty}_0 \frac{(r/r_{\rm co})^n dr}{[(r/r_{\rm co})^n+\chi]^{3/2}
[1+(r/r_{\rm co})^n]^{1/2}}  \, .
\label{eq:korn}
\end{eqnarray}
We compare this to the numerical results below.

\subsection{Comparison With Numerical Simulations}

Figure \ref{fig:circulationx10},
shows the time evolution of the total circulation, $\Gamma$, for 
several different models.
For all the models, the evolution is 
qualitatively similar: $\Gamma$ increases monotonically until 
the intercloud shock reaches the rear of the cloud, then decreases with time
as the shock transmitted into the cloud interacts with the reflected 
shock on the $z$-axis.  Thereafter, the total circulation
 increases again.
The first peak corresponds to 
$\Gamma_{\rm shock}$ because at this time the incident shock has passed
over the cloud and reached the $z$-axis at the rear of the cloud.
The subsequent decrease in $\Gamma$ corresponds to the positive
contribution due to the supersonic vortex ring, $\Gamma_{\rm ring}$.
The interaction of the shocked cloud 
with the post-shock flow continues to generate the circulation
$\Gamma_{\rm post}$, so that the total circulation subsequently increases
with time.

     For clouds with relatively large density gradients
[ $(n, \chi)=(8, 10)$ 
and $(8, 100)$], the values of $\Gamma_{\rm shock}/r_{\rm co}v_b$ are predicted 
from eq. (\ref{eq:gamma_shock}) as $-1.77$ and $-2.58$, respectively, 
which are within about 15\% and 30\% of the peak values of the total
circulation of $-2.01$ and $-3.78$, respectively.
\citet{PKornreich00}'s model (see eq. \ref{eq:korn})
predicts $\Gamma_{\rm KS}=-1.30$ and $-2.18$ for the two cases, which
is somewhat less accurate than our model.
Our numerical results show that
the circulation associated with the supersonic vortex ring is 
about 1.02 for $(n, \chi)=(8, 10)$, which is close
to our analytic estimate of 0.866. For $(n,\chi)=(8, 100)$,
the numerical and analytic results are
1.40 and 0.953, respectively, again in good agreement.
The analytic model predicts that the
circulation due to the post-shock flow is 
$\Gamma_{\rm post}\simeq -(0.48, 4.4) r_{\rm co}v_b$ for
the models with $\chi=10$ and 100, respectively, which is
in reasonable agreement with our final results.
At the end of the computation, the total 
circulation is continuing to increase. However, 
we expect the rate of increase
to decrease with time as the cloud becomes comoving with the post-shock
flow
so that the circulation will not increase much thereafter.  
Lower resolution calculations confirm this prediction.

The components of $\Gamma_{\rm post}$ and $\Gamma_{\rm ring}$ 
can be clearly seen in the spatial distribution of the circulation 
shown in Fig. \ref{fig:circulationx10}, 
in which the vorticity is integrated from the bottom of
the computation box up to $z$,
The cumulative circulation at $t=2.86 \, \tcc$, 
shown in Fig. \ref{fig:vorticityn8x10}f, 
decreases with $z$, reaching a minimum value of $-2.3$ $r_{\rm co}v_b$,
which corresponds to $\Gamma_{\rm shock}$.
As $z$ increases further, the circulation increases
due to the contribution of the supersonic vortex ring, 
$\Gamma_{\rm ring}= 1.0 r_{\rm co}v_b$.
For clouds with relatively steep density gradients (like $n=8$),
the spatial distribution of vorticity generated by the shock-cloud
interaction is qualitatively similar to
that of the uniform cloud (see Fig. 16 of KMC), i.e., 
immediately after the initial shock passage, the vorticity generated by
the interaction is confined to a thin layer between the cloud and
intercloud medium (Fig. \ref{fig:vorticityn8x10}b).
In the subsequent evolution, the vorticity is concentrated in the
slip surface, as can be seen in Fig. \ref{fig:vorticityn8x10}c.

     For clouds with small density gradients, the analytic
model continues to work well for the circulation
produced by the shock.
In this case, to estimate the circulation from the analytic model, 
we adopt the reduced initial cloud radii used in \S \ref{subsec:drag}.
For $(n, \chi)=(2, 10)$, our simulations show that
the first peak in the circulation is $-3.10 r_{\rm co}v_b$,
compared with $\Gamma_{\rm shock}=-3.41 r_{\rm co}v_b$ from our analytic model.
For $(n,\chi)=(2, 100)$ the corresponding values
of the circulation are $-12.6 r_{\rm co}v_b$ and 
$-14.1 r_{\rm co}v_b$, respectively.
Again, the analytic predictions are within about 12 \% of
our numerical results.
In contrast, equation (\ref{eq:gamma_ring}) overestimates 
the vorticity production by the supersonic vortex ring for the model
with $n=2$ and $\chi=10$
by almost a factor 3 (our numerical results give 
$\Gamma_{\rm ring}\simeq 0.591 r_{\rm co}v_b$, whereas the analytic model
predicts $\Gamma_{\rm ring} = 1.66 r_{\rm co}v_b$).
In this model, the supersonic vortex
ring is not as powerful as for the models with larger density gradients.
As mentioned in \S \ref{subsec:n=2}, 
the vortex ring propagates more slowly than the triple
point that moves with almost the same velocity as the incident shock $v_b$.
This slower propagation speed of the supersonic vortex ring appears
to be special to the n=2 and $\chi=10$ model. 
For example, for the model with $n=2$ and $\chi=100$,
a powerful supersonic vortex ring is formed after shock passage.
The numerical result shows $\Gamma_{\rm ring}\simeq 5.50 r_{\rm co}v_b$, 
whereas the analytic solution predicts $\Gamma_{\rm ring} = 5.25
r_{\rm co}v_b$, about 5 \% less than our numerical result.

For smooth clouds with $n=2$, the analytic model 
of \citet{PKornreich00} predicts
$\Gamma_{\rm KS}/r_{\rm co}v_b=-3.98$ and $-14.5$
for $\chi=10$ and 100, respectively.
These values are in reasonable agreement with the numerical
results given above, 
although once again the error is somewhat larger than for our analytic model.

For smooth clouds, the vorticity generated by the initial interaction
with a shock extends 
throughout the cloud envelope
(see Fig. \ref{fig:vorticityn2x10}b).
However, this radially-extended vorticity shifts towards 
the slip surface with time.
Eventually, the vorticity is concentrated in the slip
surface, and is subsequently converted to random motions of small fragments
as small-scale fragmentation proceeds.

We note that the evolution of the total circulation for 
the uniform clouds ({\it dotted curves} in Fig. \ref{fig:circulationx10})
are apparently different from those of KMC (Figs. 13 and 15 of KMC)
after the incident shock reaches the $z$-axis at the rear of the cloud.
This discrepancy occurs 
because the contribution of the supersonic vortex ring 
is always included in our calculations, 
whereas KMC did not include it after
the ring left the high-resolution part of the grid.
If the contribution due to the supersonic vortex ring is removed
from our results, the
evolution of the circulation is in good agreement with KMC's results.

In summary, the analytic model constructed in \S \ref{sec:vorticity} 
agrees well with our numerical results.
Recently, \citet{PKornreich00} claimed 
that the vorticity production in smooth clouds
is not related to the baroclinic term, but
is purely kinematic in origin; 
the preshock density distribution, 
which varies radially through the cloud, 
naturally generates the velocity gradient in the postshock cloud. 
However, the agreement between our theory
and our simulations have clearly demonstrated that 
the vorticity is indeed produced by the baroclinic term 
in eq. (\ref{eq:vorticity}) during the shock-cloud interaction.
Their analytic model for vorticity production by the shock  
is somewhat less accurate than ours, since they did not consider
the vorticity produced in the post-shock flow or in the supersonic
vortex ring.

\section{Discussion}
\label{sec:discussion}

\subsection{Effect of a Smooth Cloud Boundary on Cloud Morphology:
  Implications for Observations}

In this paper, we have studied the interaction of shocks with interstellar
clouds with smooth density profiles.
We showed that after several cloud crushing
times, a slip surface with a significant density jump
develops in front of the core of the shocked cloud, 
irrespective of the initial density distribution. 
Thus, the cloud is deformed into a head-tail structure
before subsequent cloud destruction 
(see e.g., Figs. \ref{fig:snapn8}c and \ref{fig:snapn2}d).
This suggests that the density profiles of observed shocked 
clouds should have significant 
density jumps in the direction of the source of the
large scale shock.
One might think that the effects of the initial density profile 
on cloud morphology would be obscured as the slip surface develops,
but in fact information on the density profile
remains imprinted on the shocked cloud until significant
instabilities develop.

Figure \ref{fig:mass ratio} shows the time evolution 
of the ratio of the masses $m_{0.5}$ and $m_{0.1}$ for the models with 
$(n, \chi, M) = (2, 100, 10)$ and (8, 100, 10), 
where  $m_{0.5}$ and $m_{0.1}$ represent the masses
contained within the regions with $\rho\ge 0.5 \, \rho_{\rm max}$ and 
$\rho\ge 0.1\, \rho_{\rm max}$, respectively.
Prior to cloud destruction,
the mass ratio $m_{0.5}/m_{0.1}$ remains approximately constant,
but varies significantly with $n$.
For example, in the early stages of the shock-cloud interaction,
the mass ratio stays at $m_{0.1}/m_{0.5}\gtrsim 0.5$ for clouds
with relatively steep density gradients ($n \gtrsim 8$), whereas
it is only about 0.1 for smooth clouds ($n\sim 2$).
This suggests that observations of this mass ratio may be able to 
reveal the initial density profiles of the preshock clouds 
provided the clouds
have not entered the shock destruction stage;
the initial mass ratios $\calr ^{-1}$ listed in Table \ref{tab:model}
could be directly compared with the observed values.

Our simulations can be compared with the observations of young and
middle-aged SNRs such as Cas-A and Cygnus Loop.
Two important caveats in making any such comparison are that
(1) our simulations are non-magnetic, whereas magnetic
fields are believed to be important in the interstellar medium,
and (2) almost all of our simulations are adiabatic, whereas
radiative losses are important in shocked interstellar gas.
Recently, \citet{DPatnaude02} and \citet{CDanforth01}
performed detailed investigation on the shock-cloud interactions 
observed in the Cygnus Loop 
(the XA and southwestern region of the Cygnus Loop).
Their observations do not show any signs of
the Kelvin-Helmholtz instability at the observed cloud surface, and 
thus they concluded that the interaction of the shock and cloud is in
the initial transient stage.  
However, if the preshock cloud had a 
smooth
density gradient at the cloud boundary, 
then the Kelvin-Helmholtz instability 
does not grow in the shock compression or the
early post-shock expansion stages.
If the mass ratio $m_{0.5}/m_{0.1}$ 
(or an equivalent one) could be estimated from the observations, 
then one might be able to infer something about the structure
of the initial cloud.

\subsection{Velocity Dispersions of Shocked Clouds}

As shown above, the shock-cloud interaction produces vorticity inside the
cloud.  Much of the vorticity is concentrated at the slip surface, which
is subject to the Kelvin-Helmholtz
instability, and as a result the shocked cloud is distorted, 
producing vortex rings.
Vortical motions generate substantial velocity dispersions.  
Recently, velocity dispersions generated by 
the shock-cloud interactions have attracted a great deal of attention 
as a mechanism for replenishing turbulent motions in the ISM 
\citep[e.g.,][]{PKornreich00,MMacLow03}. 
Here, we discuss how velocity dispersions generated by shock passage 
evolve with time.
First, we compute the synthesized spectra of turbulent motions
integrated along the $z$-axis \citep[see e.g.,][]{EFalgarone94}, 
and then calcuate the dependence of the velocity dispersion on
region size, the so-called linewidth-size relation.

\subsubsection{Synthesized Spectra of Shocked Clouds}

Figures \ref{fig:spectra}a through  \ref{fig:spectra}c 
show the evolution of the synthesized spectra of a shocked cloud
with $(n,\chi, M) = (8,10,10)$.
The synthesized spectra have been computed 
by averaging density-weighted histograms of the
line-of-sight velocity over the entire cloud component.
As mentioned in the next subsection, the velocity dispersions
are typically subsonic for adiabatic shocks, 
and thus thermal broadening would
smear the motions of small cloud fragments. 
Thus, to understand how the shocked cloud fragments due to
hydrodynamic instabilities, we focus on the non-thermal motions.
The line-of-sight is assumed to be parallel to the $z$-axis.
At $t=4 \, t_{\rm cc}$, the synthesized spectrum has a fairly 
smooth profile with a peak at 
$0.6v_{i1}$, where $v_{i1}$ is the velocity of the shocked intercloud
medium,
and a tail at  
higher velocities.  
At this stage, the cloud has not been destroyed, although 
it has been deformed into a head-tail structure 
(see Fig. \ref{fig:snapn8}d).  
The velocity component near the peak corresponds to the main body of 
the shocked cloud,
while the higher velocity component corresponds to 
the tail of the shocked cloud.
By $t=8 \, t_{\rm cc},$, the cloud has broken into small fragments,
mainly due to the Kelvin-Helmholtz instability
(see Fig. \ref{fig:snapn8}f).
Each fragment has a different velocity, which produces
many spikes in the synthesized profile. 
The velocity dispersion of these fragments is estimated to be of 
the order of 0.1 $v_b$, 
somewhat less than the sound speed behind the cloud shock
($\sim 0.18 v_b$).
The synthesized spectrum has a profile similar to panel (b) 
for $6 \, t_{\rm cc} \lesssim t \lesssim 10 \, t_{\rm cc}$.
By $t=13.2 \, t_{\rm cc}$ (Fig. \ref{fig:spectra} c),
these spikes have become fainter and the peak velocity
is shifting towards the post-shock ambient velocity 
as the cloud is accelerated and mixed with 
ambient medium by the interaction with the postshock flow.
The component near the peak velocity
has a velocity dispersion 
$\sim 0.05 v_b$.
The synthesized spectrum also has a plateau extending towards small
velocities, which corresponds to the cloud components 
near the $z$-axis that were destroyed 
recently.
This plateau becomes smaller with time.

     Figures \ref{fig:spectra}d through  \ref{fig:spectra}f 
show the evolution of the synthesized spectra for a  shocked cloud
with $(n,\chi, M) = (2,10,10)$.
The overall evolution of these synthesized spectra 
is similar to that for the $n=8$ case.
Prior to the onset of the Kelvin-Helmholtz instability, 
the spectrum is more-or-less continuous 
and has a peak at low velocities, extending towards
higher velocities.
By $t=12 \, t_{\rm cc}$, the cloud has begun to be destroyed and a number of
spikes appear in the spectrum.  
The synthesized spectrum has a profile similar to panel (e) 
in the time interval of $10 \, t_{\rm cc} \lesssim t \lesssim 20 \, t_{\rm cc}$.
At the late stage of cloud destruction, 
the peak of the spectrum shifts towards the velocity of the post-shock 
ambient medium.  The evolution of the spectrum is slower than that of 
the $n=8$ case because of the smaller relative velocity between the cloud 
and intercloud medium.

The dissipation of random motions seems to be slow 
compared to the cloud-crushing time.
The velocity dispersion decreases by about 30 \% 
from the maximum after $t=t_{\rm dest}$ for $n=8$, 
while for $n=2$, the velocity dispersion does not
decrease significantly after $t=t_{\rm dest}$. 
On the other hand, for the uniform cloud (KMC),
the velocity dispersion decreases by about 40 \% by the end of
computation ($10 \, \tcc$).  
This reflects the fact that smoother clouds tend to evolve more slowly.
The velocity dispersions would decrease more if further evolution 
is followed for smoother clouds.
In the next subsection, we discuss how dissipation of random motions 
depend on the size, based on three-dimensional simulations.

\subsubsection{The Linewidth-Size Relation (3D simulation)}

Observations of the ISM indicate that interstellar clouds have
nonthermal internal motions that are inferred to be
supersonic turbulence.
The observed velocity dispersions are found 
to scale with region size $R$ as a power law, $\Delta v \sim R^p$,
with $p=0.4-0.6$ \citep[e.g.,][]{RLarson81, PMyers87, Dame86}, 
although the power index $p$ can depend on the 
environment \citep[e.g.,][]{PCaselli95}. 
In this paper, we have shown that 
the interaction of
shocks with interstellar clouds can generate
random motions in the cloud material,
but only by completely destroying the clouds.
Here, we calculate the dependence of the velocity dispersion on
region size, the so called linewidth-size relation, 
from our numerical results,
and discuss whether random motions generated by the shock-cloud interaction
agree with the observed scaling law.
Because the characteristics of hydrodynamic turbulence 
depend strongly on the spatial dimension,  
we perform a three-dimensional calculation with an intermediate 
spatial resolution $R_{60}$ for the nonradiative cloud 
with $(n, \chi, M)=(8,10,10)$.
We confirmed that all the global quantities in the three-dimensional
calculation
coincide with those of the two-dimensional calculation 
with the same spatial resolution to within 10\%
in the time interval $0\le t \le 10 \, \tcc$.

In Fig. \ref{fig:velocity-size}, 
we represent the linewidth-size relation 
for the model with $(n, M, \chi, \gamma) = (8, 10, 10, 5/3)$
at three different times: $t=6 \, \tcc$, $8 \, \tcc$, and $10 \, \tcc$.
The abscissa denotes the beam size, while the ordinate denotes 
the velocity dispersions of the cloud material.  
The figure shows the results of observing the shocked cloud 
from two different directions, 
(1) along the $z$-axis
(perpendicular to the incident shock plane; shown by solid curves), 
and (2) perpendicular to the
$z$-axis (in the plane of the incident shock;
shown by dotted and dashed curves).
In the latter case, we consider two different positions for the 
center of the observing beam: at
the center of gravity, $z_G$, and downstream from that point, 
$z_G + r_{\rm co}$, where the cloud is more disrupted than at $z = z_G$.

By $t=6 \, \tcc$, the cloud has begun to experience 
significant destruction due to hydrodynamic instabilities, 
but at this stage the main axial core is still recognized on the
$z$-axis (see Fig. \ref{fig:snapn8}).
The solid and dotted curves have peaks at $\sim 0.3r_{\rm co}$ 
and $\sim 0.6r_{\rm co}$, respectively. These peaks correspond 
to the components near the main axial core.
In the downstream observing beam, much of
the cloud is mixed with the intercloud material, and
the velocity dispersion (dashed curve) is almost independent of scale.
This part of the cloud has been destroyed 
due to the nonlinear growth of small-scale 
Kelvin-Helmholtz instabilities. 
This indicates that the turbulent motions are generated mainly in
the cloud envelope that has been destroyed, and in the cloud core,
the turbulent generation delays.
By $t=8 \, \tcc$, the velocity dispersions in all the observing beams 
become relatively flat, indicating that the entire cloud is being
destroyed due to the Kelvin-Helmholtz instabilities.
Thereafter, the velocity dispersions in the small scale 
dissipate, as can be seen in Fig. \ref{fig:velocity-size}(c).
As a result of dissipation,
 at the time of $t= 10 \, \tcc$, if the velocity dispersions are approximated
by power-laws as $\Delta v \sim R^p$, the power index is estimated as 
$p=0.2 - 0.3$, although
the power index depends on the direction and beam position.
At later times, $p$ increases somewhat since the small-scale 
fluctuations damp out more rapidly.
Many numerical studies show that interstellar turbulent motions
dissipate in a dynamical time, $L/\delta v$, 
where $L$ is the size of the region (Elmegreen \& Scalo 2004).  
At $t\simeq 8 \, \tcc$, we find $L/\delta v\simeq 
0.53 \tcc$ and $2.9 \tcc$ 
at $L/r_{\rm co}\approx 0.1$ and 1.0, respectively,
where we have averaged over three different beams.
The velocity dispersion at small scales, $r\approx 0.1r_{\rm co}$,
dissipates significantly from $t=8 \, \tcc$ (Fig. \ref{fig:velocity-size}b) 
to $t=10 \, \tcc$ (Fig. \ref{fig:velocity-size}c), whereas
at large scales, $r\approx 1r_{\rm co}$, turbulent dissipation is
less efficient.  
It appears that there has not been enough time
to establish a self-consistent turbulent cascade 
to maintain the turbulent motions 
on small scales.

Our simulations show that at the early stage of cloud destruction, 
small-scale fluctuations dominate,
which makes the linewidth-size relation flat.
Subsequently, these small-scale fluctuations tend to damp out
in a few cloud crushing times,
leading to a linewidth that increases with scale.
The linewidth-size relations obtained from our simulations 
are time-dependent and have not reached their asymptotic values. 
At late times, the cloud material dissolves into the ambient medium due
to mixing, and therefore, it is difficult to distinguish them well.
The turbulent cascade could also influence the evolution 
of the linewidth-size relations,
although our resolution
is inadequate to study this precisely. 
Sytine et al. (2000) performed convergence tests on decaying turbulence
with periodic boundary conditions and suggested that at least $1024^3$ 
grids are needed to capture the characteristics of turbulent fields
accurately.  However, 
in our 3D simulation with $R_{60}$, 
the effective radius of
the shocked cloud expands by a factor of two by $t=t_{\rm dest}$,
so that the shocked cloud is covered by only about $256^3$ grids
at the epoch of cloud destruction.
Thus, to discuss the time evolution of the linewidth-size relation
more accurately, higher resolution runs may be needed.

\subsection{Implications for Interstellar Turbulence}

The ISM is observed to be highly turbulent.  
Recently, \citet{CHeiles03} found that 
the mass-weighted velocity dispersion is 7.1 km s$^{-1}$
for cold neutral medium (CNM) 
clouds and 11.4 km s$^{-1}$ for warm neutral medium (WNM)
clouds.  
The one-dimensional rms Mach number of internal motions 
for individual CNM clouds is $\sim 1.7$.
The mean temperatures of the CNM and WNM are measured to be 
$\sim 70$ K and $\sim 8000$ K, respectively.
Under the assumption of pressure equilibrium between the CNM and the 
WNM, the typical density contrast between CNM and WNM clouds
is of order $\chi=100$. 
\citet{CHeiles03} argue that the CNM
clouds are organized into large sheets.
More recently, Stanimirovic \& Heiles (2005) have found
very small CNM clouds, with column densities $\sim 10^{18}$ cm$^{-2}$.
In this subsection, we discuss how our results on the shock-cloud 
interaction might be related to these observations.

Interstellar shocks are produced by several events such as 
supernova explosions and stellar winds from massive stars.
Among them, the  injection rate of energy into the ISM through supernovae is 
expected to exceed those of other sources (e.g., Castor 1993).
Thus, we concentrate on the effects of supernova remnants 
(SNRs) 
and attempt to estimate the magnitude of the 
generated velocity dispersions.
Most of the shock-cloud interactions 
are expected to be at low Mach numbers, 
because the shock speeds of SNRs rapidly decrease with expansion,   
although the typical Mach number is highly uncertain.
In the following, we adopt the typical Mach number of $M\simeq 3$ as a
representative value.
For shocks at $M\simeq 3$, the distance from the origin of a blast wave 
$R_c$ is estimated to be of the order of 100 pc \citep{DCioffi88}.
In this case, the small cloud approximation is applied for 
clouds with radii of the order of 1 pc 
or less.
(see \S \ref{subsec:time scales}).
This size is comparable to that of typical CNM clouds 
observed by \citet{CHeiles03}.

As shown above, the velocity dispersion induced by
the shock-cloud interaction is of the order of 
\begin{equation}
\delta v \simeq 0.074 v_b 
= 2.2 M_3 \chi_{2}^{1/2} C_{c0,\, 5}~~~~ {\rm km}\, {\rm s}^{-1} \ ,
\end{equation} 
for radiative clouds with $\gamma=1.1$,  
where $M_3\equiv M/3$, $\chi_{2}\equiv \chi/10^2$, and
$C_{c0,\, 5}\equiv C_{c0}/10^5$ cm s$^{-1}$.
The above one-dimensional velocity dispersion 
[$\delta v^2=(2\delta v_r ^2 + \delta v_z^2)/3$]
is estimated as the mean value of the $n=8$ and $n=2$ clouds at $t=t_m$. 
For $T= 70$ K, the one-dimensional velocity dispersion 
is computed as about 1.4 km s$^{-3}$.
This magnitude of turbulence is comparable to the observed 
internal velocity dispersions of CNM clouds, $\delta v \simeq 1.1$
km s$^{-1}$, \citep{CHeiles03}.

When one observes the ISM, it is difficult to directly distinguish
the structures along the line-of-sight from observations.
Approximating the thickness of the CNM
to be $b = N_{\rm HI}/n_{\rm HI}$, 
\citet{CHeiles03} infer that the CNM is in thin sheets
with typical aspect ratios in the range of 100 $-$ 300, 
where $N_{\rm HI}$ and $n_{\rm HI}$ are the column density along the
line-of-sight and the volume density, respectively.
If one observes a shocked cloud from the side and models it 
as a sheet of length $2c$, width $2a$, and thickness $b$, 
then the thickness $b$ is evaluated as
\begin{equation}
b=\frac{N_{\rm HI}}{n_{\rm HI}} = 
\frac{m_{\rm cl,0}}{(2a \times 2c)\left<\rho_1\right>}
= \frac{\pi a_0^3 M^2}{3 ac} \ ,
\end{equation}
where we consider isothermal shocks.
Then, the inferred aspect ratio is given by
\begin{equation}
2c/b = 6ac^2 M^2/ (\pi a_0^3) \ .  
\end{equation}
For a shocked cloud, the above aspect ratio tends to be large
and increases with time because the length $c$ increases with time.
For example, for a shocked cloud with $n=8$, $\chi=100$, and $M=3$, 
the aspect ratio $2c/b$ reaches about 2000 at $t\simeq t_{\rm dest}$.
This implies that some of the observed CNM clouds 
with large aspect ratios could be shocked clouds torn up into
small fragments, instead of thin sheets.
Small cold clouds recently observed by Stanimirovic \& Heiles (2005)
could be related to this case.
Since we don't take into account the magnetic fields, 
we cannot compare our result quantitatively with observations. 
The omission of the magnetic fields most likely leads to 
overestimation of the shocked density and therefore the aspect
ratio. Although shocks in the real ISM are not isothermal as we 
assumed above, 
deviations from isothermality behind the shocks would cause 
only quantitative, not qualitative, changes in the result, provided
that the cloud was initially CNM.

In our simulations, the vorticity generated by the shock-cloud 
interaction is converted into random motions of small cloud 
fragments produced by the Kelvin-Helmholtz and 
Rayleigh-Taylor instabilities.
In that case, the CNM may consist of small shreds in a coherent flow. 
We note that this mechanism of turbulence generation 
is different from that of \citet{PKornreich00}, 
who proposed that the vorticity produced by the
shock-cloud interaction is confined {\it inside the shocked cloud} and
eventually converted into the turbulent motions observed in
the interstellar clouds.  They assumed that the smooth clouds
are not destroyed significantly by shocks and that the 
density profile of the cloud is more or less maintained after shock passage.
They suggest that the velocity dispersions generated 
by shock-cloud interaction can be scaled as 
$ \Delta v /v_b = n^{1/2}(\rho_1-\rho_0)/(2\rho_1) \chi^{-0.5} 
\propto \chi^{-0.5}$.
However, our simulations have demonstrated that 
the density profile in the cloud changes significantly 
with time, and within a few cloud-crushing times a slip surface 
forms in front of the core of the cloud.
In fact, the velocity dispersions are estimated as 
$\sim 0.1 v_b$, which is more or less independent of $\chi$.
Furthermore, prior to complete cloud destruction,
 turbulent motions produced by the shock-cloud interaction 
 are not induced in the cloud core, but in the cloud envelope.
For example, for the model with $(n, \chi, M)=(2, 100, 10)$, 
the circulation in the cloud core ($\Gamma_{\rm co}$),
which has a density greater than $1.96 \, \rho_{c0}$ 
[$\approx 4 \, \rho_0(r=r_{\rm co})$] is 
$\Gamma_{\rm co}/r_{\rm co} v_b = 0.089$ at $t\simeq 1.5 \, \tcc$.
This value is only about 0.7 \% of the total circulation 
($\Gamma_{\rm tot}/r_{\rm co} v_b\sim 12$), indicating that  
most of the vorticity is in the cloud envelope.
(If we use eq. 40 of \citet{PKornreich00},
which gives the distribution of vorticity as a function of radius,
the circulation generated by the shock passage is estimated 
as $\Gamma_{\rm KS}/r_{\rm co} v_b = 0.039$
in the cloud core of $r\le r_{\rm co}$ 
and $\Gamma_{\rm KS}/r_{\rm co} v_b = 14.5$ 
in the whole region of $r\le \infty$.
These estimates from the KS theory also show
that the shock-cloud interaction is ineffective
at inducing turbulent motions 
in the cloud core.)
For other models with larger $n$, the ratios of $\Gamma_{\rm co}$
to $\Gamma_{\rm tot}$ are much smaller than for $n=2$.

In summary, the shock-cloud interaction can produce the magnitude 
of the velocity dispersions observed in the typical CNM clouds with
the size of $\sim 1$ pc.
However, turbulent motions produced by the shock-cloud interaction 
are directly associated with cloud destruction  and, 
in the absence of a magnetic field at least, 
do not significantly affect the cloud core,
 contrary to \citet{PKornreich00}'s view.
The interaction of a shock with a cold H~I cloud is expected to 
lead to the production of a spray of small H~I shreds which
have a coherent velocity distribution.  
Such structures could be related to the small cold clouds 
recently observed by Stanimirovic \& Heiles (2005).
However, we cannot make a realistic estimate 
for the size of the smallest shreds to compare with their observations,
because our calculations do not include
magnetic fields and thermal conduction
(see e.g., Koyama \& Inutsuka 2002 for the effect of thermal conduction).

\subsection{Implications for Star Formation}

Under some circumstances, compression due to interstellar shocks 
is expected to make a shocked cloud self-gravitating even if self-gravity is
unimportant in the preshock cloud, thereby triggering 
star formation \citep{BElmegreen77}.
\citet{PFoster96} and \citet{HVanhala98} numerically studied the interaction
of radiative shocks with self-gravitating spherical clouds, 
showing that gravitational collapse can be induced by shock
compression when radiative cooling is efficient in the shocked clouds.  
In particular, \citet{PFoster96} studied the interaction of 
a critical Bonnor-Ebert sphere and a thin pulse-like
wave that mimics an interstellar shock or wind, under the assumption
that the gas is isothermal.
They found a critical momentum of this wave above which 
gravitational collapse is induced after the interaction.
Unfortunately, their result cannot be directly compared with
ours because their initial conditions are different from ours; 
e.g., their initial wave is not satisfied with the 
shock jump condition.
\citet{HVanhala98} performed SPH simulations of 
the interaction of a planar shock and a spherical,
centrally concentrated
molecular cloud core, 
taking into account the effect of radiative cooling. 
They found that shocks with intermediate speeds of 20 $-$ 45 km s$^{-1}$
are capable of triggering gravitational collapse since
the effective $\gamma$ of the shocked cloud remains close to unity.
In contrast, stronger shocks 
inhibit gravitational collapse because the main coolants 
such as CO and H$_2$ are destroyed for velocities
above about 40-50 km s$^{-1}$, driving the effective $\gamma$
above 4/3.  
This critical velocity for shock destruction of the molecules
is consistent with that of \citet{Draine93}.
More recently, \citet{PHennebelle03, PHennebelle04} studied the evolution
of molecular cloud cores compressed by external pressure
and demonstrated that a compression wave triggers 
collapse from the outside in.

Here we first show that only radiative shocks
can trigger gravitational instability in shocked clouds.
We show that turbulence behind the shock has
a strong stabilizing effect. Shocked clouds
are compressed into a sheet-like configuration,
and we analyze the condition for gravitational
instability for this case following the treatment
of \citet{BElmegreen78}.
Since we have not included the effects of magnetic fields
in our simulations, we neglect them here. Magnetic fields
tend to stabilize against gravitational collapse, so
the results below give necessary but not sufficient
conditions for gravitational instability.

\subsubsection{Maximum critical mass}

Define the dimensionless cloud mass as (McKee \& Holliman 1999)
\beq
\mu\equiv m_{\rm cl}\left[
	\frac{(C_{cs}/\gamma_c^{1/2})^3}
	{(G^3\rho_{cs})^{1/2}}\right]^{-1},
\label{eq:mu}
\eeq
where $\rho_s$ and $C_{cs}$ are the density and
adiabatic sound speed at the surface of the cloud.
A stable isothermal cloud has $\mu<1.18$, the value
for a Bonnor-Ebert sphere.
More generally, a stable cloud has $\mu<\mu_{\rm cr}$, where
$\mu_{\rm cr}$ is a number that, 
for polytropes ($P\propto \rho^{\gamma_p}$), depends
on both the polytropic index $\gamma_p$ and
the adiabatic index $\gamma$ \citep{CMcKee99}.
The maximum mass of a cloud in hydrostatic equilibrium
(i.e., the cloud mass corresponding to $\mu_{\rm cr}$ in
eq. \ref{eq:mu}) is
\begin{equation}
m_{\rm cr}\propto \left(\frac{P_s}{\rho_s^\gamma}\cdot
       \rho_s^{\gamma-\frac 43}\right)^{3/2} \ ,
\label{eq:be2}
\end{equation}
The first factor is proportional to the exponential of the 
entropy, and therefore always increases in shocks
unless they are sub-isothermal (i.e., the post-shock
temperature is less than the pre-shock temperature). Since
the density also increases in shocks, we conclude that
shocks increase the critical mass---i.e., they stabilize
clouds against gravitational collapse---for $\gamma\geq \frac 43$.
This is not surprising since clouds that are approximately
locally adiabatic are stable against collapse for $\gamma>\frac 43$
in any case.

 For $\gamma<\frac 43$, it is possible that the 
effect of the increased density might overcome the effect of
the increased entropy in the expression for the critical mass. 
In fact, in the real ISM, radiative cooling can sometimes reduce
the entropy of the postshock gas below the preshock value
(a sub-isothermal shock).
In that case, the above critical mass is likely to decrease 
below the preshock value, if turbulent motions generated by 
shock passage do not contribute to cloud support significantly.

	In order to treat sub-isothermal shocks, we write 
the shock jump conditions as 
\begin{eqnarray}
\rho_1 v_1 &=& \rho_0 v_0 \  , \\
\rho_1 (v_1^2 + \epsilon C_{\rm iso,\,0}^2) &=& \rho_0 (v_0^2 + 
C_{\rm iso,\,0}^2) \ , 
\end{eqnarray}
where 
$C_{\rm iso}\equiv (P/\rho)^{1/2}$ is the isothermal sound speed and
$\epsilon \equiv C_{\rm iso,\, 1}^2/C_{\rm iso,\, 0}^2$. 
Thus, $\epsilon = 1$ corresponds to an isothermal shock
and $\epsilon<1$ to a sub-isothermal shock; normal adiabatic
shocks have $\epsilon > 1$.
In shocked CNM clouds, 
the temperature might decrease to about 10 K 
if molecules can form behind the shock; in that case,
$\epsilon\sim 0.1$. 
>From the above relations, the post-shock density is
\begin{equation}
\rho_1 = \frac{\miso^2}{\epsilon} F \rho_0 \ ,
\label{eq:jumpdensity}
\end{equation}
where $\miso\equiv v_0/C_{\rm iso,\, 0}$ is the isothermal Mach number and
\begin{equation}
F = \left[\frac{1+\miso^{-2}+\sqrt{(1+\miso^{-2})^2
-4\epsilon \miso^{-2}}}{2}\right] \ .
\end{equation}
The factor $F$ approaches unity either for $\epsilon\rightarrow 1$
or for $\miso\rightarrow \infty$ and $\epsilon/\miso^2\rightarrow 0$.
In general, the factor $F$ is about unity 
for $|1-\epsilon|/\miso^2\ll 1$, which is true for
sub-isothermal shocks provided the isothermal Mach number 
$\miso$ is not small;
e.g., for $\epsilon=0.1$ and $\miso=3$, $F\simeq 1.1$.
Thus, the post-shock density can be approximated by
$\rho_1\simeq \miso^2 \rho_0 /\epsilon$ for
isothermal and sub-isothermal shocks.

	We can obtain an upper bound on $\epsilon$ such
that the shocked cloud is gravitationally unstable by
assuming that the shocked cloud material is as localized
as possible--i.e., it is spherical. Recall that
$\mu$ is the ratio of the cloud mass to the critical
mass (eq. \ref{eq:mu}), so that
\beq
\frac{\mu}{\mu_0}=\left(\frac{P_0^{3/2}}{\rho_0^2}\right)
	\left(\frac{\rho_1^2}{P_1^{3/2}}\right),
\label{eq:muratio1}
\eeq
where $\mu_0$ is the value of $\mu$ for the pre-shock
cloud.
Now, $P_1\simeq \miso^2 P_0$ and, for $F\simeq 1$,
$\rho_1\simeq(\miso^2/\epsilon)\rho_0$ from 
equation (\ref{eq:jumpdensity}). A necessary condition
for gravitational instability is 
$\mu>\mu _{\rm cr}$, 
which
requires
\beq
\epsilon <(\mu_0\miso/\mu_{\rm cr})^{1/2}.
\label{eq:epmax}
\eeq 
Under the assumption that the pre-shock cloud is
gravitationally stable, we have 
$\mu_0<\mu_{\rm cr}$, so the value of $\epsilon$ in this equation
satisfies the condition that $F\simeq 1$
(i.e., $\epsilon/\miso^2\ll 1$) for $\miso\ga 3$.
This equation suggests that arbitrarily strong radiative shocks
can induce gravitational collapse, but two effects
prevent this: non-thermal motions behind the shock (considered
immediately below) and inadequate time for the
gravitational instability to develop (considered in
\S \ref{sec:grav}.
	
The shock-cloud interaction generates nonthermal motions behind the shock,
both random and systematic, that
tend to stabilize the cloud against gravitational collapse.
In a radiative shock, these motions will dominate the
thermal motions behind the shock so that equation
(\ref{eq:muratio1}) can be rewritten as
\beq
\frac{\mu}{\mu_0}\simeq\left(\frac{C_{c,\,\rm iso,\, 0}}{\delta v_1}
\right)^3\left(\frac{\rho_1}{\rho_0}\right)^{1/2}.
\eeq
Since $\delta v_1$ increases with scale, it must
be evaluated on the same scale as $\mu/\mu_0$.
Our calculations are
two-dimensional, so we estimate the one-dimensional 
velocity dispersion as
$\delta v^2=(2\delta v_r ^2 + \delta v_z^2)/3$.
Behind the shock, we express $\delta v$ as
a fraction of the shock velocity, $\delta v_1 = \kappa v_b
\simeq \kappa \left<\chi\right>^{1/2}v_0$, so that
\beq
\frac{\mu}{\mu_0}=\frac{F^{1/2}}{\kappa^3\langle\chi\rangle^{3/2}\miso^2
\epsilon^{1/2}},
\eeq
where we used equation (\ref{eq:jumpdensity}) 
to evaluate the density jump.
As remarked above, $\delta v_1$ and therefore
$\kappa$ increase with scale. When
averaged over the entire shocked cloud, 
our numerical calculations give $\kappa \simeq 0.04-0.07$
for $t\simeq (1.5- 2) \tcc$ for a uniform cloud with $\chi =100$.
Therefore, in typical CNM clouds, which have $\chi\sim 100$, 
the condition for instability 
($\mu>\mu_{\rm cr}$) 
restricts the
Mach number to be
\beq
\miso<\frac{F^{1/4}\mu_0^{1/2}}
{\kappa_{-1}^{3/2}\langle\chi_2\rangle^{3/4}\epsilon^{1/4}\mu_{\rm cr}^{1/2}},
\eeq
where $\kappa_{-1}\equiv\kappa/0.1$ and $\chi_2\equiv
\chi/100$. For isothermal and sub-thermal shocks
($0<\epsilon\leq 1$), we have $1\leq F<2$.
We conclude that turbulence behind the shock stabilizes
gravitational instability on the scale of the cloud
for dense clouds ($\chi_2\ga 1$) unless the 
shock is relatively weak. Even for a very sub-thermal
shock ($\epsilon=0.1$ so that $F\simeq 2$) with
$\kappa=0.05$, instability is possible only if $\miso<6$.
Magnetic fields, which we have not considered, also
tend to stabilize the clouds against collapse.

\subsubsection{Gravitational stability of a shocked layer}
\label{sec:grav}

	We turn now to a more detailed study of the gravitational
instability of a shocked cloud, taking into account its geometry.
The passage of a shock through
a cloud that has an approximately uniform density 
(i.e., $n\ga 4$)
flattens it in the direction of shock propagation.
The gravitational stability of a pressurized,
isothermal sheet of gas
was analyzed by Elmegreen \& Elmegreen (1978).
They showed that such a sheet is gravitationally unstable
and fragments to form clumps,
which may be filamentary.
When the incident shock is very strong, the sheet is
confined by a strong external pressure
and exhibits effectively incompressible behavior.
In this case, the most unstable mode has
a wavelength that scales with the
disk thickness and can be much 
shorter than the Jeans length 
\citep{BElmegreen78}.  
Such a perturbation can have only a modest growth 
because the total mass is smaller than the critical 
Bonnor-Ebert mass.  
Only longer wavelengths, which grow more slowly,
can contain more than a Bonnor-Ebert mass and lead
to star formation. 
Three-dimensional numerical calculations by
\citet{Umekawa02} showed that the stable clumps or filaments
formed by the fastest growing modes
can merge with themselves owing to mutual gravitational attraction,
resulting in formation of more massive spherical cores 
\citep[see also][]{Umekawa99};
this is presumably equivalent to the growth of longer
wavelength instabilities.

We assume that the mass of the pre-shock 
cloud is significantly less than  the Bonnor-Ebert mass,
so that it has approximately constant density---i.e.,
$n\rightarrow \infty$.
Correspondingly, its temperature, and therefore sound speed,
are approximately constant, both before and after the
passage of the shock. \cite{BElmegreen78} analyzed the
gravitational instability of an isothermal,
planar layer of gas. They showed that the
parameter that governs the properties of the
instability is
\beq
A=\left(1+\frac{P_1}{(\pi/2)G\Sigma_1^2}\right)^{-1/2},
\eeq
where $\Sigma_1$ is the surface density of the shocked sheet
and $P_1$ is the pressure in the shocked intercloud
medium and therefore at the surface of the shocked cloud.

	The pressure in the midplane of
the shocked cloud, $P_{\rm mid}$, is greater than that
at the surface, $P_1$, due to self gravity:
$P_1=(1-A^2)P_{\rm mid}$ \citep{BElmegreen78}. 
We now show that
shocked clouds generally have $A\ll 1$, so that we can approximate
the shocked sheet as isobaric.
The surface density of the initial cloud,
which is assumed to be uniform, is given in
terms of $\mu_0$ (the value of $\mu$ for the
pre-shock cloud) by
\beq
\Sigma_0 = 0.83\left(\frac{P_0}{G}\right)^{1/2}\mu_0^{1/3}.
\eeq
The shock compresses the cloud in the transverse direction
by a factor $f$ (which is $\sim 0.8$ for $\gamma=1.1$ and 
$\chi=100$), 
so that $\Sigma_1=f^{-2}\Sigma_0$.
Approximating the pressure of the shocked cloud by
$P_1\simeq M^2 P_0$, we then find that
\beq
\frac{P_1}{(\pi/2)G\Sigma_1^2}=0.93\frac{f^4 M^2}{\mu_0^{2/3}}.
\eeq
For $M\ga 3$ and $\mu_0<1$, this is large; 
as a result,
the parameter $A$ is small and the cloud is approximately
isobaric.

\cite{BElmegreen78} showed that not all
unstable modes lead to gravitational collapse.
For that to occur, two conditions must be satisfied: 
First, the mode
must be large enough to capture
a Bonnor-Ebert mass, which requires that the wavenumber $k$
satisfies
\beq
kH<0.91A^{1/2}(1-A^2)^{1/4}\simeq 0.91 A^{1/2},
\eeq
where $H=C_{c,1}(2\pi G\rho_{\rm mid})^{-1/2}$ is the 
gravitational scale height of the shocked cloud.
The corresponding e-folding time for the instability
is 
\beq
t_g\simeq (4\pi G\rho_{\rm mid} kHA)^{-1/2}.
\eeq
Second, the mode must grow fast enough that
it becomes nonlinear during the age of the shock.
Our results show that the density of the shocked
cloud drops by
a factor of 2 in $1.5-2 \tcc$ for $10\la\chi\la 100$,
so we require that the growth time $t_g$ be less than
$2\tcc$.
In units of the cloud-crushing time,
\beq
\tcc=\frac{r_{\rm co}}{MC_{c0}}=0.62\frac{\mu_0^{1/3}}{M(G\rho_{c0})
	^{1/2}},
\eeq
the growth time of an unstable mode that encompasses
at least a Bonnor-Ebert mass is then
\beq
\frac{t_g}{\tcc}=0.47 f^{3/2}
\left(\frac{\rho_{c0}}{\rho_{c1}}\right)^{1/2}
	\frac{M^{7/4}}{\mu_0^{7/12}}.
\eeq

	The condition for gravitational collapse to
be possible is $t_g\la 2\tcc$ as discussed above.
Setting $F=1$ in equation (\ref{eq:jumpdensity}), this leads to
\beq
M\la 7 \frac{\mu_0^{7/9}}{f^2\epsilon^{2/3}},
\label{eq:maxM}
\eeq
so that only shocks of moderate strength can
initiate collapse. 
As discussed above, when the nonthermal motions behind
the shock are taken into account, the maximum Mach number
for collapse may be reduced further.
The conclusion that shocked clouds are stable against
collapse unless the Mach number is small
differs from that of \citet{HVanhala98}
since they considered a cloud that is initially highly
centrally concentrated, whereas we have assumed that
the initial cloud is not self-gravitating and is therefore
approximately uniform.

\section{Conclusion}
\label{sec:conclusion}

We have studied the interaction of shock waves with interstellar clouds, 
using a second-order, axisymmetric hydrodynamic code 
based on the Godunov method with 
local 
adaptive mesh refinement.
This problem has been studied previously by KMC, who assumed
that there is a discontinuous density jump between the cloud
and the intercloud medium. In reality, one expects that
the density distribution across the cloud boundary will be smooth
due to thermal conduction, non-equilibrium cooling effects, or
self-gravity
\citep {PKornreich00}. Here
we have generalized the treatment of KMC to
include the effects of smooth cloud boundaries on the 
shock-cloud interaction. 
We have considered a steady planar shock that strikes an initially spherical 
cloud with a density distribution that is flat near the 
center and decreases with radius as a power-law in the envelope.
Our model can be specified by four numbers: 
$n$, the power-law index of the density
distribution in the envelope; $\chi$, the density contrast between the cloud
center and the intercloud medium; $M$, the Mach number of the
shock; and $\gamma$, the ratio of specific heats.

Our numerical simulations have shown that 
the clouds are completely destroyed, primarily by the 
Kelvin-Helmholtz instability, irrespective of the initial density 
distributions, although the smoothness of the initial 
density distribution affects the timescale of cloud destruction. 
For a cloud with $n=2$ (the smoothest case we considered), 
the cloud destruction time is about 3 times longer than that 
of a uniform cloud.
The timescale of the cloud destruction is closely related to
the formation of a density discontinuity 
(the slip surface)  at the upstream side 
of the shocked cloud.  
The slip surface is subject to the Kelvin-Helmholtz and 
Rayleigh-Taylor instabilities, and therefore significant 
cloud destruction starts there.
Since the growth rates for these
instabilities are larger for perturbations
with shorter wavelengths in the linear regime and since the growth of 
perturbations with wavelengths shorter 
than the thickness of the shear layer is greatly
suppressed, the growth of the instabilities (and accordingly 
cloud destruction and mixing) are impeded prior to the formation
of the slip surface.
It therefore takes more time for shocks to destroy clouds
with smoother density gradients.

We have constructed simple analytic models describing  
the time evolution of the cloud velocity and the circulation
that agree well with our numerical results.
The vorticity is produced by the baroclinic term in the
vorticity equation, and can be divided into four components:
(1) $\Gamma_{\rm shock}$, which is generated by the initial shock-cloud
interacion, (2) $\Gamma_{\rm post}$, which is continuously produced by
the interaction between the shocked cloud and post-shock intercloud
flow, (3) $\Gamma_{\rm ring}$, which is produced in the intercloud
region by the supersonic vortex ring, and 
(4) $\Gamma_{\rm cloud}$, which is the circulation produced in the cloud
and is smaller than other components by a factor of $\chi^{-1/2}$.
For smooth clouds with $n=2$, the first component is generated 
throughout the cloud envelope 
after the initial shock passage, as discussed 
by \citet{PKornreich00}. 
However, most of the vorticity generated inside the cloud
shifts towards the upstream side of the cloud, 
being concentrated in the slip surface, and eventually 
is converted into random motions of small fragments
that are detached from the main cloud.

Vortical motions due to hydrodynamic instabilities 
produce substantial velocity dispersions in the shocked
cloud.  The vortical motions generated in the cloud are 
converted into random motions of small fragments shredded from
 the parent clouds.
Over the wide range of initial model parameters
considered here, the one-dimensional velocity dispersions
are typically $\delta v \sim 0.1 v_b$, 
which is estimated as about 2 km s$^{-1} $
for $T\simeq 70$ K, $\chi\simeq10^2$, and $M\simeq 3$.
Such velocity dispersions
are consistent with the observed internal motions of about 1.5 km s$^{-1}$
in the CNM \citep{CHeiles03}. 
Recently, \citet{CHeiles03} suggest that the CNM is in thin sheets with
large aspect ratios, by assuming that the thickness along the
line-of-sight
 is estimated to be $b=N_{\rm HI}/n_{\rm HI}$. 
If we apply the same procedure to estimate the 
aspect ratios, then the shocked clouds torn up into small fragments
also have large aspect ratios, and therefore, some of the observed CNM
clouds could be explained by the shocked clouds, instead of thin sheets.

At the early stage of cloud destruction, the small-scale fluctuations
are dominant due to the nonlinear growth of the Kelvin-Helmholtz
instability. Thus, the linewidth-size relation 
in a shocked cloud is expected to be
relatively flat.
Thereafter, the small-scale fluctuations tend to damp
gradually with time, leading to a linewidth that increases with size.

Under some circumstances, shock compression may induce 
gravitational collapse, leading to triggered star formation.
We have discussed how the critical Bonnor-Ebert mass changes due 
to shock passage.  The critical Bonnor-Ebert mass can decrease
only when $\gamma <4/3$.  
Radiative shocks are often assumed to be isothermal,
but in fact they are generally sub-isothermal.
We have determined the density jump across a sub-isothermal
shock and shown that, in the absence of significant nonthermal motions
behind the shock, star formation can be induced in an
initially uniform cloud by a shock only if it is of moderate strength.
Our results indicate that the postshock gas
has significant nonthermal motions, and this 
can further limit the onset of gravitational collapse.

We would like to thank M. R. Krumholz for making his 
valuable IDL codes available to us and for stimulating 
discussion on interstellar turbulence.
We also thank R. K. Crockett, D. Gies, and J. Walters 
for a number of valuable conversations.
F. N. acknowledges the support of the JSPS
Postdoctoral Fellowship for Research Abroad.
He is also grateful to the Astronomy Department, 
University of California at Berkeley 
for the kind hospitality during his stay.
CFM gratefully acknowledges the support of the 
Miller Institute for Basic Research and of the NSF
through grant AST-0098365.
Part of this work was supported by NASA ATP grant 
NAG5-12042 (RIK and CFM) and 
under the auspices of the US 
Department of Energy at the Lawrence Livermore National laboratory 
under contract W--7405--Eng--48 (RIK and RTF).
Computations were made  possible
by the NSF San Diego Supercomputer Center through
NPACI program grant UCB267 (RIK et al.).

\clearpage

\begin{deluxetable}{lllllllll}
\tablecolumns{9}
\tablecaption{Model Parameters \label{tab:model}}
\tablewidth{\columnwidth}
\tablehead{
  \colhead{Model}     & \colhead{$n$}          & \colhead{$\chi$}
 &\colhead{$M$}   &\colhead{$a_0/r_{\rm co}$}  &\colhead{$\cal R$} 
 & \colhead{$h_{\rm bd}/r_{\rm co}$} 
& \colhead{$\left<\rho_{\rm cl}\right>/\rho_{i0}$} 
& \colhead{Resolution} 
}
\startdata
\sidehead{$\gamma = 5/3$, sphere}
\cline{1-9}
AS1   & 2 & 10  & 1.5 & 2.61 & 6.58 & 1.22 & 2.91 & $R_{120}$\\
AS2   & 2 & 10  & 10  & 2.61 & 6.58 &1.22 & 2.91 & $R_{30}, R_{60}, R_{120}, R_{240}, R_{480}$\\
AS3   & 2 & 10  & 100 & 2.61 & 6.58 &1.22& 2.91  & $R_{120}$\\
AS4   & 2 & 100  & 10 & 8.50 & 34.7 &1.02 & 3.58  & $R_{120}$\\
AS5   & 4 & 10 & 10 & 1.49 & 1.91 & 0.611& 4.01 & $R_{120}$\\
AS6   & 4 & 100 & 10 & 2.23 &2.67  &0.510 & 8.57  & $R_{120}$\\
AS7   & 8 & 10 & 1.5 & 1.15 & 1.23 &0.306  & 5.62 &  $R_{120}$\\
AS8   & 8 & 10 & 10 & 1.15  & 1.23 &0.306  & 5.62 &  $R_{30}, R_{60}, R_{120}, R_{240}, R_{480}, R_{960}$\\
AS9   & 8 & 10 & 100 & 1.15 & 1.23 &0.306 & 5.62  & $R_{120}$\\
AS10   & 8 & 10 & 1000 & 1.15 & 1.23&0.306 & 5.62  &  $R_{120}$\\
AS11   & 8 & 100 & 10 & 1.27 & 1.28 &0.255 & 23.0  &  $R_{120}$\\
AS12   & 16 & 10 & 10 & 1.04 & 1.07 &0.153 & 7.22  &  $R_{120}$\\
AS13   & 24 & 10 & 10 & 1.02 & 1.03 &0.102 & 7.97  & $R_{120}$\\
AS14   & $\infty$ & 10  & 10 & 1.00 & 1.00 &0.00& 10.0 & $R_{30}, R_{60}, R_{120}, R_{240}$\\
AS15   & $\infty$ & 10  & 100 & 1.00 & 1.00&0.00 & 10.0 & $R_{120}$\\
AS16   & $\infty$ & 100  & 10 & 1.00 & 1.00&0.00 & 100.0 & $R_{120}$\\
\cline{1-9}
\sidehead{$\gamma = 1.1$, sphere}
\cline{1-9}
IS1   & 2 & 10  & 10 & 2.61 & 6.58  &1.22& 2.91  & $R_{120}$\\
IS2   & 8 & 10  & 10 & 1.15 & 1.23  &0.306& 5.62  & $R_{120}$\\
\cline{1-9}
\sidehead{$\gamma = 5/3$, cylinder}
\cline{1-9}
AC1   & 2 & 10  & 10 & 2.75 & 2.78 &1.22&3.47 & $R_{120}$\\
AC2   & 8 & 10  & 10 & 1.25 & 1.11 &0.306 & 6.59  & $R_{120}$
\enddata
\tablecomments{
In the eighth column, $\left<\rho_{\rm cl}\right>$ denotes the mean
cloud velocity.
In the sixth column, $\cal R$ denotes the cloud mass 
normalized with $m_0=4\pi r_{\rm co}^3 \rho_{c0}/3$ and 
$\pi r_{\rm co}^3 \rho_{c0}$ for spherical and cylindrical
clouds, respectively. To estimate the cloud mass 
of a cylindrical cloud, we set the vertical height to be 
equal to the core radius $r_{\rm co}$.
In the last column, $R_n$ means a grid resolution of $n$
cells per core radius measured at the finest level of refinement (level 3).
}
\end{deluxetable}

\begin{deluxetable}{llllllllllll}
\tabletypesize{\scriptsize}
\tablecolumns{12}
\tablecaption{Dependence on $n$ ($\chi = 10, M=10, R_{120}$)
\label{tab:dependence chi = 10}}
\tablehead{
\colhead{Model} & \multicolumn{2}{c}{AS2} &  
\multicolumn{2}{c}{AS5}  & \multicolumn{2}{c}{AS8}  
& \multicolumn{2}{c}{AS12}  
& \multicolumn{2}{c}{AS13}  
& \colhead{AS14}  \\
\colhead{Parameter} & \multicolumn{2}{c}{$n=2$} &  
\multicolumn{2}{c}{$n=4$}  & \multicolumn{2}{c}{$n=8$}  
& \multicolumn{2}{c}{$n=16$}  
& \multicolumn{2}{c}{$n=24$}  
& \colhead{$n=\infty$}  
}
\startdata
$t_{\rm slip}/\tcc $  & \mct{4.91} & \mct{1.98}  & \mct{1.71}  &
 \mct{0.722}  & \mct{0.335}   & 0.00 \\
$t_{\rm drag}/\tcc$ & \mct{7.51} & \mct{3.75}  & \mct{3.26}  &
 \mct{3.15}  & \mct{2.90}   & 2.52  \\ 
$t_{\rm dest}/\tcc$ & \mct{10.2}  & \mct{7.40}  & \mct{5.91}  &
\mct{5.34}  & \mct{4.47}   & 3.45  \\ 
$t_{\rm mix}/\tcc$ & \mct{4.73}  & \mct{8.04}  & \mct{6.59}  &
\mct{6.31}  & \mct{6.17}   & 4.91  \\ 
$t_m/\tcc $         & \mct{12.9}        & \mct{5.96}  & \mct{4.82}  &
 \mct{4.54}  & \mct{4.24}   & 3.80 \\ \cline{1-12}\\
$a/r_{\rm co} $              & 3.04    & (3.08) & 2.07 & (1.98)  & 1.73 & (1.66) &
 1.75 & (1.75)  & 1.76  & (1.75)  & 2.21 \\ 
$c/r_{\rm co} $              & 3.45    & (2.67) & 1.85 & (1.05)  & 1.48 & (1.29) &
 1.62 & (1.51)  & 1.79  & (1.75)  & 1.63 \\ 
$c/a $                & 1.12 & (0.865) & 0.893 & (0.529)  & 0.855 & (0.775) &
 0.928 & (0.862)  & 1.02  & (0.998)  & 0.738 \\ 
$\left<\rho\right>/\left<\rho_0\right>$ 
     & 2.55 & (1.59) & 2.20 & (1.97)  & 1.90 & (1.84) &
 1.65 & (1.59)  & 1.48  & (1.42)  & 1.24 \\ 
$\left<v_z\right>/v_b$      & 0.173 & (0.124) & 0.182 & (0.186)  & 0.200 & (0.206) &
 0.173 & (0.170)  & 0.162  & (0.162)  & 0.130 \\ 
$\delta v_\varpi/v_b$   & 0.0758 & (0.0840) & 0.0953 & (0.0788)  & 0.102 & (0.0840) &
 0.113& (0.105)  & 0.116  & (0.113)  & 0.115 \\ 
$\delta v_z/v_b$   & 0.0990 & (0.0703) & 0.0870 & (0.0701)  & 0.158 & (0.119) &
 0.164 & (0.159)  & 0.181  & (0.176)  & 0.182 \\ 
$-\Gamma/r_{\rm co}v_b$   & 2.38 &  & 0.501 &   & 1.25 &  &
 1.20 &  & 1.17  &   & 1.50 \\ 
\enddata

\tablecomments{All the time-dependent quantities are estimated at $t=t_m$.
The initial cloud mass $m_{\rm cl,0}$ is adopted as the threshold mass
in computing the global quantities, except that the values in parentheses 
are measured with respect to the initial core mass,
 $m_{\rm co, 0}$.
Note that the radii $a$ and $c$ are normalized by $r_{\rm co}$.
If the radius $a$ is normalized by its initial value $a_0$,
then the radius of the shocked cloud decreases monotonically with $n$.
See \S \ref{subsec:dependence}.
}
\end{deluxetable}

\begin{deluxetable}{llllllll}
\tabletypesize{\scriptsize}
\tablecolumns{8}
\tablecaption{Dependence on $n$ ($\chi = 100, M=10, R_{120}$)
\label{tab:dependence chi = 100}}
\tablehead{
\colhead{Model} & \multicolumn{2}{c}{AS4} &  
\multicolumn{2}{c}{AS6}  & \multicolumn{2}{c}{AS11} & \colhead{AS16}  \\
\colhead{Parameter} & \multicolumn{2}{c}{$n=2$\tablenotemark{a}} &  
\multicolumn{2}{c}{$n=4$}  & \multicolumn{2}{c}{$n=8$} & \colhead{$n=\infty$}  
}
\startdata
$t_{\rm slip}/\tcc$ & \mct{3.50} & \mct{1.63}  & \mct{1.10}  & 0.00 \\
$t_{\rm drag}/\tcc$ & \mct{10.5} & \mct{5.22}  & \mct{5.01}  & 3.59 \\
$t_{\rm dest}/\tcc $ & \mct{12.1} & \mct{6.51} & \mct{6.50}  & 3.02 \\ 
$t_{\rm mix}/\tcc $ &\mct{12.0} & \mct{6.24} &\mct{5.69}  & 4.72 \\ 
$t_m/\tcc$ & \mct{$>15.0$}        & \mct{7.05}  & \mct{6.57}  & 4.01 \\ 
\cline{1-8}\\
$a/r_{\rm co} $              & 8.14   & (0.385) & 3.42 & (1.91)  & 3.16 & (2.83) & 3.15 \\
$c/r_{\rm co} $              & 26.6  & (4.45) & 15.3 & (3.41)  & 11.2 & (9.50) & 3.15 \\
$c/a $                & 3.27 & (11.5) & 4.48 & (1.79)& 3.57 & (3.35) & 2.65\\
$\left<\rho\right>/\left<\rho_0\right>$ 
     & 2.01 & (0.240) & 0.959 & (0.390)  & 0.245 & (0.289) & 2.65 \\
$\left<v_z\right>/v_b$ & 0.239 & (0.286) & 0.181 & (0.314)  & 0.184 & (0.217) & 0.240 \\
$\delta v_\varpi/v_b$  & 0.0571 & (0.0582) & 0.0810 & (0.0418)  & 0.102 & (0.0840) & 0.0987 \\
$\delta v_z/v_b$   & 0.142 & (0.0989) & 0.0870 & (0.0701)  & 0.175 & (0.106) & 0.236 \\
$-\Gamma/r_{\rm co}v_b$   & 17.0 &  & 10.4 &   & 8.03 &  & 6.51 \\
\enddata

\tablenotetext{a}{The time-dependent quantities are evaluated at 
$t = 16.2 \, \tcc$.}

\tablecomments{
The time-dependent values are estimated at $t=t_m$.
The initial cloud mass $m_{\rm cl,0}$ is adopted as the threshold mass
to compute the global quantities, except in the values in parentheses 
which are the global quantities measured in the initial core mass,
 $m_{\rm co,}$.  
}
\end{deluxetable}

\begin{deluxetable}{lllllllllllll}
\tabletypesize{\scriptsize}
\tablecolumns{13}
\tablecaption{Dependence on $M$ ($\chi = 10, R_{120}$)\label{tab:mach}}
\tablehead{
\colhead{Parameter} & \multicolumn{6}{c}{$n=8$\tablenotemark{a}} &  
\multicolumn{6}{c}{$n=2$} \\ \cline{1-13} \\
\colhead{Model}           & 
\multicolumn{2}{c}{AS7} & \multicolumn{2}{c}{AS9} & 
\multicolumn{2}{c}{AS10} & 
\multicolumn{2}{c}{AS1} & \multicolumn{2}{c}{AS2} & \multicolumn{2}{c}{AS3}  \\
\colhead{$M$}           & 
\multicolumn{2}{c}{1.5} & \multicolumn{2}{c}{100} & 
\multicolumn{2}{c}{1000} & 
\multicolumn{2}{c}{1.5} & \multicolumn{2}{c}{10} & \multicolumn{2}{c}{100}  
}
\startdata
$v_{c0}/v_b$ & 
\mct{0.417} & \mct{0.750}  & \mct{0.750}  & \mct{0.417} & \mct{0.742}
  & \mct{0.750} \\ 
$t_{\rm drag}/\tcc$ & 
\mct{17.2}  & \mct{3.25}  & \mct{3.25} & \mct{21.4}  & \mct{7.51}  
& \mct{6.92} \\ 
$t_{\rm dest}/\tcc$ & 
\mct{12.2}  & \mct{5.88}  & \mct{5.87} & \mct{32.8}  & \mct{10.2}  
& \mct{10.1} \\ 
$t_{\rm mix}/\tcc$ & 
\mct{15.8}  & \mct{6.52}  & \mct{6.60} & \mct{25.1}  & \mct{4.73}  & 
\mct{4.92} \\ 
$t_m/\tcc $ & 
\mct{7.51}  & \mct{5.15}   & \mct{5.13}  & \mct{$>$ 32.7}  
 & \mct{12.9}  & \mct{13.6}  \\ \cline{1-13}\\
$a/r_{\rm co} $            & 
 1.82  & (1.74)  & 1.80  & (1.71) & 1.80 & (1.78) &
  3.89 & (4.02)  & 3.04 & (3.08) &3.18  & (3.24) \\ 
$c/r_{\rm co} $            & 
 1.49  & (1.18)  & 1.60  & (1.48)  & 1.60 & (1.50) & 
 5.20 &(4.39) & 3.45  &(2.67) & 3.46 & (2.32) \\ 
$c/a $          & 
 0.816  & (0.680) & 0.889 & (0.864) & 0.888 & (0.845)
 &
 1.34 & (1.09) & 1.12 & (0.865) & 1.09 & (0.716)  \\ 
$\left<\rho\right>/\left<\rho_0\right>$ &
1.33   & (1.37)  & 1.86  & (1.70)  & 1.86  & (1.72) &
 1.25 & (0.712) & 2.55 & (1.59) & 2.55 & (1.50)  \\ 
$\left<v_z\right>/v_b$ & 
0.193   & (0.197)  & 0.182 &(0.193) & 0.182 &(0.193) &
  0.114 & (0.0904) & 0.173  & (0.124) & 0.159 & (0.125)  \\ 
$\delta v_\varpi/v_b$ &
0.0629   & (0.0508) & 0.104 & (0.0900) &0.104&(0.0908)& 
 0.0475 & (0.0410) & 0.0758 & (0.0840) & 0.0781 &  (0.0732) \\ 
$\delta v_z/v_b$    &
0.0879   & (0.0646) & 0.172 & (0.146)&0.164&(0.140)& 
 0.0738 & (0.0654) & 0.0990 & (0.0703) & 0.101 &  (0.0711) \\ 
$-\Gamma/r_{\rm co}v_b$     & 
 1.55   &  & 1.22  &  & 1.22  &  & 2.23   &    & 2.38 & & 2.37& \\ 
\enddata

\tablenotetext{a}{see Table \ref{tab:dependence chi = 10} for 
the global quantities of $\chi=10$}

\tablecomments{
The time-dependent values are estimated at $t=t_m$.
The initial cloud mass $m_{\rm cl,0}$ is adopted as the threshold mass
to compute the global quantities, except in the values in parentheses 
which are the global quantities measured in the initial core mass,
 $m_{\rm co,}$.  
}
\end{deluxetable}

\begin{deluxetable}{lllll}
\tabletypesize{\scriptsize}
\tablecolumns{5}
\tablecaption{Radiative Clouds ($\chi = 10, M=10, R_{120}$)
\label{tab:radiative}}
\tablehead{
\colhead{Model} & \multicolumn{2}{c}{IS1} & \multicolumn{2}{c}{IS2}  \\
\colhead{Parameter} & \multicolumn{2}{c}{$n=2$} &  
\multicolumn{2}{c}{$n=8$}  
}
\startdata
$t_{\rm drag}/\tcc$ & \mct{0.500}        & \mct{1.56}        \\ 
$t_{\rm dest}/\tcc$ & \mct{7.23}        & \mct{6.71}        \\ 
$t_{\rm mix}/\tcc$ & \mct{6.10}        & \mct{5.56}        \\ 
$t_m/\tcc $         & \mct{$>8.87$}        & \mct{4.99}       \\ \cline{1-5}\\
$a/r_{\rm co} $              & 2.41    & (1.29) & 1.11  & (0.830) \\ 
$c/r_{\rm co} $              & 6.74
    & (8.19)    & 4.54  & (3.20) \\
$c/a $                & 2.79   & (6.36)    & 1.15  & (3.86) \\
$\left<\rho\right>/\left<\rho_0\right>$ 
                      & 10.8    & (6.12)    & 7.05 & (5.65) \\
$\left<v_z\right>/v_b$ &0.129   & (0.0649)  & 0.164 & (0.196) \\ 
$\delta v_\varpi/v_b|$  & 0.0541 & (0.0500)  &0.0553  & (0.0532) \\ 
$\delta v_z/v_b$       & 0.0989  & (0.149)  & 0.106 & (0.0793) \\ 
$-\Gamma/r_{\rm co}v_b$       & 0.140  &      & 0.327 &         \\
\enddata
\tablecomments{
We note that the circulation associated with the supersonic vortex ring,
 normalized with $r_{\rm co}v_b$, contributes more for radiative clouds 
than for nonradiative clouds.  
Since this has the opposite sign from the other contributions to
the circulation,
the values for the total circulation listed above 
are smaller than those in Table 
\ref{tab:dependence chi = 10}. However, 
the peak values of the circulation are close to those of the
 nonradiative cases. 
}
\end{deluxetable}

\begin{deluxetable}{lllll}
\tabletypesize{\scriptsize}
\tablecolumns{5}
\tablecaption{Cylindrical Clouds ($\chi = 10, M=10, R_{120}$)
\label{tab:cylinder}}
\tablehead{
\colhead{Model} & \multicolumn{2}{c}{AC1} & \multicolumn{2}{c}{AC2}  \\
\colhead{Parameter} & \multicolumn{2}{c}{$n=2$\tablenotemark{a}} &  
\multicolumn{2}{c}{$n=8$}  
}
\startdata
$t_{\rm drag}/\tcc$ & \mct{2.04}        & \mct{2.38}        \\ 
$t_{\rm dest}/\tcc$ & \mct{17.1}        & \mct{6.25}        \\ 
$t_{\rm mix}/\tcc$ &  \mct{18.7}        & \mct{11.3}        \\ 
$t_m/\tcc $         & \mct{$>$ 19.0}
        & \mct{9.14}        \\ \cline{1-5}\\
$a/r_{\rm co} $              & 3.43    & (3.53) & 2.23 & (2.23) \\ 
$c/r_{\rm co} $              & 3.00    & (2.84)    & 3.24 & (3.15) \\
$c/a $                & 0.876   & (0.805)    & 1.45 & (1.41) \\
$\left<\rho\right>/\left<\rho_0\right>$ 
                      & 2.46    & (2.00)    & 1.50 & (1.37) \\
$\left<v_z\right>/v_b$ & 0.0915  & (0.0975)  & 0.111 & (0.111) \\ 
$\delta v_\varpi/v_b$  & 0.0758  & (0.0635)  & 0.0893 & (0.0893) \\ 
$\delta v_z/v_b$       & 0.0885  & (0.0662)  & 0.136 & (0.136) \\ 
$-\Gamma/r_{\rm co}v_b$       & 2.03  &      & 2.24  &         \\
\enddata

\tablenotetext{a}{For $n=2$, the time-dependent global quantities are 
evaluated at $t= 12.9 \, \tcc$, which is the maximum expansion time for
 the spherical case.}

\tablecomments{
The time-dependent values are estimated at 
$t=t_m$.
The initial cloud mass $m_{\rm cl,0}$ is adopted as the threshold mass
to compute the global quantities, except in the values in parentheses 
which are the global quantities measured in the initial core mass,
 $m_{\rm co,}$.  
}
\end{deluxetable}

\clearpage

\begin{figure}
\plotone{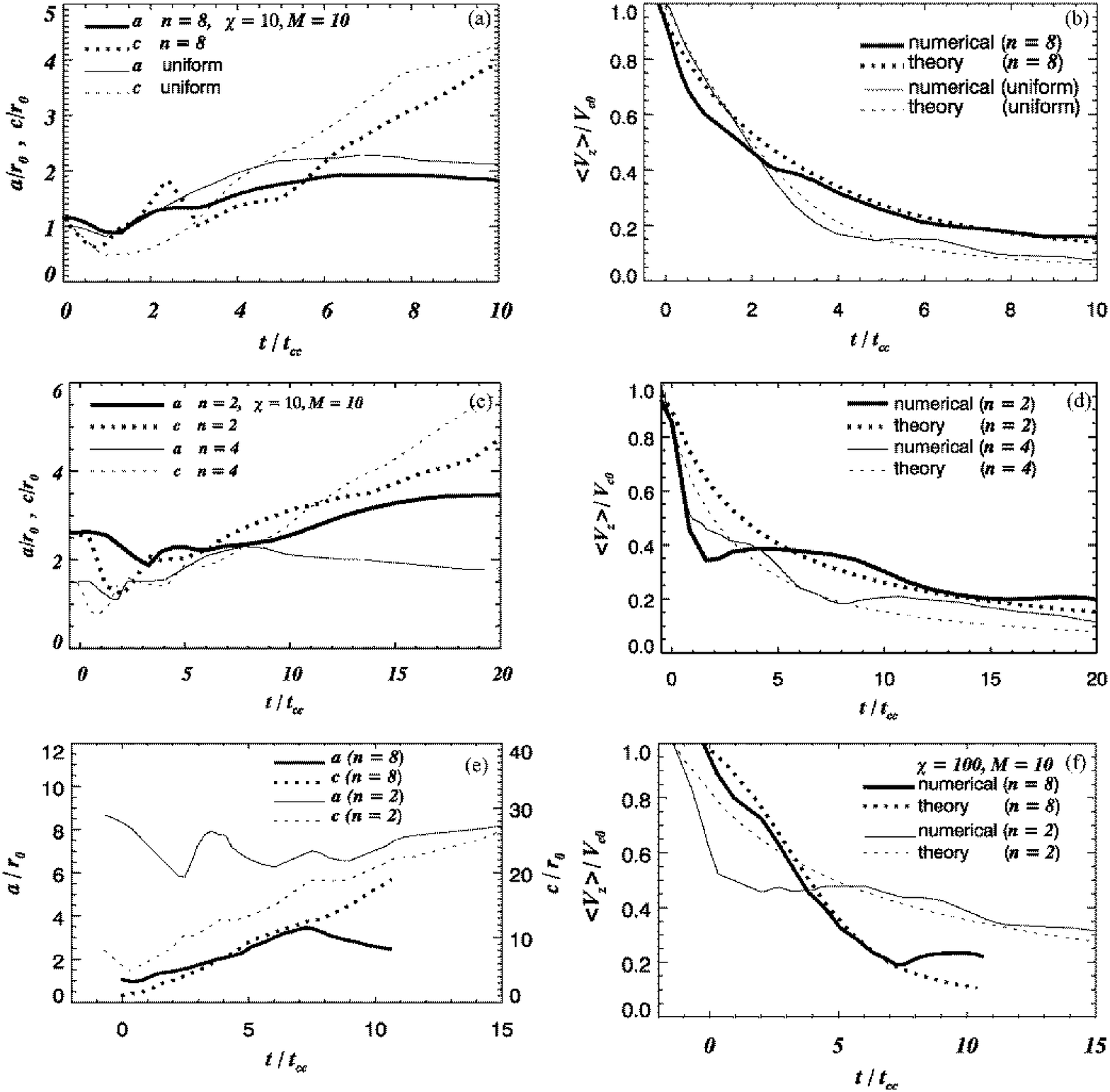}
\caption{{\it (a) and (b)}: Time evolution of the cloud shape and 
 the cloud velocity for two different models with 
$(n,\chi, M) = (8, 10, 10)$ ({\it thick curves}) 
and $(\infty, 100, 10)$ ({\it thin curves}).
Panel (a) shows the time evolution of the rms radial radius $a$ ({\it
 solid curves}) and axial radius $c$ ({\it dotted curves}).
Panel (b) shows the time evolution of the mean cloud velocity 
({\it solid curves}). For comparison, the analytic solution is indicated 
by {\it dotted curves}.
{\it (c) and (d)}:Same as the upper panels but for the models with
$(n,\chi, M) = (2, 10, 10)$ and (4, 10, 10).
{\it (e) and (f)}:Same as the upper panels but for the models with
} 
\label{fig:cloud shape1}
\end{figure}

\begin{figure}
\plottwo{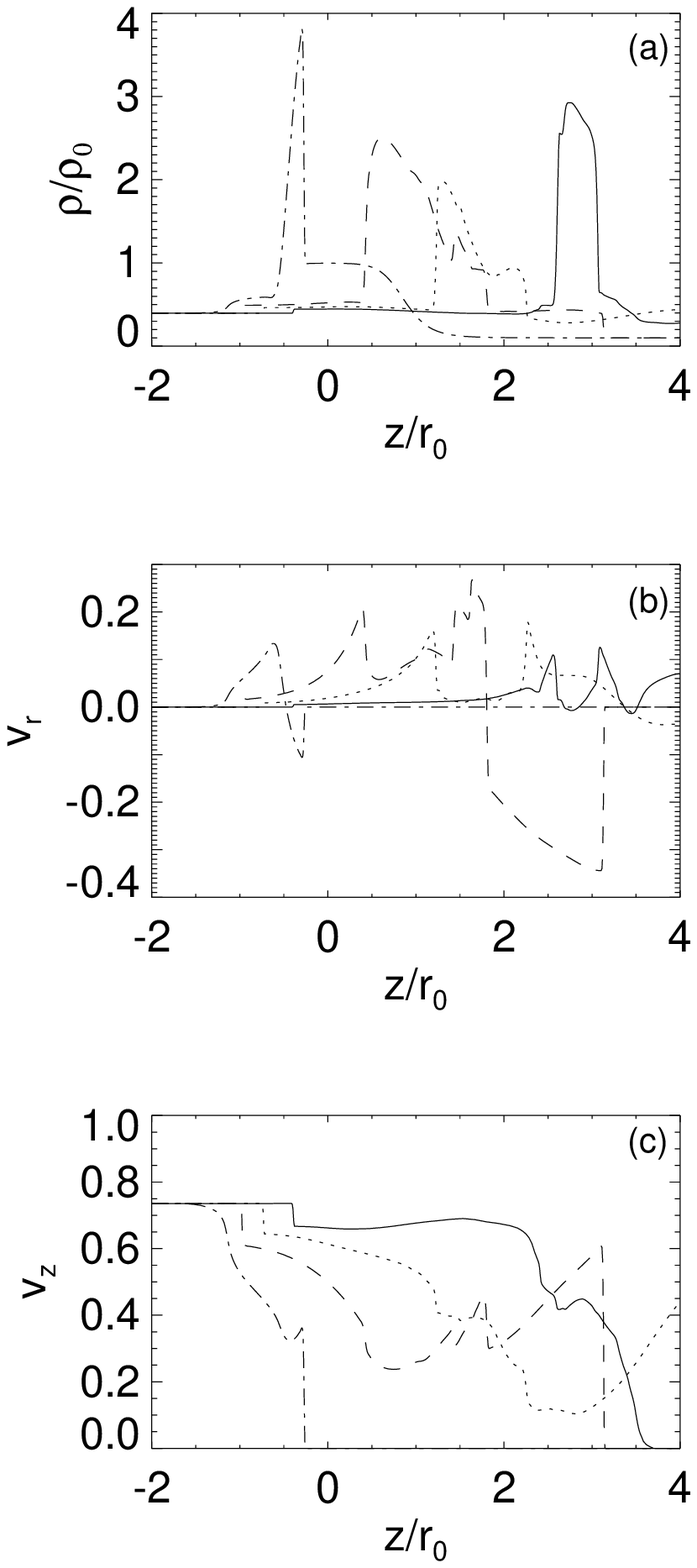}{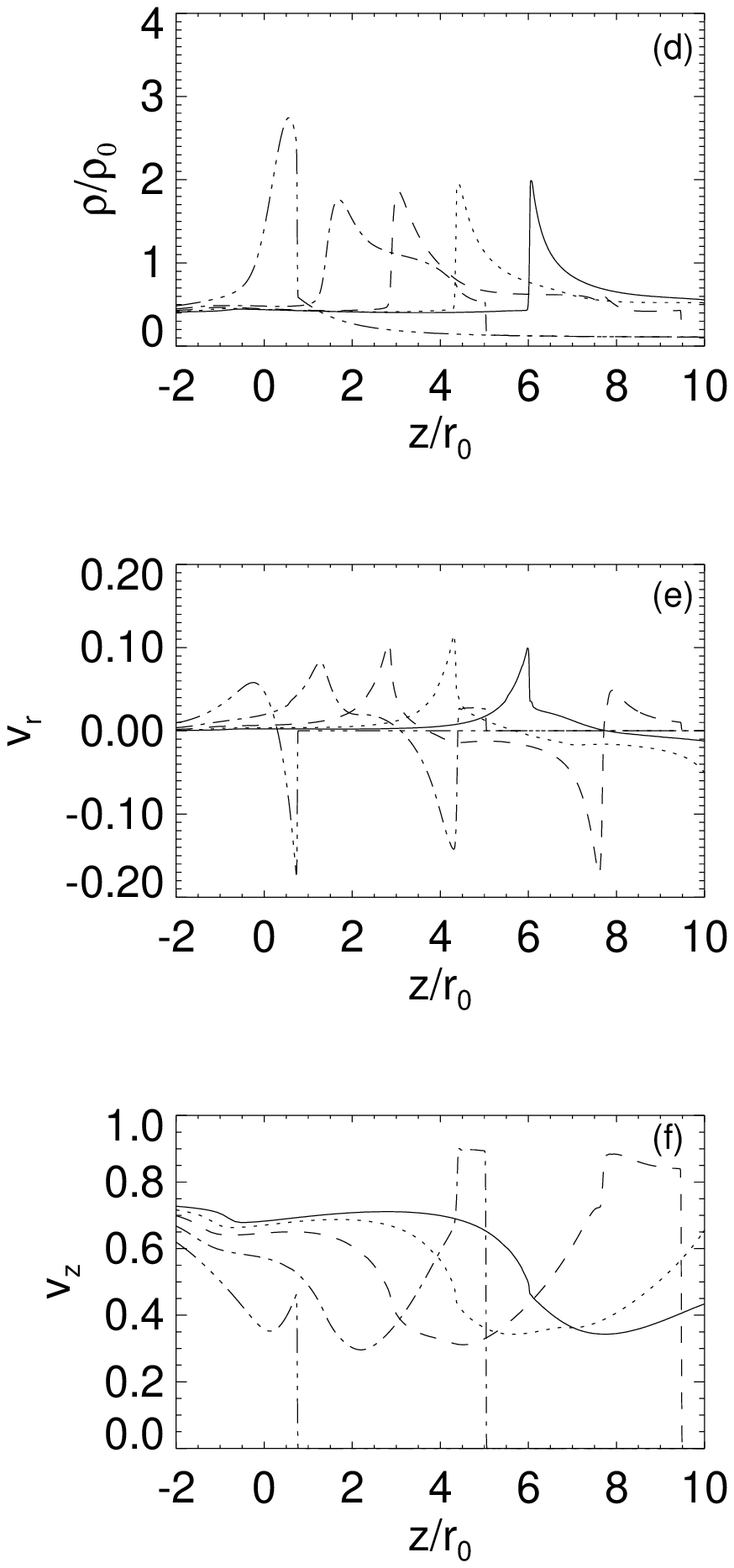}
\caption{
Distributions of (a) density, (b) $v_r$, and (c) $v_z$, along
 $r=0.5r_{\rm co}$, for $n=8$, $\chi=10$, and  $M=10$.
Solid, dotted, dashed, and dash-dot curves indicate
the profiles at $t=0.249 \, \tcc$, $1.31 \, \tcc$, $2.13 \, \tcc$, and
$3.04 \, \tcc$, respectively.
Distributions of (d) density, (e) $v_r$, 
and (f) $v_z$, along $r=0.5 \, r_{\rm co}$, for $n=2$, $\chi=10$, and $M=10$.
Solid, dotted, dashed, dash-dot, and dash-dot-dot-dot curves indicate
the profiles at $t=1.49 \, \tcc$, $2.82 \, \tcc$, $4.05 \, \tcc$, 
$5.18 \, \tcc$, and $6.41 \, \tcc$, respectively.
}
\label{fig:slip}
\end{figure}

\begin{figure}
\plottwo{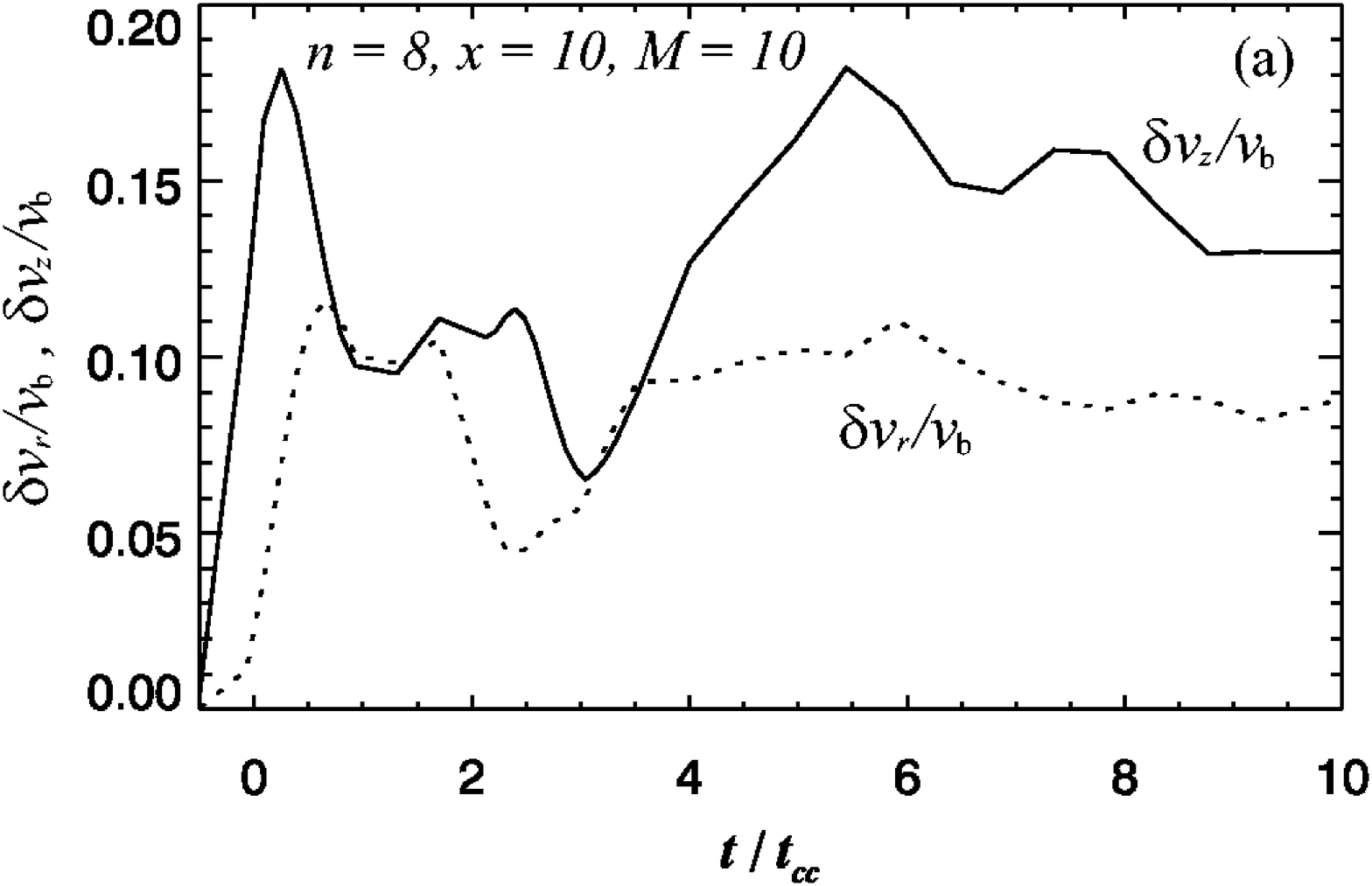}{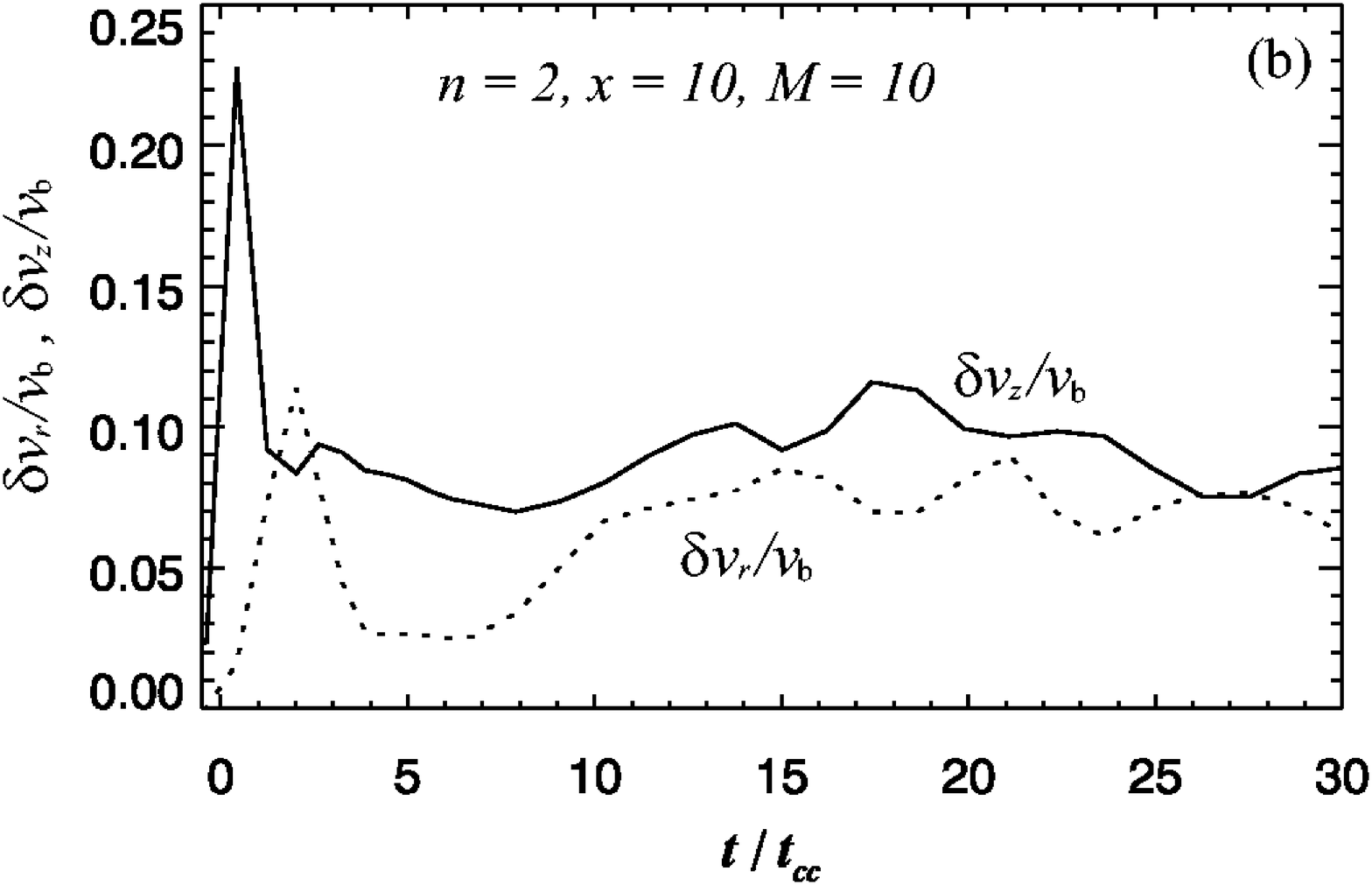}
\caption{Time evolution of the radial (dotted curve)
and axial (solid curve) velocity dispersions for
(a) $n=8$, $\chi=10$, and $M=10$ and (b) $n=2$, $\chi=10$, and $M=10$.}
\label{fig:dispersion}
\end{figure}

\begin{figure}
\plotone{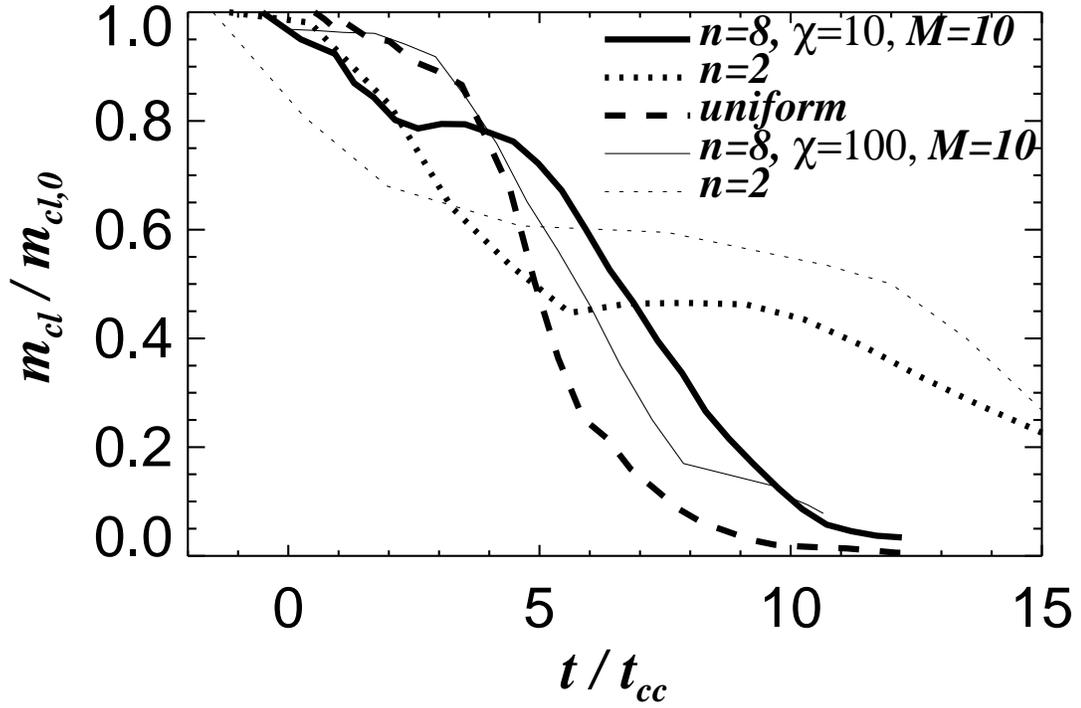}
\caption{Time evolution of fractional masses with densities greater than 
2 $\rho_{\rm i1}$, normalized by the initial cloud mass.
} 
\label{fig:mixing}
\end{figure}

\begin{figure}
\plotone{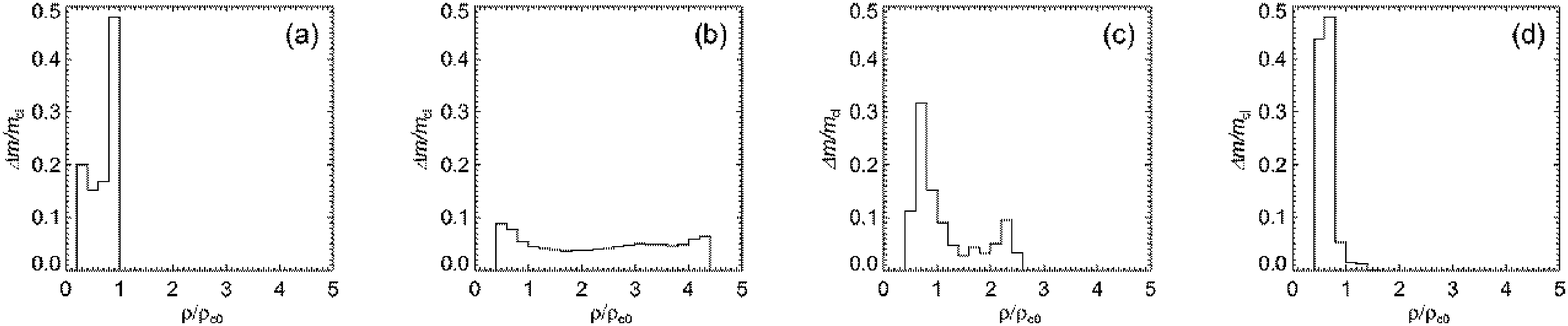}
\plotone{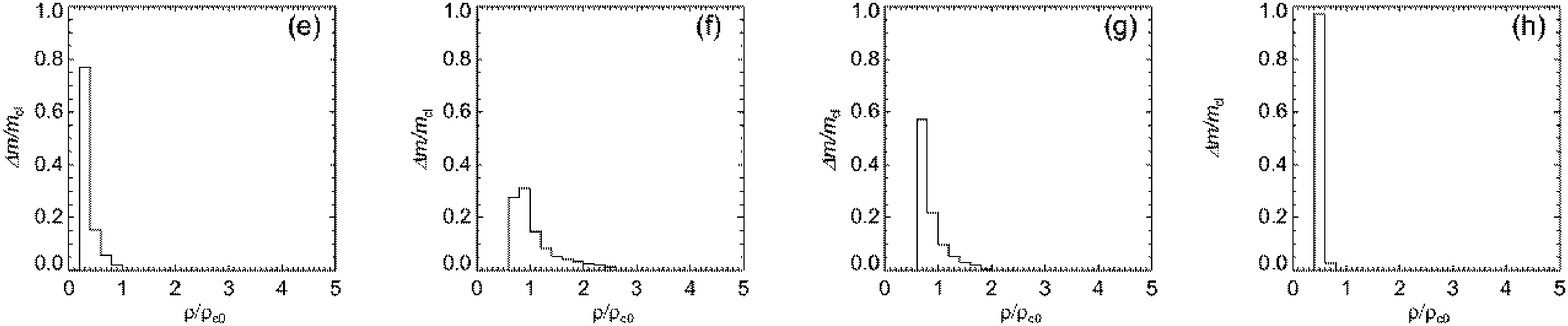}
\caption{Evolution of the mass fraction for the models with
($n,\chi, M$)=(8, 10, 10)
at times (a) $t=0$, (b) $1.06 \, t_{\rm cc}$, (c) $6.04 \, t_{\rm cc}$, 
and (d) $8.00 \, t_{\rm cc}$, 
and for the models with ($n,\chi, M$)=(2, 10, 10)
at times (e) $t=0$, (f) $1.40 \, t_{\rm cc}$, (g) $7.28\, t_{\rm cc}$, 
and (h) $29.5\, t_{\rm cc}$. 
The histograms indicate the fraction of mass contained within a
 corresponding density bin with a width of 0.2 $\rho_{c0}$,
which is normalized by the total cloud mass $m_{\rm cl,0}$. }
\label{fig:mass fraction}
\end{figure}

\begin{figure}
\plottwo{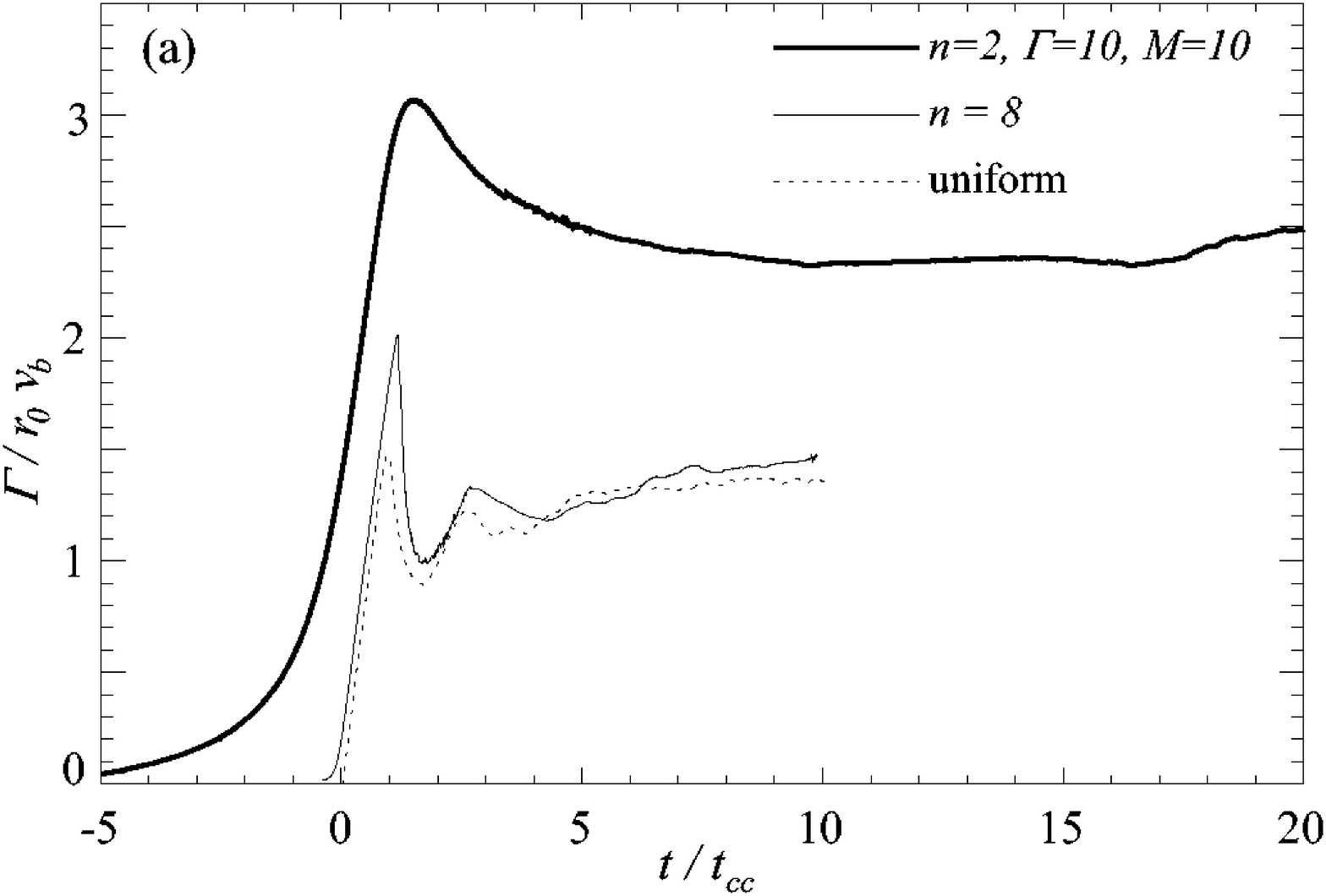}{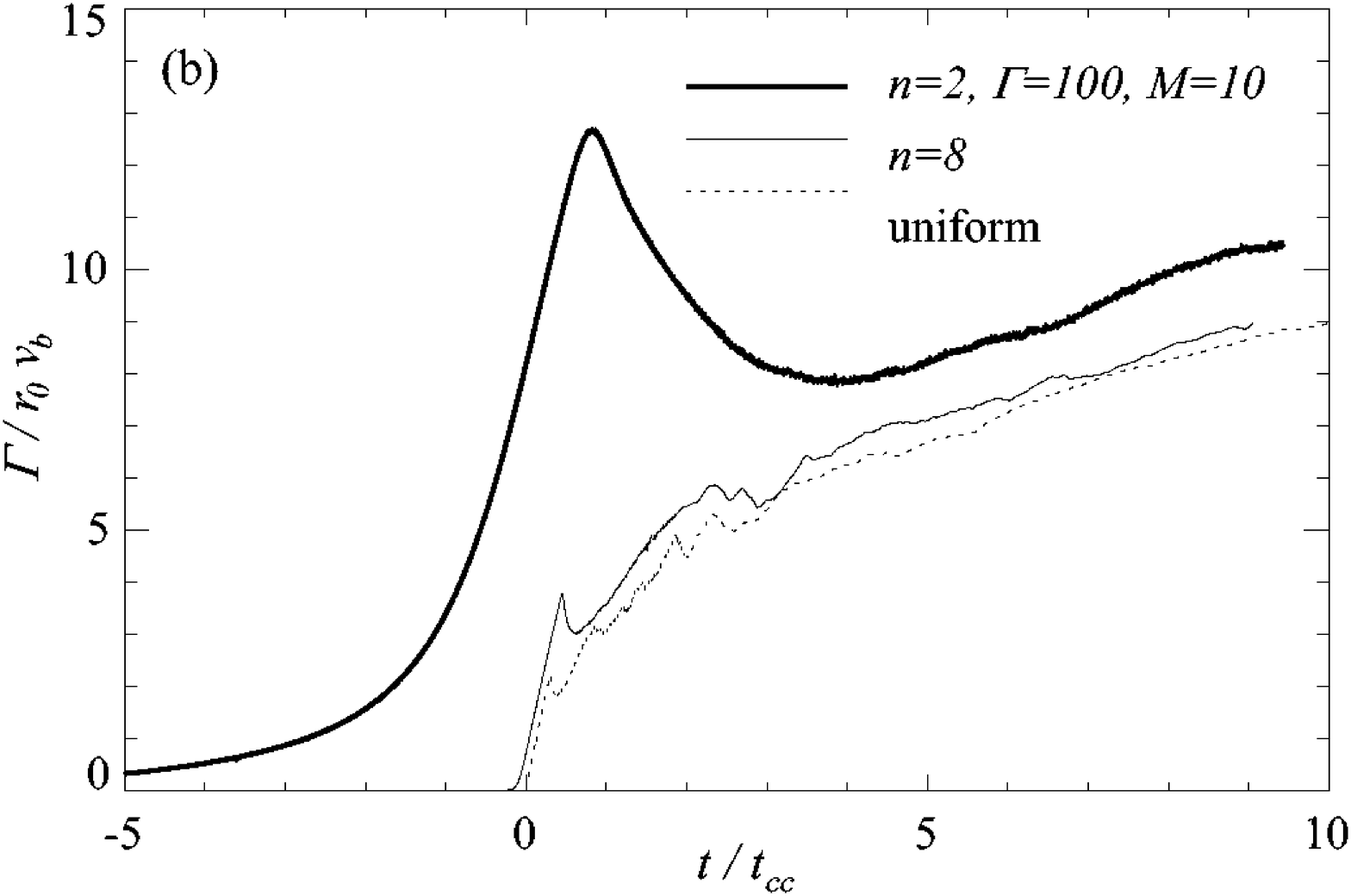}
\caption{Time evolution of circulations
for clouds with (a) $\chi=10$  and 
clouds with (b) $\chi=100$.
Thick solid, thin solid, and dashed curves denote the time evolution of the 
total circulation for 
the models with $n=2$, 8, and $\infty$, respectively.}
\label{fig:circulationx10}
\end{figure}

\begin{figure}
\plotone{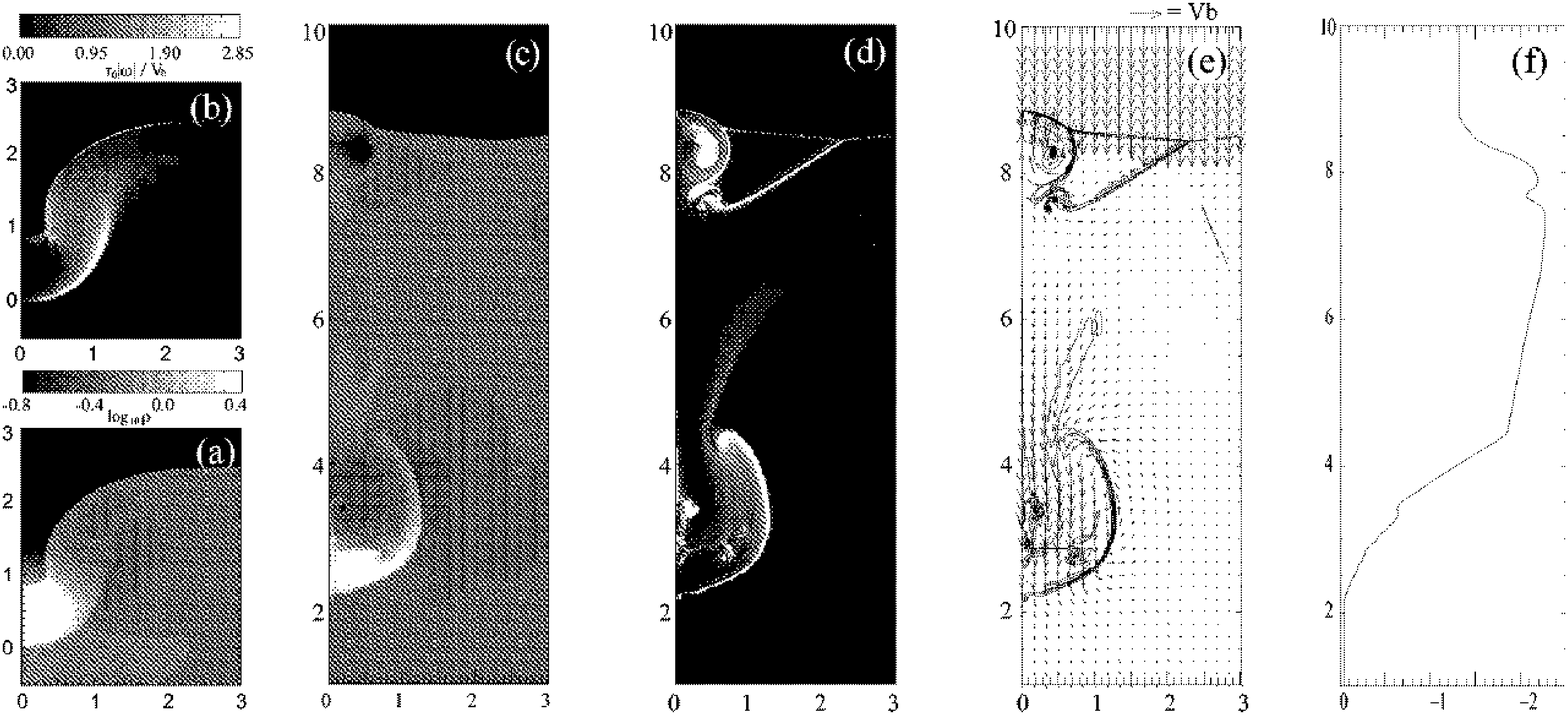}
\caption{
Density and vorticity distributions 
for the model with $(n, \chi, M)= (8, 10, 10)$.
Panels (a) and (b): the density distribution and vorticity
 distribution at $t=0.930 \, \tcc$.
Panels (c), (d), and (e): the density distribution, vorticity distribution,
and the velocity distribution at $t=2.86 \, \tcc$.
The velocity vectors are measured in the frame that moves with 
the postshock flow. 
Panel (f): cumulative circulation at $t=2.86 \, \tcc$.
}
\label{fig:vorticityn8x10}
\end{figure}

\begin{figure}
\plotone{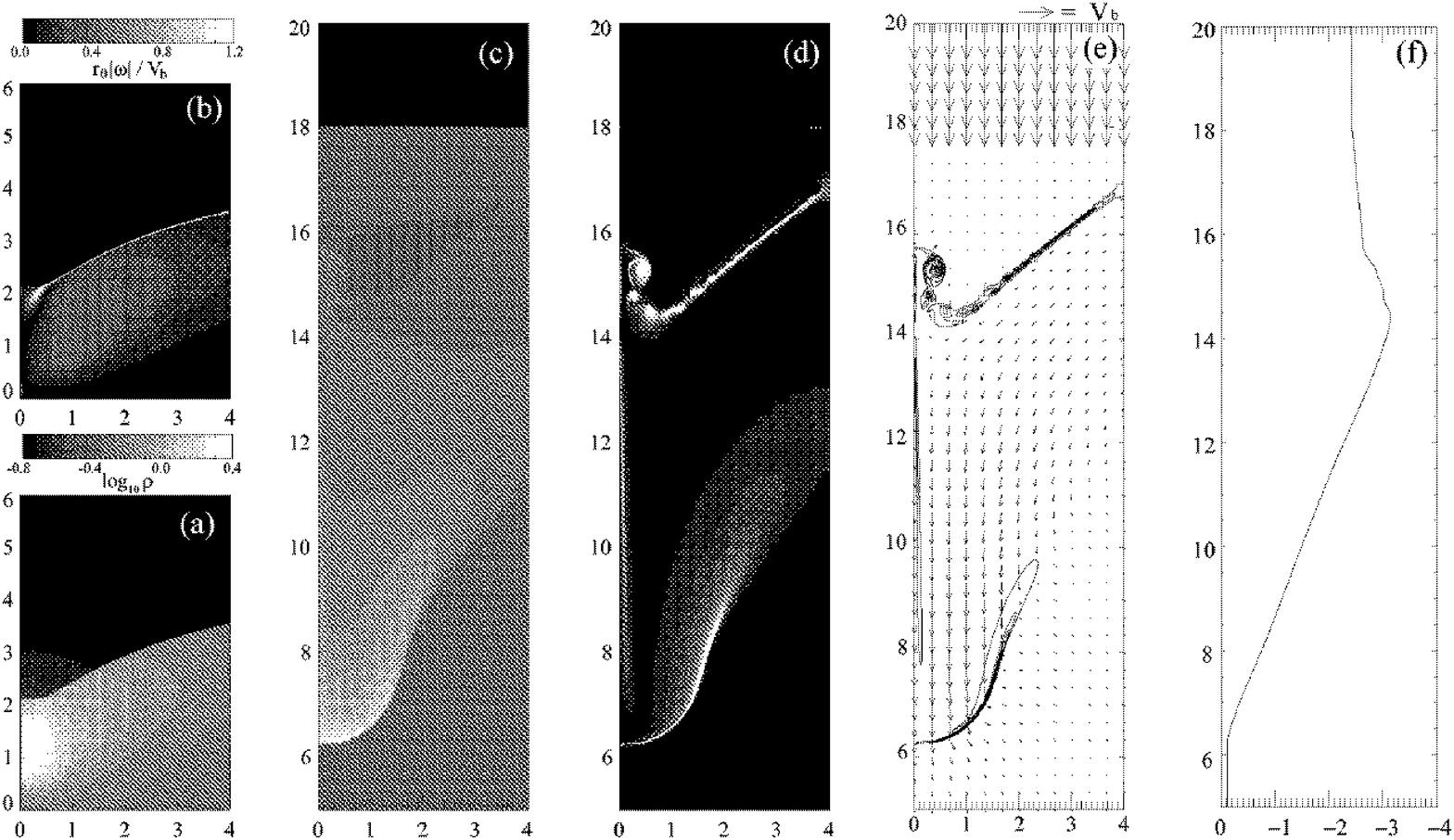}
\caption{Density and vorticity distributions 
for the model with $(n, \chi, M)= (2, 10, 10)$.
Panels (a) and (b): the density distribution and vorticity
 distribution at $t=2.02 \tcc$.
Panels (c), (d), and (e): the density distribution, 
vorticity distribution,and the velocity distribution at $t=6.65 \, \tcc$.
The velocity vectors are measured in the frame that moves with 
the postshock flow. 
Panel (f): cumulative circulation at $t=6.65 \, \tcc$. 
}
\label{fig:vorticityn2x10}
\end{figure}

\begin{figure}
\plotone{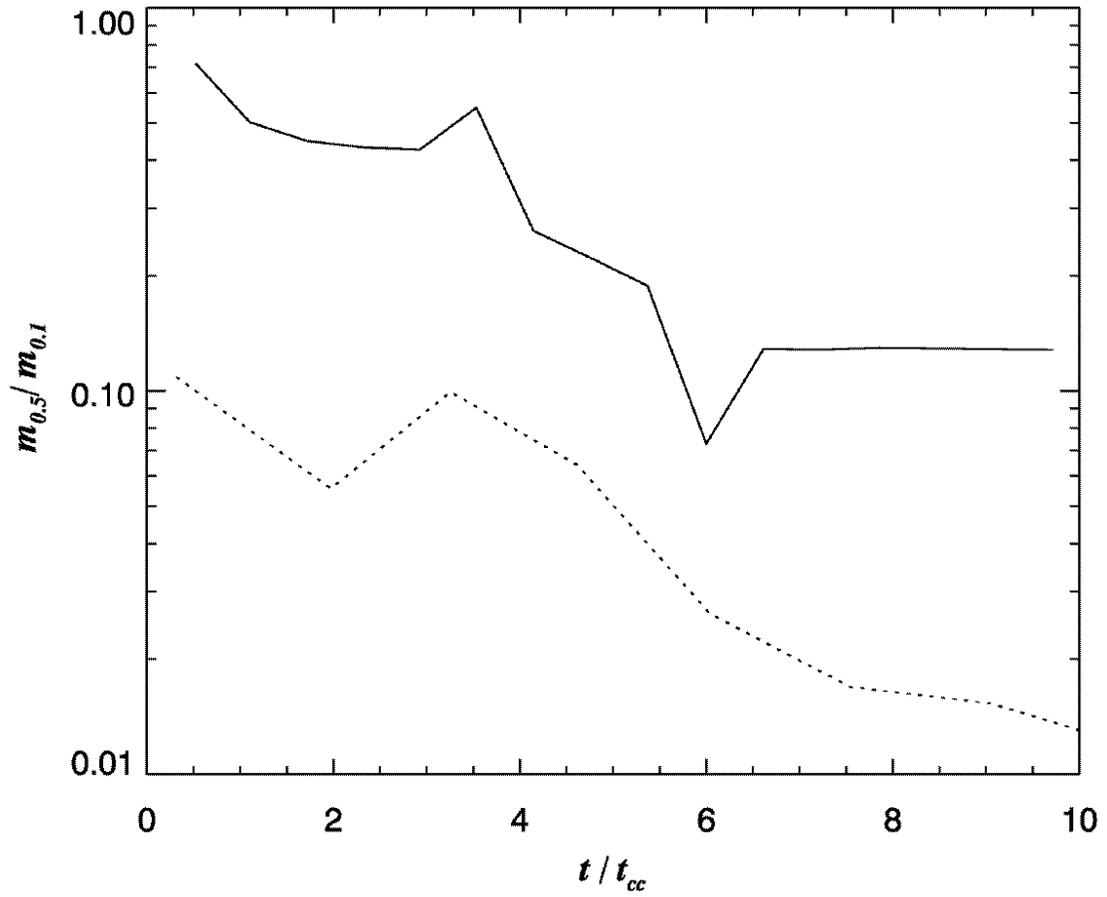}
\caption{Evolution of the mass ratio for the models with 
$(n, \chi, M) = (8, 100, 10)$ and 
(2, 100, 10), which are plotted with 
solid and dotted curves, respectively.}
\label{fig:mass ratio}
\end{figure}

\begin{figure}
\plotone{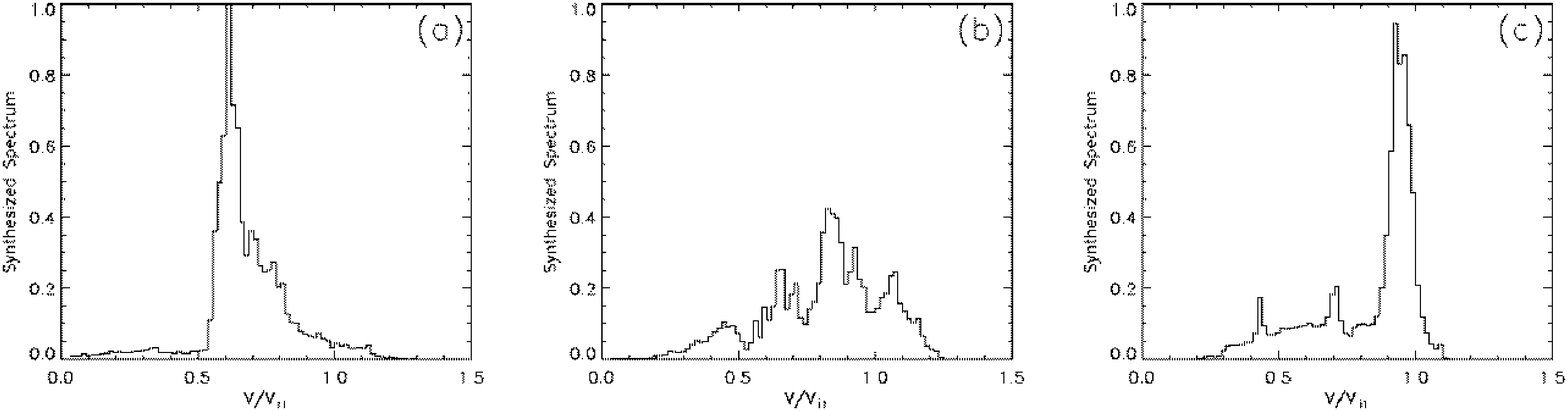}
\plotone{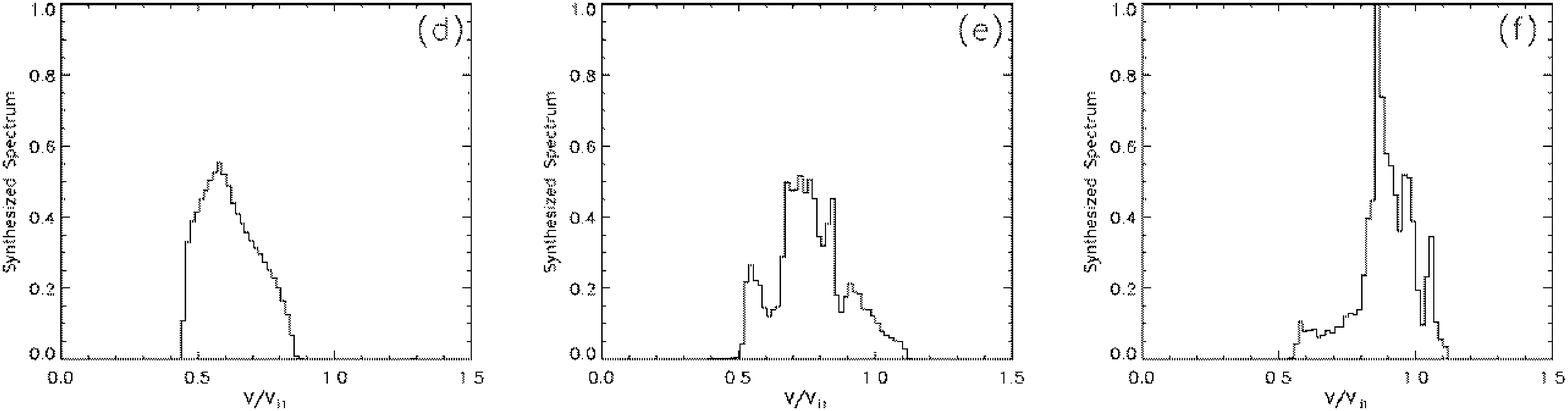}
\caption{Synthesized spectra for the model with 
$(n, \chi, M) = (8, 10, 10)$ at (a) t = 4.00 $t_{\rm cc}$, 
(b) $8.00 \, t_{\rm cc}$, 
and (c) $13.2 \, t_{\rm cc}$,
and for the model with $(n, \chi, M) = (2, 10, 10)$ 
at (d) t = 6.00 $t_{\rm cc}$, 
(e) $12.0 \, t_{\rm cc}$, 
and (f) $20.1 \, t_{\rm cc}$,
The spectra are normalized by the maximum at (a) and (d) for $n=8$ and
 2, respectively.
The histograms denote the mass contained within a corresponding
 velocity bin with a width of 0.015$v_{i1}$. 
The abscissa denotes the velocity normalized by the velocity of the 
postshock ambient medium $v_{i1}$.
The integration is done over the entire cloud under the assumption
that the line-of-sight is parallel to the $z$-axis.
}
\label{fig:spectra}
\end{figure}

\begin{figure}
\includegraphics[width=5cm]{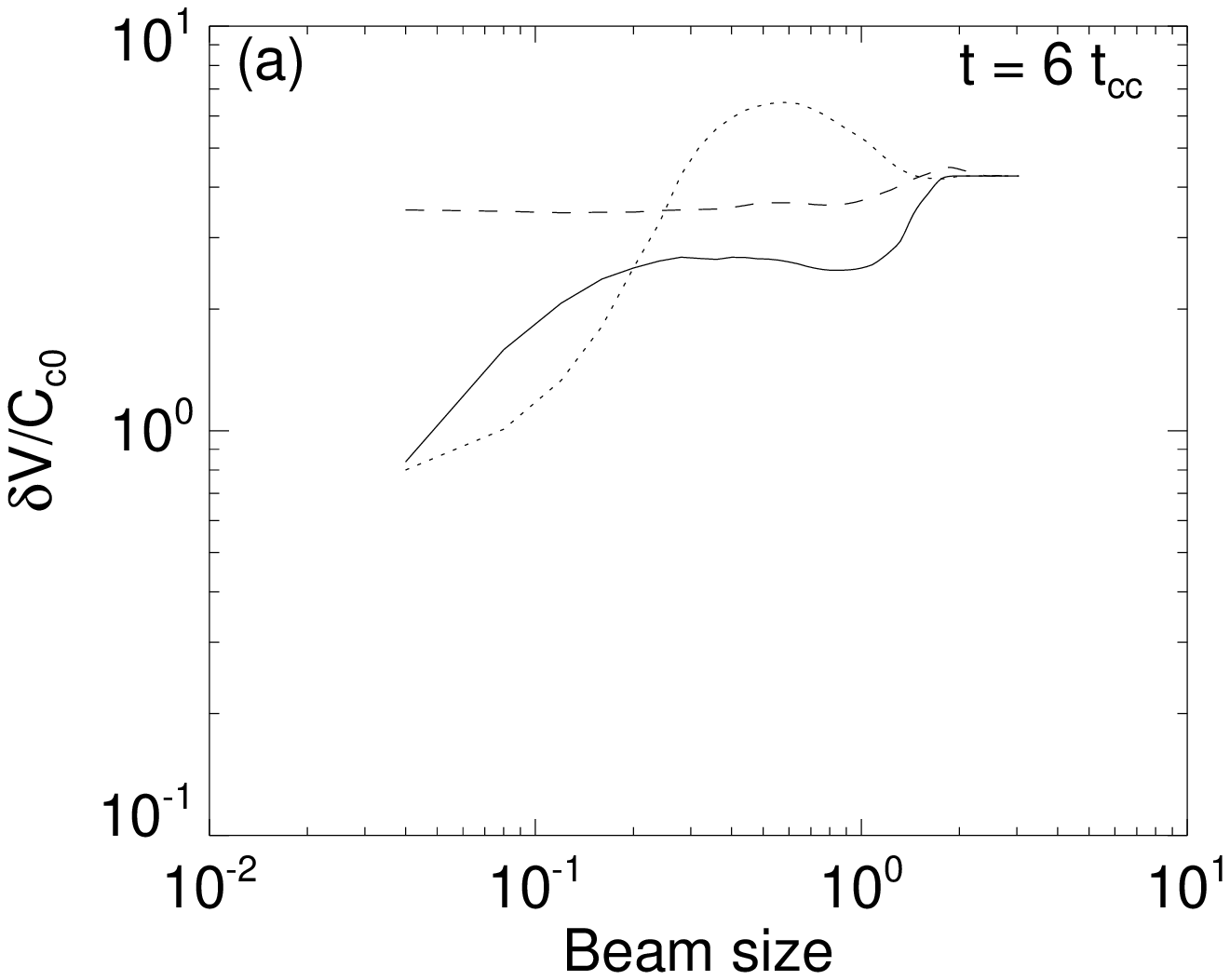}
\includegraphics[width=5cm]{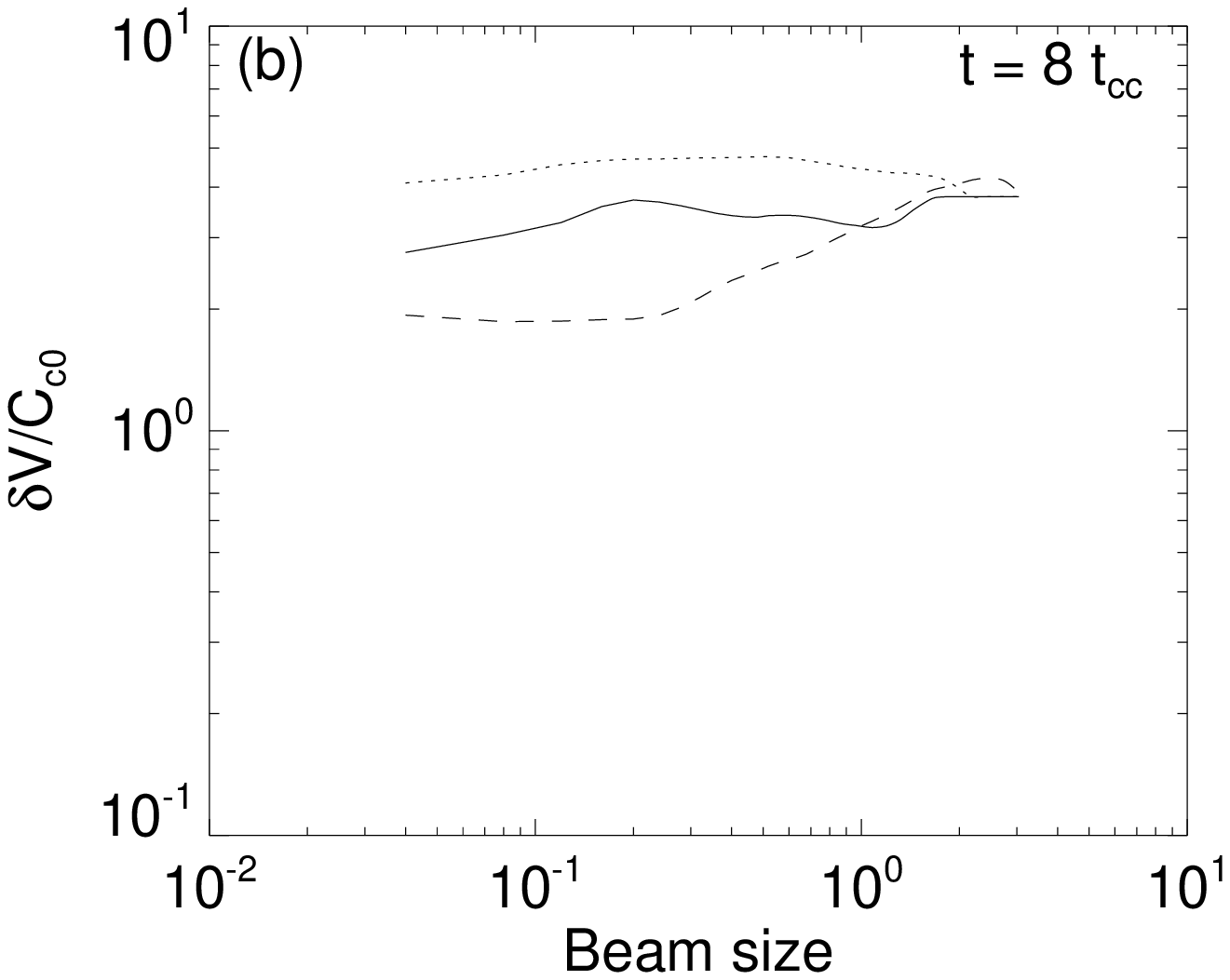}
\includegraphics[width=5cm]{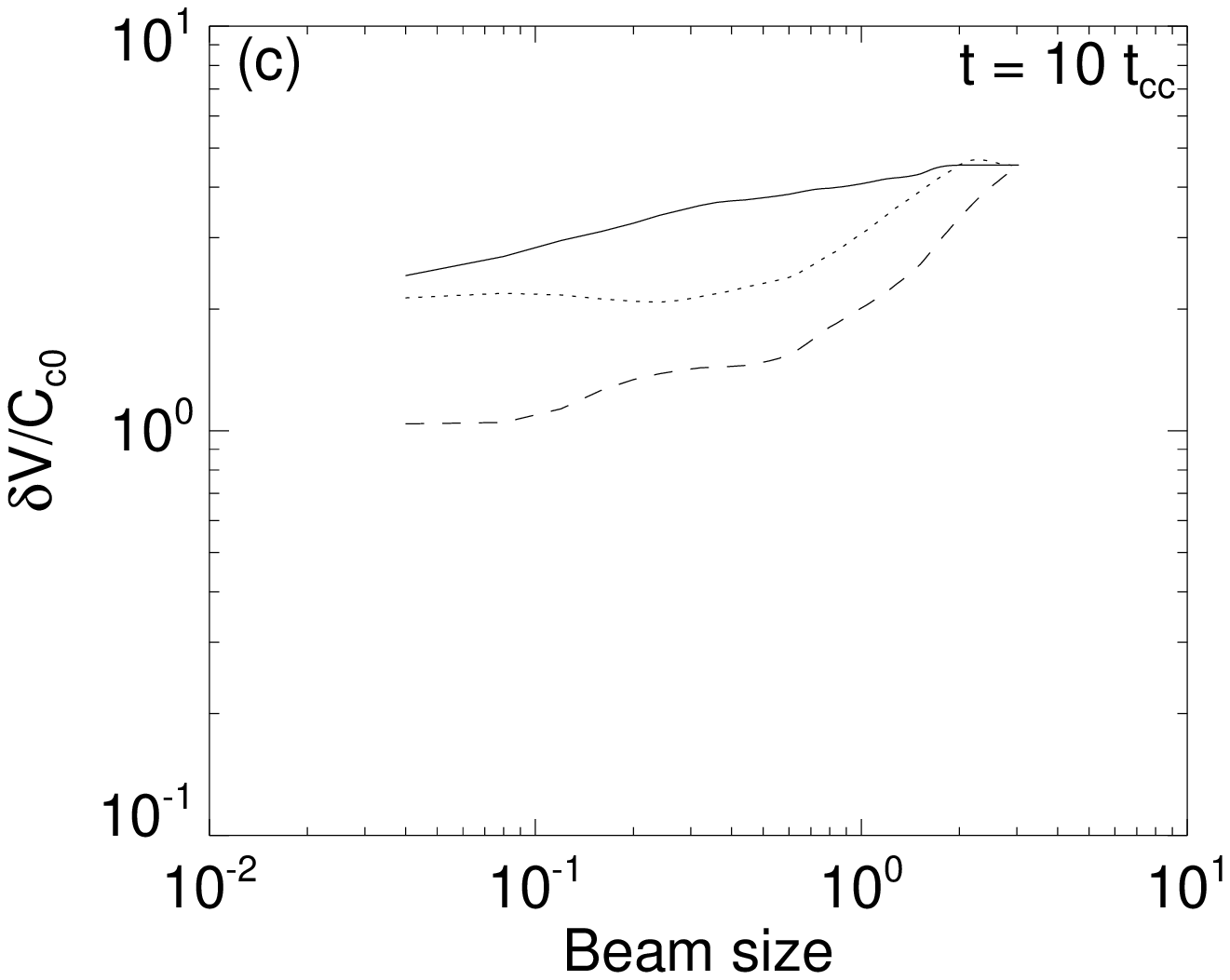}
\caption{linewidth-size relation in shocked clouds with
$(n,\chi, M)=(8,10,10)$ at three different times:
(a) $t=6 \, \tcc$, (b) $8 \, \tcc$, and (c) $10 \, \tcc$. 
This model is followed with the three-dimensional simulation
with an intermediate resolution $R_{60}$.
The abscissa denotes the beam size, while the ordinate 
is the velocity dispersion estimated in the beam with a size $R$.
The velocity dispersions are measured in 
units of the sound speed at the center of the preshock cloud, $C_{c0}$.
Solid curves show the velocity dispersions measured from the
direction whose line-of-sight is identical to the $z$-axis.
Dotted and dashed curves show the velocity dispersions measured from
the direction whose line-of-sight is perpendicular to the
$z$-axis.  For dotted and dashed curves, the beam center is put
on the center of gravity, $z_G$, and $z_G + r_{\rm co}$, respectively.
}
\label{fig:velocity-size}
\end{figure}

\end{document}